\documentclass[preprint]{aastex}
\usepackage{emulateapj5}
\usepackage{natbib}
\newcommand{\HI}{\ion{H}{1}}
\newcommand{\HII}{\ion{H}{2}}
\newcommand{\NII}{[\ion{N}{2}]}
\newcommand{\kms}{\mbox{km~s$^{-1}$}}
\newcommand{\Msol}{\mbox{M$_\odot$}}

\newcommand{\tgas}{\ensuremath{\tau_{\rm gas}}}
\newcommand{\mjybm}{\mbox{mJy~bm$^{-1}$}}
\newcommand{\jybm}{\mbox{Jy~bm$^{-1}$}}
\newcommand{\Tsyse}{\mbox{$T_{\rm sys}^*$}}
\newcommand{\siggas}{\mbox{$\Sigma_{\rm gas}$}}
\newcommand{\sigsfr}{\mbox{$\Sigma_{\rm SFR}$}}
\newcommand{\sighi}{\mbox{$\Sigma_{\rm HI}$}}
\newcommand{\sightwo}{\mbox{$\Sigma_{\rm H_2}$}}

\journalinfo{The Astrophysical Journal, in press}
\slugcomment{Received 2001 September 17; accepted 2001 December 5}

\shorttitle{Gas and Star Formation in Spiral Galaxies}

\begin{document}

\title{The Relationship Between Gas Content and Star Formation 
in Molecule-Rich Spiral Galaxies}

\author{Tony Wong\altaffilmark{1} and Leo Blitz}
\affil{Astronomy Department and Radio Astronomy Laboratory, 
University of California, Berkeley, CA 94720}
\altaffiltext{1}{Present address: CSIRO Australia Telescope 
National Facility, PO Box 76, Epping NSW 1710, Australia;
Tony.Wong@csiro.au}

\begin{abstract}

We investigate the relationship between \HI, H$_2$, and the star
formation rate (SFR) using azimuthally averaged data for seven
CO-bright spiral galaxies.  Contrary to some earlier studies based on
global fluxes, we find that \sigsfr\ exhibits a much stronger
correlation with \sightwo\ than with \sighi, as \sighi\ saturates at a
value of $\sim$ 10 \Msol\ pc$^{-2}$ or even declines for large
\sigsfr.  Hence the good correlation between \sigsfr\ and the total
(\HI+H$_2$) gas surface density \siggas\ is driven by the molecular
component in these galaxies, with the observed relation between
\sigsfr\ and \sightwo\ following a Schmidt-type law of index $n_{\rm
mol}$$\approx$0.8 if a uniform extinction correction is applied or
$n_{\rm mol}$$\approx$1.4 for a radially varying correction dependent
on gas density.  The corresponding Schmidt law indices for \siggas\
vs.\ \sigsfr\ are 1.1 and 1.7 for the two extinction models, in rough
agreement with previous studies of the disk-averaged star formation
law.  An alternative to the Schmidt law, in which the gas depletion
timescale is proportional to the orbital timescale, is also consistent
with the data if radially varying extinction corrections are applied.
We find no clear evidence for a link between the gravitational
instability parameter for the gas disk ($Q_g$) and the SFR, and
suggest that $Q_g$ be considered a measure of the gas fraction.  This
implies that for a state of marginal gravitational stability to exist
in galaxies with low gas fractions, it must be enforced by the stellar
component.  In regions where we have both \HI\ and CO measurements,
the ratio of \HI\ to H$_2$ surface density scales with radius as
roughly $R^{1.5}$, and we suggest that the balance between \HI\ and
H$_2$ is determined primarily by the midplane interstellar pressure.
These results favor a ``law'' of star formation in quiescent disks in
which the ambient pressure and metallicity control the formation of
molecular clouds from \HI, with star formation then occurring at a
roughly constant rate per unit H$_2$ mass.

\end{abstract}



\section{Introduction}

Understanding the factors which control the star formation rate (SFR)
is of fundamental importance to studies of the interstellar medium
(ISM) and galaxy evolution.  The most important factor, of course, is
the availability of cold neutral gas, especially molecular gas.  Over
the past several decades, observations of the CO(1--0) line have
established a clear link between star formation in the Galaxy and
molecular gas \citep[e.g.,][]{Loren:73}, and even in environments
where star formation is observed without CO emission, such as dwarf
irregular galaxies, a plausible case can be made that molecular clouds
are weak CO emitters due to low metal abundances \citep{Wilson:95,Taylor:98}.  
On the other hand, \citet{KC:89} has shown that the disk-averaged
\sigsfr\ is much better correlated with \sighi\ than with \sightwo\
inferred from CO observations, a surprising result given the very
different radial profiles of \HI\ and H$\alpha$ emission in galaxies.
Similar conclusions have been reached by other studies based on global
averages \citep{Deharveng:94,Boselli:94,Casoli:96}.  Thus, it is not
obvious whether the \HI\ or H$_2$ (as traced by CO) content will be a
better predictor of the SFR on large scales.  \citet{KC:89} argues that
it is the {\it total} (\HI+H$_2$) gas surface density which shows the
best correlation with the SFR per unit area.

A quantitative prescription for the SFR has been provided by
\citeauthor{KC:98a} (1998a, hereafter K98), who finds that the total gas
surface density, averaged over the optical disk, is related to the SFR
surface density by a \citet{Schmidt:59} law,
\begin{equation}
\sigsfr \propto (\siggas)^n\;.
\label{eqn:schmidt}
\end{equation}
where $n \approx 1.4$.  Kennicutt's formulation of the star formation
law has been widely applied in ``semi-empirical'' models of galaxy
evolution \citep[e.g.,][]{Boissier:99,Tan:99,vdB:00}.  While the
Schmidt law appears to be valid at high gas surface densities,
\citet{KC:89} and \citet{Martin:01} have presented evidence for a
threshold density for star formation based on the gravitational
instability parameter $Q$.  Below the threshold density, which
observationally is of order 5--10 \Msol\ pc$^{-2}$, the Schmidt law
breaks down and star formation is strongly suppressed.  The general
applicability of a $Q$ threshold, however, remains controversial
\citep{Thornley:95,Hunter:98,Ferguson:98b}.


\begin{table*}[b]
\begin{center}
\caption{Properties of the Sample Galaxies\label{tbl:props}}
\bigskip
\begin{tabular}{cccccccc}
\hline\hline
Name & RA (2000) & DEC & Morph. & $B_T^0$\tablenotemark{(a)} & 
$R_{25}$\tablenotemark{(b)} & $V_{\odot}$\tablenotemark{(c)} & Dist. \\
& hh:mm:ss & dd:mm:ss & Type & (mag) & (arcsec) & (km/s) & (Mpc) \\[0.5ex]
\hline\hline
NGC 4321 & 12:22:54.9 & 15:49:21 & SAB(s)bc & 9.98 & 220 & 1571 
	& 16\tablenotemark{(d)}\\
NGC 4414 & 12:26:27.1 & 31:13:24 & SA(rs)c? & 10.62 & 110 & 716 
	& 19\tablenotemark{(e)}\\
NGC 4501 & 12:31:59.2 & 14:25:14 & SA(rs)b  & 9.86 & 210 & 2281 
	& 16\tablenotemark{(d)}\\
NGC 4736 & 12:50:53.1 & 41:07:14 & (R)SA(r)ab & 8.75 & 340 & 308 
	& 4.2\tablenotemark{(f)}\\
NGC 5033 & 13:13:27.5 & 36:35:38 & SA(s)c  & 10.21 & 320 & 875 
	& 17.5\tablenotemark{(g)}\\
NGC 5055 & 13:15:49.3 & 42:01:45 & SA(rs)bc & 9.03  & 380 & 504 
	& 10\tablenotemark{(g)}\\
NGC 5457 & 14:03:12.5 & 54:20:55 & SAB(rs)cd & 8.21 & 870 & 241 
	& 7.4\tablenotemark{(h)}\\
\hline\hline
\end{tabular}
\end{center}
\tablenotetext{a}{Total blue magnitude, corrected for Galactic and
internal absorption and redshift, from RC3 \citep{RC3}.}
\tablenotetext{b}{Semimajor axis at 25.0 mag arcsec$^{-2}$, from RC3.}
\tablenotetext{c}{Heliocentric velocity, from NED.}
\tablenotetext{d}{Cepheid distance to NGC 4321 from \citet{Ferrarese:96}
adopted for Virgo cluster.}
\tablenotetext{e}{Cepheid distance from \citet{TurnerA:98}.}
\tablenotetext{f}{Redshift distance using $H_0$ = 75 \kms\ Mpc$^{-1}$.}
\tablenotetext{g}{Distance from Mark III data \citep{Willick:97} 
using $H_0$=75.}
\tablenotetext{h}{Cepheid distance from \citet{Kelson:96}.}
\end{table*}


As noted by K98, an alternative formulation of the star formation law,
where the star formation timescale is proportional to the orbital
timescale,
\begin{equation}
\sigsfr \propto \siggas \Omega\;,
\end{equation}
is also consistent with the disk-averaged data.  Such a law is
predicted by models in which spiral arms play an important role in
triggering star formation \citep{Wyse:89}, or in which star formation
is self-regulated to yield a constant value of $Q$ \citep{Larson:88,Silk:97}.
Distinguishing between this law and a Schmidt law requires {\it
spatially resolved} measurements of gas densities and orbital
timescales (i.e., rotation curves).

A third prescription for the SFR, arising from CO studies by J. Young
and collaborators \citep{Devereux:91,Young:96,Rownd:99}, states that
the SFR is roughly proportional to the mass of molecular gas, so that
their ratio (commonly referred to as the {\it star formation
efficiency}, SFE) is roughly constant.  In particular, \citet{Rownd:99} 
find that the SFE within the disks of spiral galaxies exhibits no
strong radial gradients, under the assumption of a constant
CO-to-H$_2$ conversion factor.  An SFE that is independent of
galactocentric radius is also consistent with studies of molecular
clouds in our Galaxy \citep{Mead:90,Wouter:88}.  If star formation in
disk galaxies can be generically decomposed into two processes, the
formation of giant molecular clouds (GMCs) from \HI\ clouds and the
formation of stars within GMCs \citep[e.g.,][]{Tacconi:86}, then the
latter process occurs at a rate that is largely independent of
location within a galaxy, and may well be universal in all disk
galaxies.

Previous studies of the star formation law in external galaxies have
tended to concentrate on large samples observed at the low resolution
provided by single-dish radio telescopes.  Although such samples are
valuable for characterizing the global properties of galaxies,
detailed study of individual objects at high spatial resolution is
essential for understanding the physical basis of the star formation
law and reconciling the various empirical descriptions.  In this paper
we undertake such an investigation, comparing recent CO data from the
BIMA Survey of Nearby Galaxies (BIMA SONG) with previously published
\HI\ and H$\alpha$ data.  Our high-resolution CO maps include
single-dish data, which is essential when accurate flux measurements
are required for extended sources.  We employ azimuthally averaged
radial profiles in order to improve the signal-to-noise ratio and
average over time-dependent effects.

The organization of this paper is as follows.  In \S\ref{obs} we
describe our galaxy sample, observing parameters, and data reduction
procedures.  \S\ref{analysis} describes how we produced radial
profiles from the intensity images and our corrections for H$\alpha$
extinction.  \S\ref{schmidt} presents our results for the relation
between \sigsfr\ and \siggas, as well as the relations between
\sigsfr\ and \sighi\ and \sightwo\ individually.  In \S\ref{crit} we
compare \siggas\ with the threshold surface density discussed by
\citet{KC:89}, and in \S\ref{hitoh2} we investigate the relationship
between the radial profiles of atomic and molecular gas.  In
\S\ref{disc} we show how our results can be interpreted in terms of a
nearly constant star formation efficiency for molecular gas,
with molecular cloud formation in turn linked to the average ISM
pressure.  Finally, in \S\ref{conc} we summarize our results.  In
subsequent papers of this series, we will analyze the CO and \HI\
kinematics of the galaxies in our sample for evidence of radial
inflows (Wong, Blitz, \& Bosma 2002, Paper II) and evaluate whether
the observed gas depletion times and oxygen abundances are consistent
with closed-box evolution (Wong \& Blitz 2002, Paper III).


\section{Observations and Data Reduction}\label{obs}

\subsection{Sample Description}

Most of the new observations presented in this paper were taken as
part of the BIMA Survey of Nearby Galaxies (BIMA SONG)
\citep{Regan:01}, an interferometric CO survey of 44 spiral galaxies
selected primarily by optical luminosity (absolute $B$ magnitude $M_B
< 11$).  Six of these galaxies, along with NGC 4501, were chosen for
this study.  All exhibit strong CO emission at a range of radii and
have available \HI\ maps from the Very Large Array (VLA) at the
resolution of the C configuration ($\sim$20\arcsec) or better.  The
insistence on high-resolution \HI\ data is essential for the CO and
\HI\ data to be used in a complementary way; although \HI\ maps at
even higher ($\sim$6\arcsec) resolution would be ideal, few such maps
are available because of the large investment of observing time needed
to obtain good sensitivity at those resolutions.  The galaxies span a
range of morphological type, from Sab to Scd.  While none are involved
in strong interactions with other galaxies, NGC 4321 (M100) and NGC
4501 are members of the Virgo Cluster and presumably interacting with
the intracluster medium.  The basic properties of the sample are
summarized in Table~\ref{tbl:props}.  The tabulated coordinates are
for the center of the galaxy as given by the NASA Extragalactic
Database (NED), except for NGC~5055 where we have adopted the peak of
the astrometrized $K$-band image of \citet{Thornley:97a}.


\vskip 0.25truein
\includegraphics[width=3.25in]{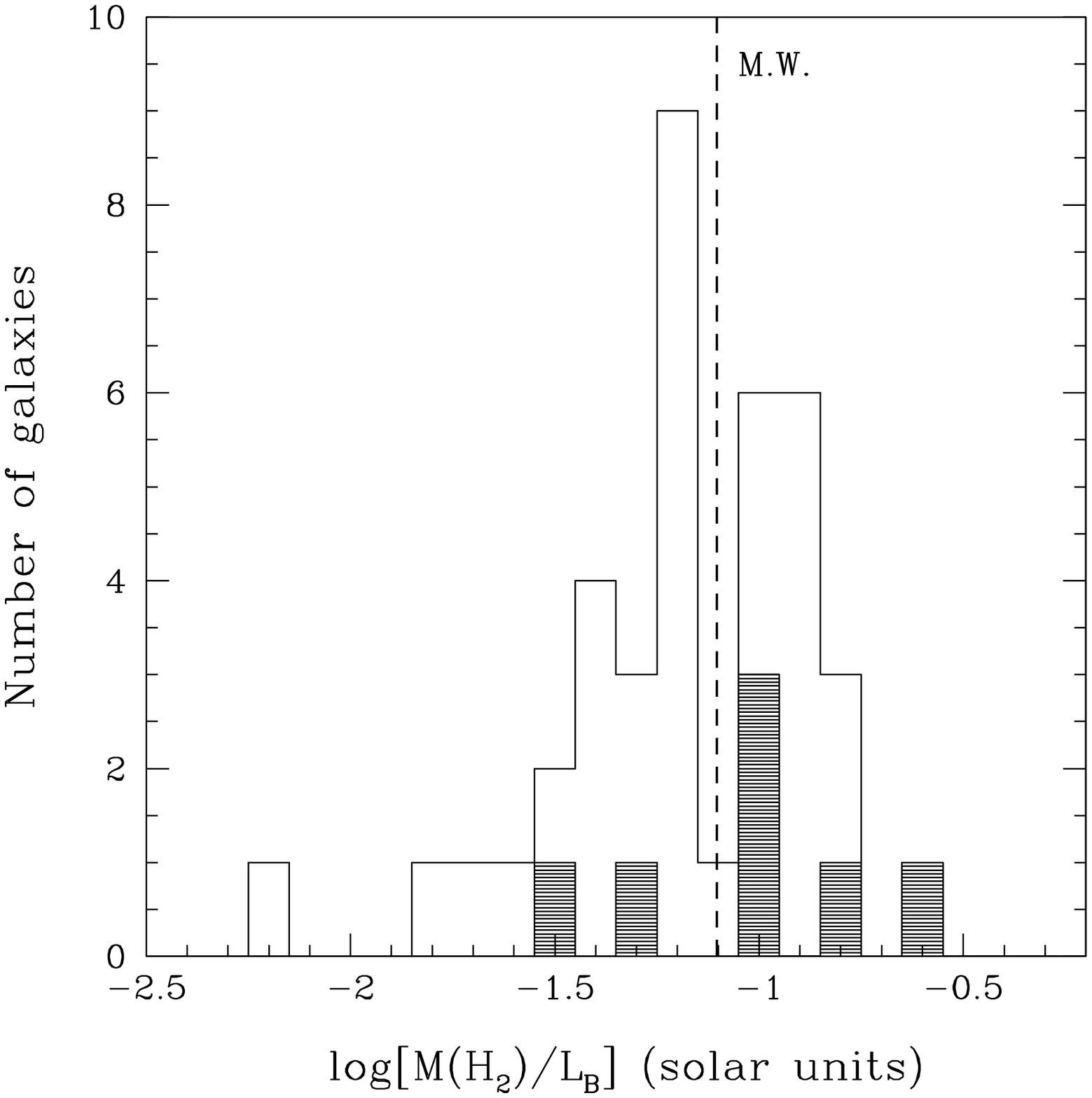}
\figcaption{
Histogram of H$_2$ masses normalized by $B$ luminosity, using data
from \citet{Young:95} and a constant $X$-factor, for our sample ({\it
shaded}) and 40 of the 44 galaxies in the optically selected BIMA SONG
sample.  The remaining 4 galaxies were not observed by \citet{Young:95} 
and probably have below-average CO fluxes.  For NGC 5457 (second
shaded box from left) we have used the CO flux measured by
\citet{Kenney:91a}, which is probably an underestimate since their map
is incomplete.  The vertical dashed line is for the Milky Way.
\label{fig:fluxhist}}
\vskip 0.25truein


Based on single-dish measurements from the Five College Radio
Astronomy Observatory (FCRAO) Extragalactic CO Survey
\citep{Young:95}, the median CO flux for the galaxies in our sample is
2760 Jy \kms, compared to a median of 1640 Jy \kms\ for the entire
SONG sample.  To determine if the difference results from our choice of
more optically luminous galaxies, Figure~\ref{fig:fluxhist} shows a
histogram of the ratio of H$_2$ mass to $B$-band luminosity, as
tabulated by \citet{Young:95}, for the galaxies in our sample (shaded)
as well as for the entire SONG sample.  We have scaled the CO fluxes
derived by \citet{Young:95} to H$_2$ masses using a conversion factor
appropriate for the Galaxy:
\begin{equation}
N_{\rm H_2} = 2 \times 10^{20}
        \left(\frac{I_{\rm CO}}{\rm K\;\kms}\right) \rm cm^{-2}\;,
\label{eqn:cotoh2}
\end{equation}
\citep{Strong:96,Dame:01}.  The median for the entire SONG sample
($-1.17$ dex in solar units) is somewhat less than the value 
for the Milky Way ($-1.1$ dex,
based on \citealt{Dame:93b} and \citealt{vdK:90}), which in turn is
less than the median for our sample ($-1.0$ dex).  Note that a difference
of 0.17 dex corresponds to a factor of 1.5.  Despite considerable
uncertainty in the FCRAO fluxes, since they are estimated from
major axis scans rather than complete two-dimensional maps, we
conclude that the galaxies in our sample are somewhat overluminous in
CO (relative to their optical luminosities) compared to bright
galaxies in general.  This is not surprising, given that we explicitly
required that CO be observed over a large portion of the galaxy disk.
Thus, one should bear in mind that properties of the ISM in the
galaxies studied here may not be representative of the field galaxy
population, or even the population of optically luminous galaxies.


\subsection{BIMA CO Observations}\label{obsbima}

Observations with the BIMA\footnote{The Berkeley-Illinois-Maryland
Association is funded in part by the National Science Foundation.}
interferometer at Hat Creek, California were conducted primarily in
the C array configuration, which gives a synthesized beam of
$\sim$6\arcsec\ FWHM, with a few additional observations (``tracks'')
taken in the D array configuration (15\arcsec\ FWHM).  The synthesized
beam gives the resolution of the resulting maps, and is determined by
the maximum spacing of the antennas: in C-array the spacings range
from 7 to 90 m (2.7--35 k$\lambda$), while in D-array the spacings
range from 7 to 30 m (2.7--11 k$\lambda$).  Several of the C-array
observations of NGC 4501, 4736, and 5033 were made in 1996 and 1997
when the BIMA array consisted of nine antennas; all other observations
were made in 1998 and 1999 with ten antennas.

In our standard frequency setup, the receivers were tuned to the CO
($J=1\rightarrow 0$) transition at 115.2712 GHz ($\lambda$=2.6 mm),
with the correlator configured to have 4 independently positioned
spectral windows, each with 100 MHz bandwidth and 64 channels.  The
spectral windows were placed side-by-side in frequency with an overlap
of $\approx$10 MHz (6 channels).  The resulting velocity range covered
was $\approx$950 \kms\ at a resolution of 4.06 \kms.  Data from the
lower sideband of the first local oscillator, at a frequency of 112
GHz, were also recorded, but no continuum emission was detected in any
of the sample galaxies.


\begin{table*}[t]
\begin{center}
\caption{Properties of the BIMA CO maps\label{tbl:bimamaps}}
\bigskip
\begin{tabular}{ccccccc}
\hline\hline
Galaxy & Weighting  & Resolution & PA\tablenotemark{(a)} & 
	$\sigma_{ch}$\tablenotemark{(b)} & 
	$R_{\rm max}$\tablenotemark{(c)}\\ 
& & (\arcsec\ $\times$ \arcsec\ $\times$ \kms) & (\arcdeg) &
	(mJy/bm) & (\arcsec)\\[0.5ex]
\hline\hline

NGC 4321 & natural & 8.85 $\times$ 5.62 $\times$ 10 
	& \phs\phn4.7 & 49 & 94\\
& 20\arcsec\ taper & 20 $\times$ 20 $\times$     10 
	& \phs\phn0.0 & 97 \\[0.25ex] \hline
NGC 4414 & robust & 6.53 $\times$ 4.88 $\times$  10 
	& \phs\phn8.2 & 42 & 50\\ 
& 13\arcsec\ taper & 17.5 $\times$ 15.3 $\times$ 10 
	& \phs72.1 & 64 & 94\\[0.25ex] \hline 
NGC 4501 & natural & 7.89 $\times$ 6.23 $\times$ 20 
	& $-80.9$ & 56 & 100\\
& 22\arcsec\ taper & 23.2 $\times$ 22.3 $\times$ 10 
	& \phs24.6 & 174 \\[0.25ex] \hline
NGC 4736 & robust & 6.86 $\times$ 5.02 $\times$  10 
	& \phs62.5 & 69 & 100\\
& 13\arcsec\ taper & 15 $\times$ 15 $\times$     10 
	& \phs\phn0.0 & 83 \\[0.25ex] \hline
NGC 5033 & natural & 7.52 $\times$ 6.49 $\times$ 20 
	& \phs82.8 & 52 & 94\\
& 18\arcsec\ taper & 19.9 $\times$ 17.0 $\times$ 20 
	& \phs89.1 & 108 \\[0.25ex] \hline
NGC 5055 & robust & 5.69 $\times$ 5.30 $\times$  10 
	& \phs13.9 & 83 & 94\\
& 12\arcsec\ taper & 12.9 $\times$ 12.8 $\times$ 10 
	& $-83.5$ & 107 \\[0.25ex] \hline
NGC 5457 & natural & 6.88 $\times$ 6.36 $\times$ 10
	& \phs72.5 & 43 & 94\\
& 15\arcsec\ taper & 15 $\times$ 15 $\times$     10 
	& \phs\phn0.0 & 80\\
\hline\hline

\end{tabular}
\end{center}
\tablenotetext{a}{Position angle of beam measured from N towards E.}
\tablenotetext{b}{RMS noise in a single emission-free channel measured 
across the inner 2\arcmin\ $\times$ 2\arcmin, in \mjybm.}
\tablenotetext{c}{Radius at which half-power point of the outermost 
pointing is reached.}
\end{table*}


At $\lambda$=2.6 mm the BIMA antennas have a primary beam FWHM of
100\arcsec, which defines the effective field of view.  During most
observing sessions a 7-point hexagonal grid centered on the galaxy
nucleus was observed, with the points separated by 50\arcsec\ (for NGC
4501 and 4736) or 44\arcsec\ (for the remaining galaxies).  In order
to maintain roughly equal sampling in the visibility plane for each
pointing center, the entire grid was observed in 7--9 minutes and then
repeated.  For a few tracks on NGC 4414, 4736, and 5033, only the
central pointing was observed.  In cases where the same galaxy was
observed as both a single pointing and a mosaic, the tracks were
imaged together as a mosaic---with the exception of NGC 4414, where
only the central pointing was used for the high-resolution map.  The
resulting values of $R_{\rm max}$ for each datacube, defined as the radius
at which one reaches the half-power point of the outermost pointing,
are given in Table~\ref{tbl:bimamaps}.


\subsubsection{Assessment of Data Quality}

The two principal factors which can affect the data quality from a
millimeter-wave interferometer are the tropospheric water vapor
column, which increases the atmospheric opacity, and fluctuations in
that column, which lead to variations in the refractive index of the
atmosphere (``seeing'') and hence variations in the phase of the
signal measured at the telescope.  Changes in atmospheric opacity are
reflected in the ``effective'' system temperature \Tsyse, which ranged
from 300--600 K during our observations, depending on the source
elevation and weather conditions.  Atmospheric phase noise can be
quantified by a frequency-independent quantity $\sigma_{path}$,
representing the RMS fluctuation in the difference in effective path
length through the atmosphere for two parallel sight lines separated
by a given distance \citep[e.g.,][]{Wright:96}.  Since mid-1998,
$\sigma_{path}$ has been measured every 10 minutes at Hat Creek by a
dedicated phase monitor, which records signals from a television
satellite using a pair of commercial satellite dishes separated by
100~m \citep{Lay:99}.  These measurements yield a fiducial
$\sigma_{path, 0}$, which was typically 200--400 $\mu$m for the
C-array observations and 800--1200 $\mu$m for the D-array observations
(the latter occurring mostly in summer).  Note that with the RMS phase
given by $\phi_{rms} = 2\pi\sigma_{path}/\lambda$, the visibility
amplitude after time-averaging is given by
\begin{equation}
\left<V\right> = V_0\, e^{-(\phi_{rms})^2/2}\,.
\end{equation}
Hence a value of $\sigma_{path, 0} = 560$ $\mu$m corresponds to
a 50\% decorrelation (i.e. loss of visibility amplitude) at $\lambda =
3$ mm for a 100~m baseline \citep{Akeson:98}, roughly the longest
baseline in the C-array.  Although the measured $\sigma_{path, 0}$ was
typically worse than this in D-array, the actual phase fluctuations
that affect the short baselines of D-array ($b \ll 100$ m) are much
smaller since $\sigma_{path} \propto b^{5/6}$ \citep{Wright:96}.


\subsubsection{Data Calibration}

The data were calibrated and reduced following the procedures
described in \citet{Regan:01}.  We give only a summary here.  An
online linelength calibration was applied to correct for variations in
the electrical path length from the antenna to the correlator.
Remaining drifts in the antenna phase gains were then corrected by
observing a nearby (within $\sim$15\arcdeg) quasar every half hour and
fitting a low-order (quadratic or cubic) polynomial to the
antenna-based self-calibration solutions.  Mars was used for amplitude
calibration except in 1998 April, when the quasar 3C273 was used.  For
all tracks both amplitude and phase gains were assumed to be
frequency-independent; this approximation was found to be adequate in
several cases by observing 3C273 using the same correlator setup as
for the source observations.

For tracks during which $\sigma_{path}$ was measured, an additional
calibration step was performed in order to compensate for the effects
of phase decorrelation on the maps.  In this procedure, implemented in
a MIRIAD task called UVDECOR written by M.\ Regan, the visibility
amplitudes for all datasets are scaled up by a baseline-dependent
correction factor, typically $\sim$1--1.2, based on the expected
amplitude loss due to phase decorrelation \citep[see][]{Regan:01}.  We
found that applying UVDECOR led to a $\sim$10\% increase in the fluxes
of the final maps.  Overall, we estimate that the flux scale of the
resulting BIMA maps is accurate to within $\sim$20\%.

Following the initial calibration, the visibilities were subjected to
three iterations of phase-only self-calibration, which in all cases
led to convergence.  Channel maps were then produced for each galaxy
by assigning weights to the calibrated visibilities as described
below, transforming to the image plane after binning to 10 or 20 \kms\
channels (to improve signal-to-noise), and deconvolving using a
variant of the CLEAN algorithm developed by \citet{Steer:84} to
improve the deconvolution of spatially extended emission.  In general,
assigning greater weight to short-baseline data results in lower
resolution but higher signal-to-noise for extended sources, as well as
reduced systematic errors because short baselines are less subject to
atmospheric phase noise.  Thus, for all galaxies we generated a
low-resolution map by applying a Gaussian taper in the visibility
plane, chosen to match the resolution of the galaxy's \HI\ cube.  In
addition, a high-resolution map was generated using either natural or
robust weighting.  Natural weighting provides the best point-source
sensitivity and was used for sources with poorer signal-to-noise
(NGC~4321, 4501, 5033, and 5457).  Robust weighting, which provides
poorer sensitivity but somewhat higher resolution, was used for
NGC~4414, 4736, and 5055.  Table~\ref{tbl:bimamaps} summarizes the
properties of the final deconvolved maps.


\subsection{Kitt Peak CO Observations}\label{obskp}

Single-dish CO observations were performed at the NRAO\footnote{The
National Radio Astronomy Observatory is a facility of the National
Science Foundation, operated under cooperative agreement by Associated
Universities, Inc.} 12m telescope on Kitt Peak, Arizona.  We refer to
these data hereafter as the ``KP data.''  At $\lambda$=2.6~mm the
half-power beamwidth (HPBW) of the telescope is 55\arcsec.  All
galaxies with the exception of NGC~4501 were observed between 1998
April and 2000 January using the On-The-Fly (OTF) technique, in which
the telescope is scanned rapidly across the sky and spectra are
written out at a very high rate (every 0.1~s).  The relative pointing
of the telescope during the map is also continuously measured,
although the absolute pointing must still be checked and adjusted at
regular intervals (once every 1--2 hr) by observing a bright continuum
source (typically Mars).  Each map covered a field roughly 6\arcmin\
$\times$ 6\arcmin\ in size and was completed in $\lesssim$15 minutes;
roughly 20--40 maps were obtained per galaxy and later co-added during
data reduction.  NGC~4501 was observed in 1997 June in
position-switched (hereafter PS) mode, where spectra are taken at
specified grid points spaced by half a beamwidth or less.  The
integration time was typically 4--8 minutes per pointing, with a total
of 27 pointings observed covering a region roughly 2\arcmin\ $\times$
2\arcmin\ in size.

The telescope was configured to observe orthogonal polarizations using
two 256-channel filterbanks (one for each polarization) at a spectral
resolution of 2 MHz (5.2 \kms).  Calibration was performed using the
chopper wheel method every 5--15 minutes; typical values of \Tsyse\
were 300--400 K.  Spectra were output on the $T_R^*$ scale, which
represents the source antenna temperature corrected for ohmic losses,
atmospheric absorption, spillover, and scattering \citep{Kutner:81}.
Background emission was measured by observing an off-source position
10\arcmin\ from the pointing center (for the PS observations) or by
averaging spectra at the edges of the OTF maps; the off-source
spectrum nearest in time was then used for background subtraction.

Data reduction was performed in the COMB software package for the PS
spectra and the AIPS and MIRIAD packages for the OTF spectra.  A
linear baseline was fitted to the signal-free channels of each
spectrum and subtracted.  FITS datacubes were then generated using a
cell size of 13\farcs5 for the PS map and 18\arcsec\ for the OTF maps.
For each galaxy, the RMS noise in the final maps ranged from 14--28 mK
in a 2 MHz channel.  The maps were converted into Jy units using a
standard telescope gain of 33 Jy~$({\rm K}[T_R^*])^{-1}$.

Prior to combination with the BIMA maps, OTF maps from different dates
were cross-correlated against each other to correct for pointing
offsets, as described in \citet{Regan:01}.  The derived offsets were
generally around 10\arcsec\ but ranged as high as 27\arcsec\ (half a
beam width) in one case.  After this internal registration of the OTF maps,
a final cross-correlation with the BIMA map (smoothed to 55\arcsec)
was performed, under the assumption that the flux resolved out by BIMA
did not significantly affect the centroid of the CO emission.  The
resulting corrections amounted to 5\arcsec--11\arcsec.  For
NGC~5457, which has a low inclination and a large angular diameter,
this assumption was found to be inappropriate due to the large amount
of flux missing from the BIMA map, and the final cross-correlation was
not performed.


\begin{table*}[t]
\begin{center}
\caption{Properties of the VLA HI Maps\label{tbl:vlaobs}}
\bigskip
\begin{tabular}{ccccl}
\hline\hline
Galaxy & Resolution & PA\tablenotemark{(a)} & 
	$\sigma_{ch}$\tablenotemark{(b)} & Reference\\ 
& (\arcsec\ $\times$ \arcsec\ $\times$ \kms) & (\arcdeg) &
	(mJy/bm)\\[0.5ex]
\hline\hline

NGC 4321 & 20 $\times$ 20 $\times$ 20.6 & \phs\phn0.0 & 0.75 
	& \citet{Knapen:93}\\
NGC 4414 & 17.5 $\times$ 15.3 $\times$ 10.3 & \phs72.1 & 0.4 
	& \citet{Thornley:97b}\\ 
NGC 4501 & 23.2 $\times$ 22.3 $\times$ 10.3 & \phs24.6 & 0.6 
	& A. Bosma (Paper II)\\
NGC 4736 & 15 $\times$ 15 $\times$ 5.15 & \phs\phn0.0 & 1.8
	& \citet{Braun:95}\\
NGC 5033 & 19.9 $\times$ 17.0 $\times$ 20.6 & \phs89.1 & 0.3 
	& \citet{Thean:97}\\
NGC 5055 & 12.9 $\times$ 12.8 $\times$ 10.3 & $-83.5$ & 0.2 
	& \citet{Thornley:97a}\\
NGC 5457 & 15 $\times$ 15 $\times$ 5.15 & \phs\phn0.0 & 1.8
	& \citet{Braun:95}\\
\hline\hline

\end{tabular}
\end{center}
\tablenotetext{a}{Position angle of beam measured from N towards E.}
\tablenotetext{b}{RMS noise in a single emission-free channel measured 
across the inner 4\arcmin\ $\times$ 4\arcmin, in \mjybm.}
\end{table*}



\begin{table*}[b]
\begin{center}
\caption{Sources for H$\alpha$ Images\label{tbl:haobs}}
\bigskip
\begin{tabular}{ccccl}
\hline\hline
Galaxy & Image & $-\log f_{\rm H\alpha}$\tablenotemark{(a)} & 
Credit & Photometry\\
& Size & & & Reference\\ [0.5ex]
\hline\hline
NGC 4321 & 11\arcmin & 11.28 & M. Regan, K. Sheth & \citet{Massey:88}\\
NGC 4414 & 11\arcmin & 11.38 & \citet{Thornley:97b} & \citet{Thornley:97b}\\
NGC 4501 & 6\arcmin & 11.14 & \citet{Martin:01} & \citet{Young:96}\\
NGC 4736 & 6\arcmin & 10.80 & \citet{Martin:01} & \citet{KK:83}\\
NGC 5033 & 12\arcmin & 11.29 & M. Regan & \citet{KK:83}\\
NGC 5055 & 11\arcmin & 10.74 & M. Regan, K. Sheth & \citet{Massey:88}\\
NGC 5457 & 20\arcmin & 10.18 & \citet{vanZee:98} & \citet{KC:88}\\
\hline\hline
\end{tabular}
\end{center}
\tablenotetext{a}{$f_{\rm H\alpha}$ is the total flux in the image,
given in erg cm$^{-2}$ s$^{-1}$.}
\end{table*}



\subsection{CO Data Combination}\label{comb}

The relative calibration between two millimeter telescopes is always
somewhat uncertain, so when combining single-dish and array data one
typically attempts to solve for a relative calibration factor, using
data from both telescopes that have been filtered to the same range of
spatial frequencies.  The factor needed to scale the KP data to match
the BIMA data was calculated for each galaxy using the MIRIAD task
IMMERGE and ranged from 1 to 1.3.  In the case of NGC~4501, a reliable
calibration factor could not be determined since the field of view of
the KP observations was not large enough, so a factor of 1 was
adopted.  We note that it is quite possible that the BIMA data should
be scaled down rather than the KP data scaled up; this is especially
likely if the effect of decorrelation has been overestimated when
applying UVDECOR.  For simplicity, we assume here that the BIMA data
have the ``correct'' flux scale and emphasize that the CO fluxes we
measure may be slightly overestimated.  Given the uncertainties in
converting CO brightness to H$_2$ column density, an overall scaling
factor of $\lesssim$1.3 is probably unimportant.

We examined two methods of combining the BIMA and KP data, one which
generates a weighted sum of the two cubes before deconvolution (the
``dirty-map'' combination method) and another which folds in the KP
data after the BIMA map has been deconvolved, by partially replacing
the inner part of the Fourier plane with the KP data.  The two methods
gave similar results to within the noise.  We adopted the latter
method, as implemented in the IMMERGE task.  The resulting datacubes
are sensitive to structures on all spatial scales larger than
$\sim$6\arcsec.  A more detailed discussion of issues surrounding the
combination of single-dish and interferometer data can be found in
Appendix B of \citet{Wong:th}.


\subsection{HI Data}\label{obshi}

The \HI\ data used in this study were obtained at the NRAO Very Large
Array (VLA) by other researchers, who kindly provided us with reduced
datacubes in FITS format.  The properties of these cubes and
references to the original papers in which they were presented are
given in Table~\ref{tbl:vlaobs}.  (Details of the NGC 4501
observations will be presented in Paper II.)  At $\lambda$=21~cm the
HPBW of the VLA is 32\arcmin.  All galaxies were observed in the C
configuration, which gives a synthesized beam of $\sim$13\arcsec\ for
uniform weighting and $\sim$20\arcsec\ for natural weighting; three of
the galaxies (NGC~4736, 5055, and 5457) were also observed at higher
resolution in the B configuration.  Data obtained in the D
configuration are included for all galaxies except NGC~4501 and 5033,
so that the resulting maps are sensitive to even very extended
($\sim$10\arcmin) emission.  The datacubes were imported into MIRIAD
and transformed to the spatial and velocity coordinate frames of the
BIMA cubes.


\subsection{H$\alpha$ Images}\label{obsha}

Continuum-subtracted H$\alpha$+\NII\ CCD images were obtained from a
variety of sources, as listed in Table~\ref{tbl:haobs}.  All images
were originally taken at the Kitt Peak National Observatory (KPNO) 
0.9-m telescope, except for NGC 5033 (Palomar 60-inch) and 5457 (KPNO
Burrell Schmidt telescope).  For NGC~4321 and 5055, the images were
obtained in 1999 April under photometric conditions, with the standard
stars Feige 34 and PG 0939+262 used for flux calibration
\citep{Massey:88}.  The other images have not been flux calibrated, so
the flux scale was determined by comparison with published H$\alpha$
fluxes in the literature, as stated in the table.  Note that the
fluxes of \citet{KK:83} have been scaled up by a correction factor of
1.16 (K98).  The uncertainty in our bootstrap calibration is at least
as large as the uncertainty in the original published fluxes
(generally $\sim$25\%).  However, this will only affect the overall
flux level, which is subject to significant uncertainties anyway when
converted into a star formation rate (\S\ref{sfrprof}).

Since most of the images contained artifacts due to bad pixels, cosmic
ray events, and foreground stars that were not cleanly subtracted
(e.g., due to changes in seeing), these were masked out interactively
using the IRAF task IMEDIT.  Artifacts isolated from emission were
replaced with zero values, while those superposed on diffuse emission
were replaced by a background substitution algorithm that fits a
smooth surface to an annular region of the image centered on the
artifact.  Remaining outliers were removed by clipping negative values
below $-5\sigma$ and positive values above a limit determined by
visually identifying the brightest \HII\ region.  Astrometry was
performed by comparing the locations of stars in the continuum image
with a DSS image using the KARMA package \citep{Gooch:95} and then
transferring the plate solution to the H$\alpha$ image; in some cases
there were compact H$\alpha$ sources visible in the DSS image and the
astrometry could be determined by direct comparison to the H$\alpha$
image.  Astrometric errors should be less than 2\arcsec.



\begin{figure*}
\epsscale{2.2}
\plottwo{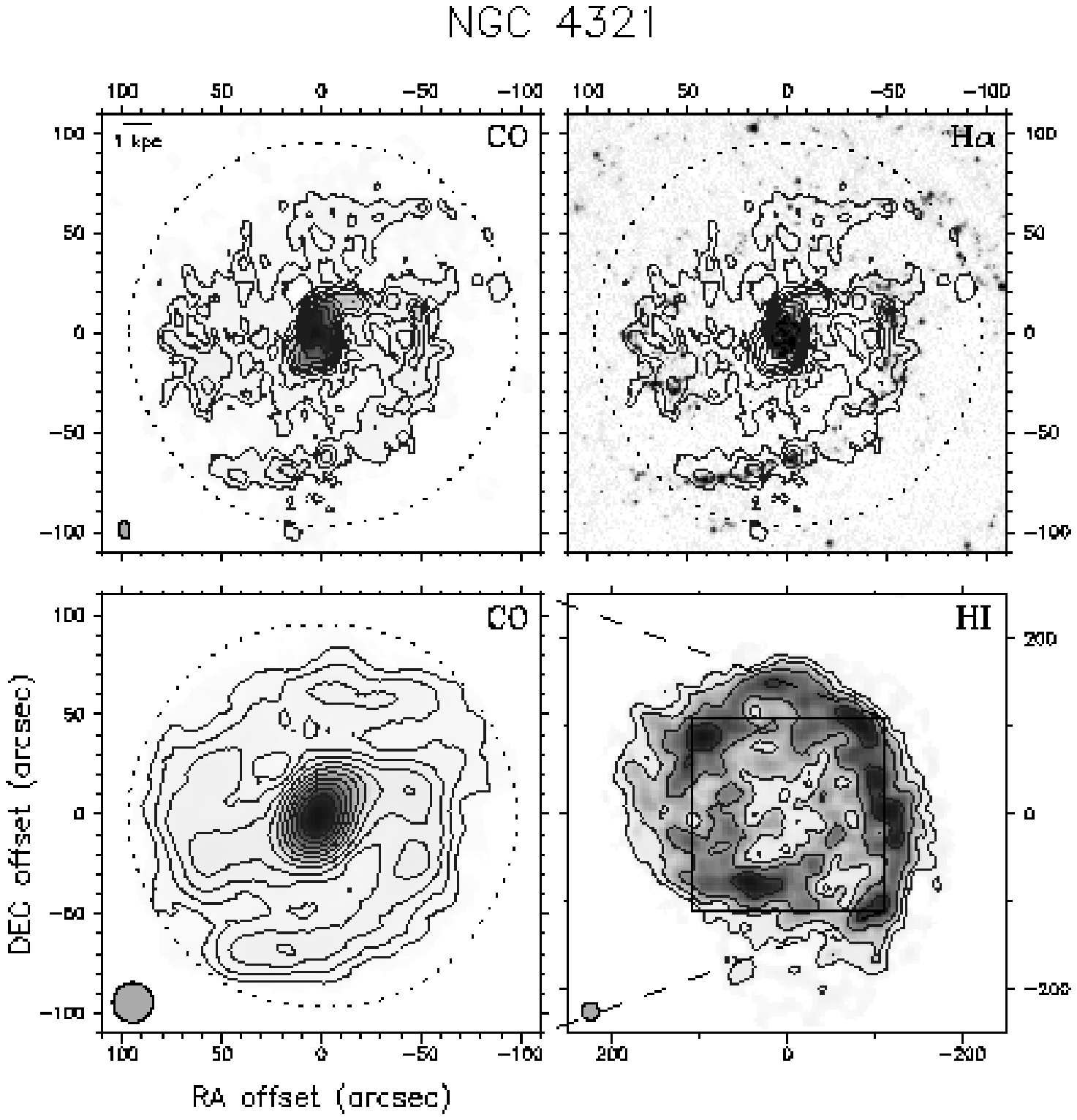}{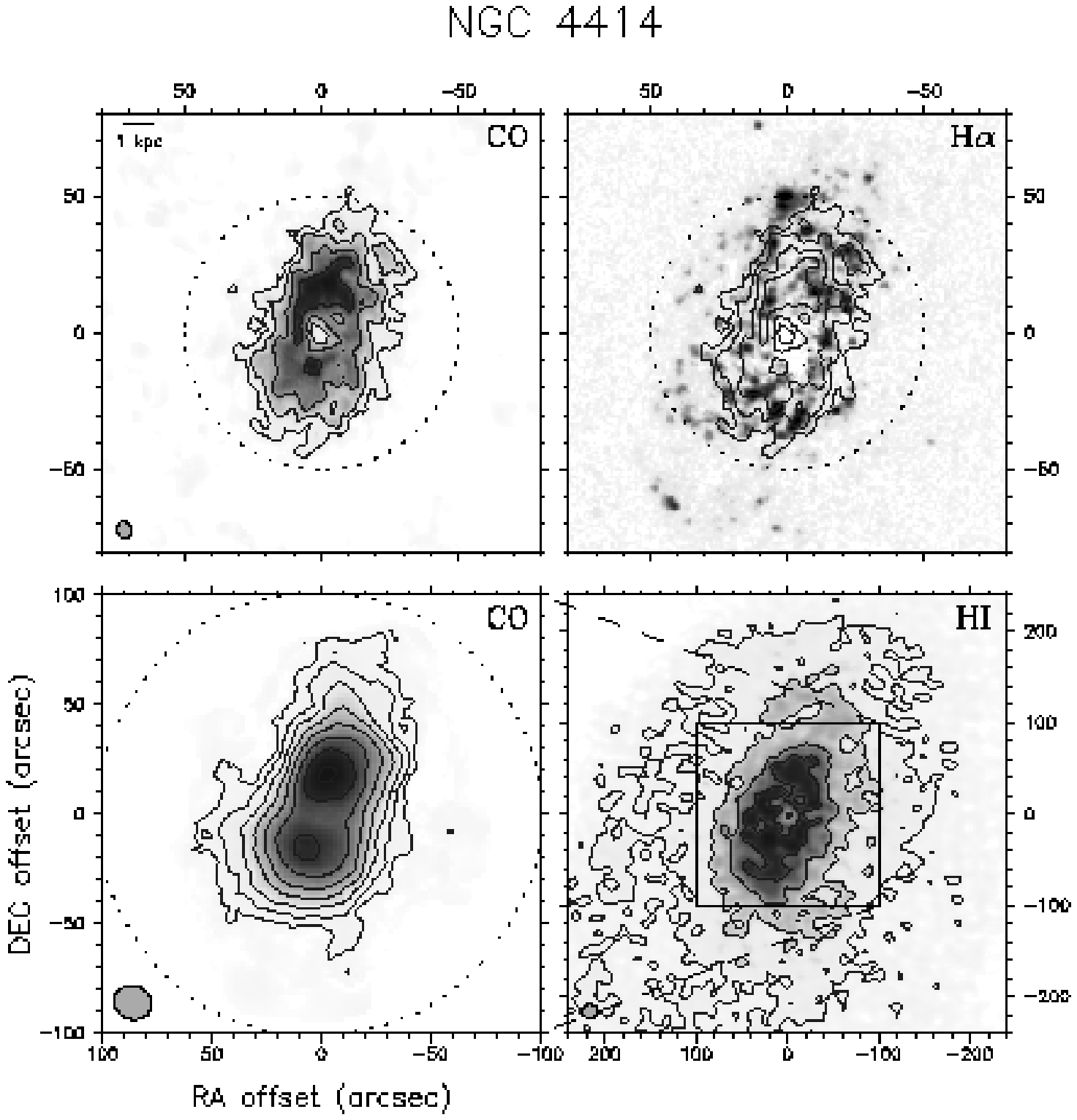}
\end{figure*}

\begin{figure*}
\epsscale{2.2}
\plottwo{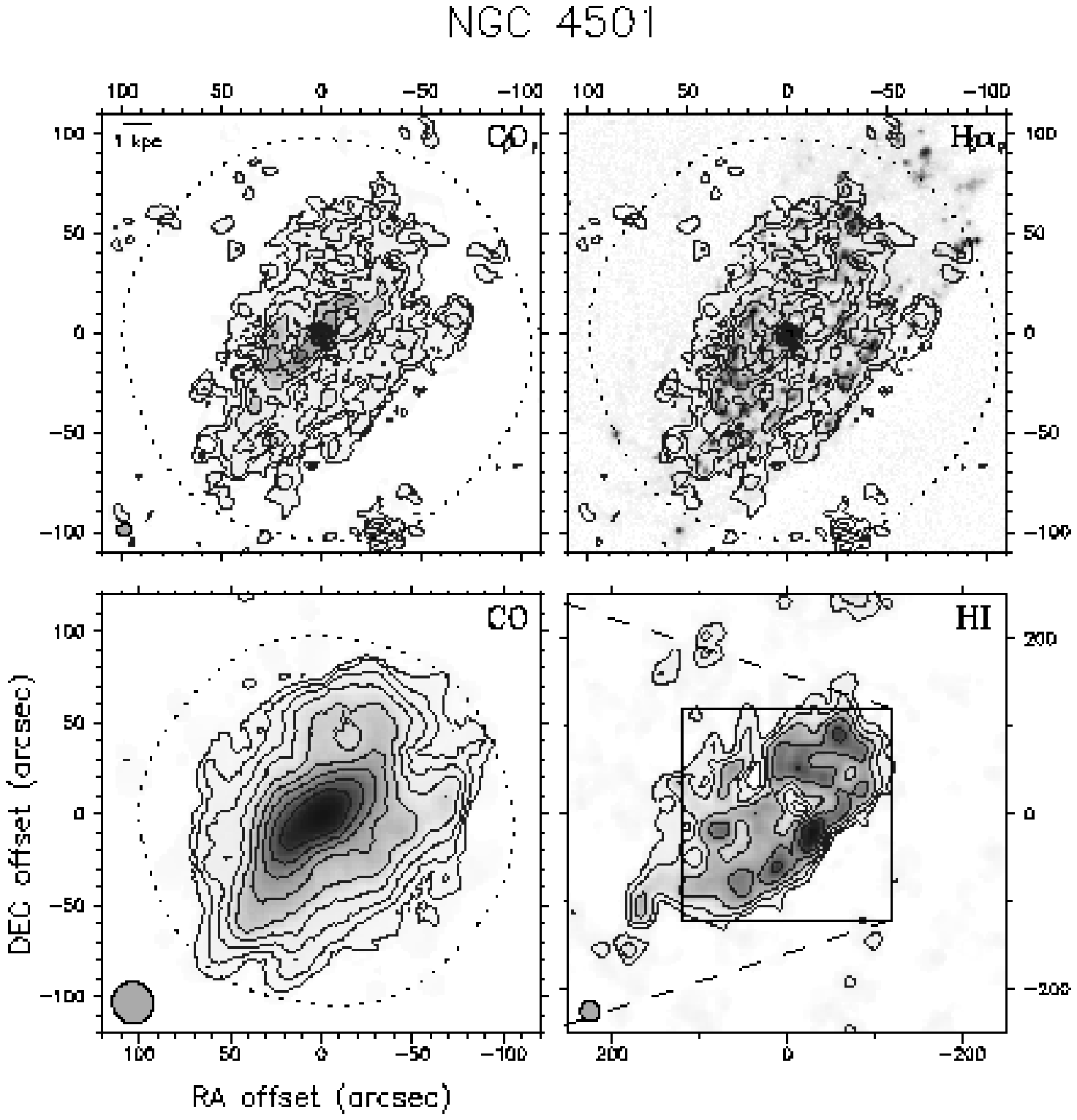}{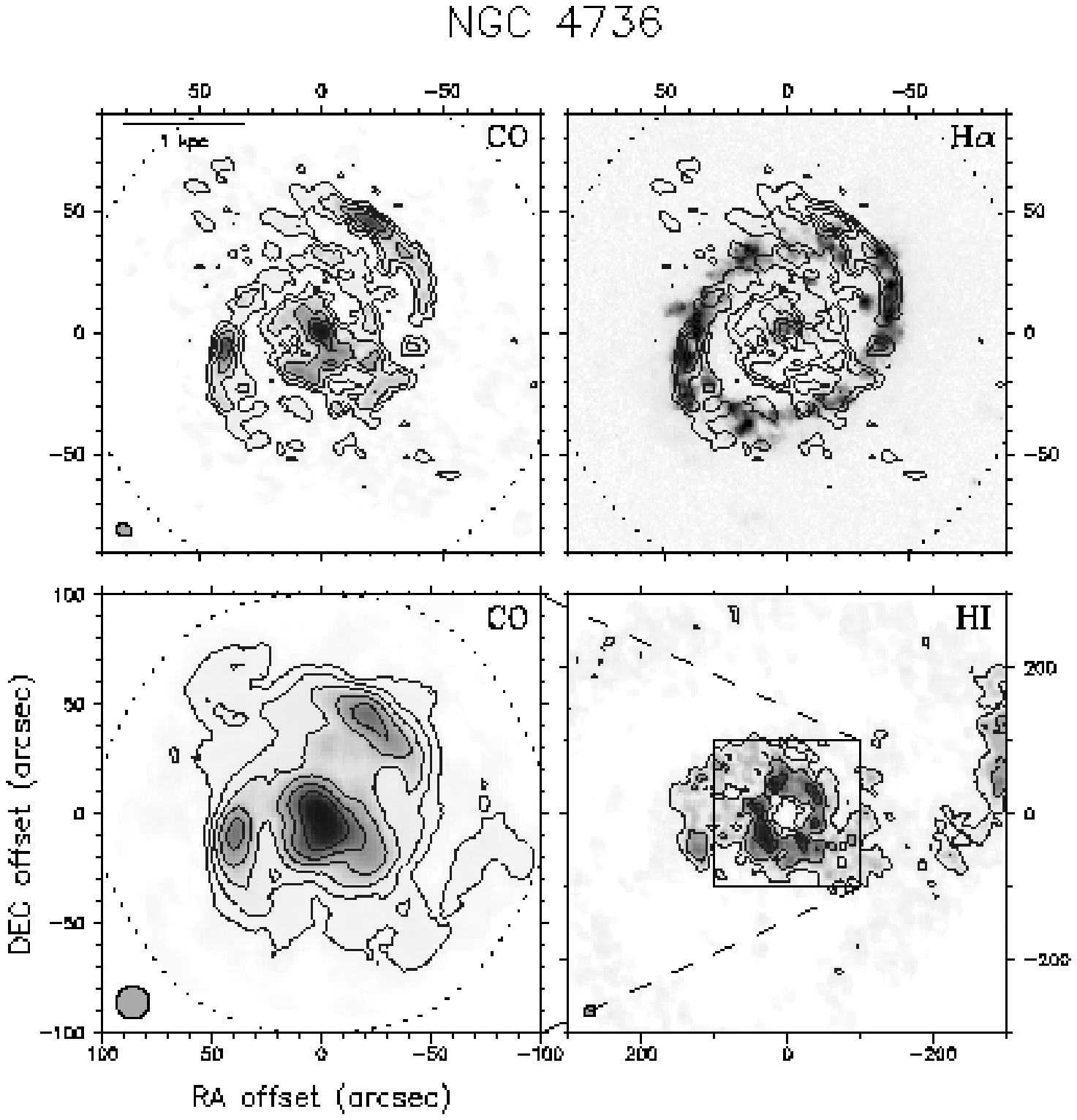}
\caption{
Intensity images for each of the seven galaxies in our sample.  
{\it Top left:} CO at full spatial resolution (robust or natural
weighting).  Contours are $k_1 n^2$ \jybm\ \kms, for integers $n \ge
2$ and $k_1$=0.75 for NGC 4321, 4414, and 4501, $k_1$=1 for NGC 4736
and 5055, $k_1$=1.5 for NGC 5033, and $k_1$=0.6 for NGC 5457.
{\it Top right:} H$\alpha$ image with CO contours overlaid.  
{\it Bottom left:} CO at the resolution of the \HI\ image.  Contour
levels are $k_2 n^2$ \jybm\ \kms, for integers $n \ge 2$ and $k_2$=2
for NGC 4321, 4414, 4501, and 4736, $k_2$=4 for NGC 5033, $k_2$=3 for
NGC 5055, and $k_2$=1 for NGC 5457.  
{\it Bottom right:} \HI\ intensity image, with contours of 0.02$n^2$
\jybm\ \kms\ ($n \ge 2$).
\label{fig:mom0maps}}
\end{figure*}

\begin{figure*}
\epsscale{2.2}
\plottwo{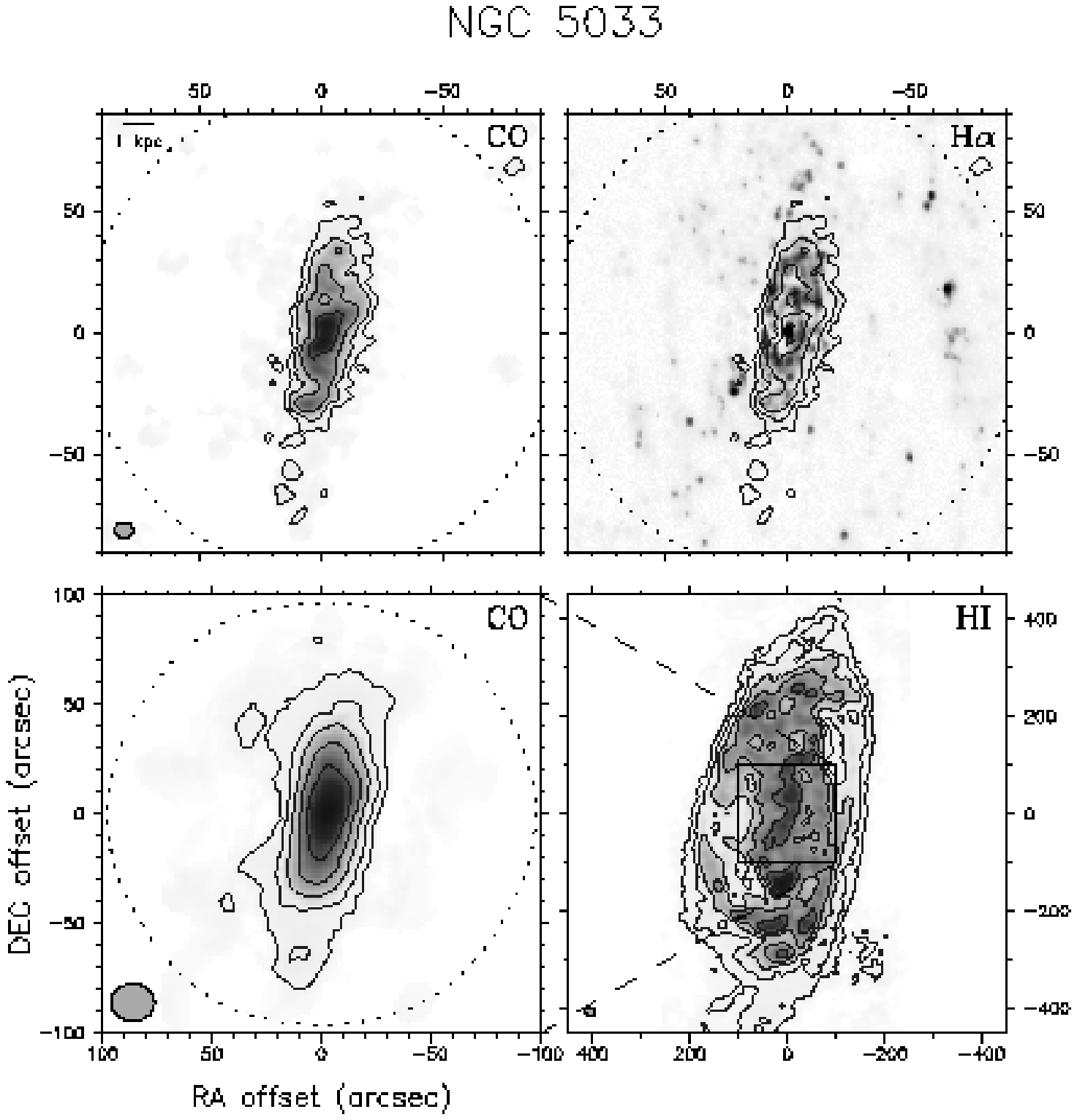}{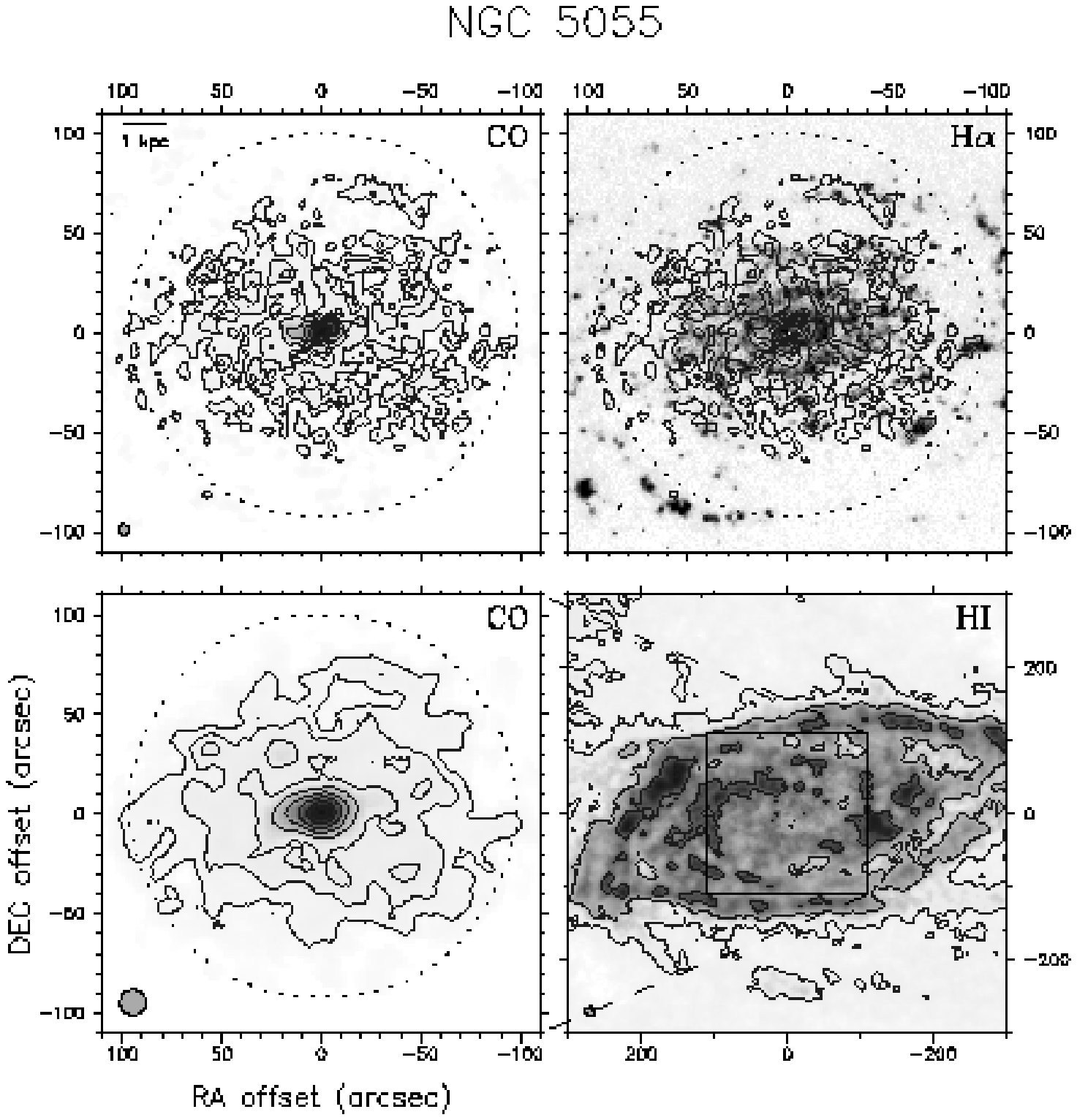}
\end{figure*}

\begin{figure*}
\epsscale{1.2}
\plotone{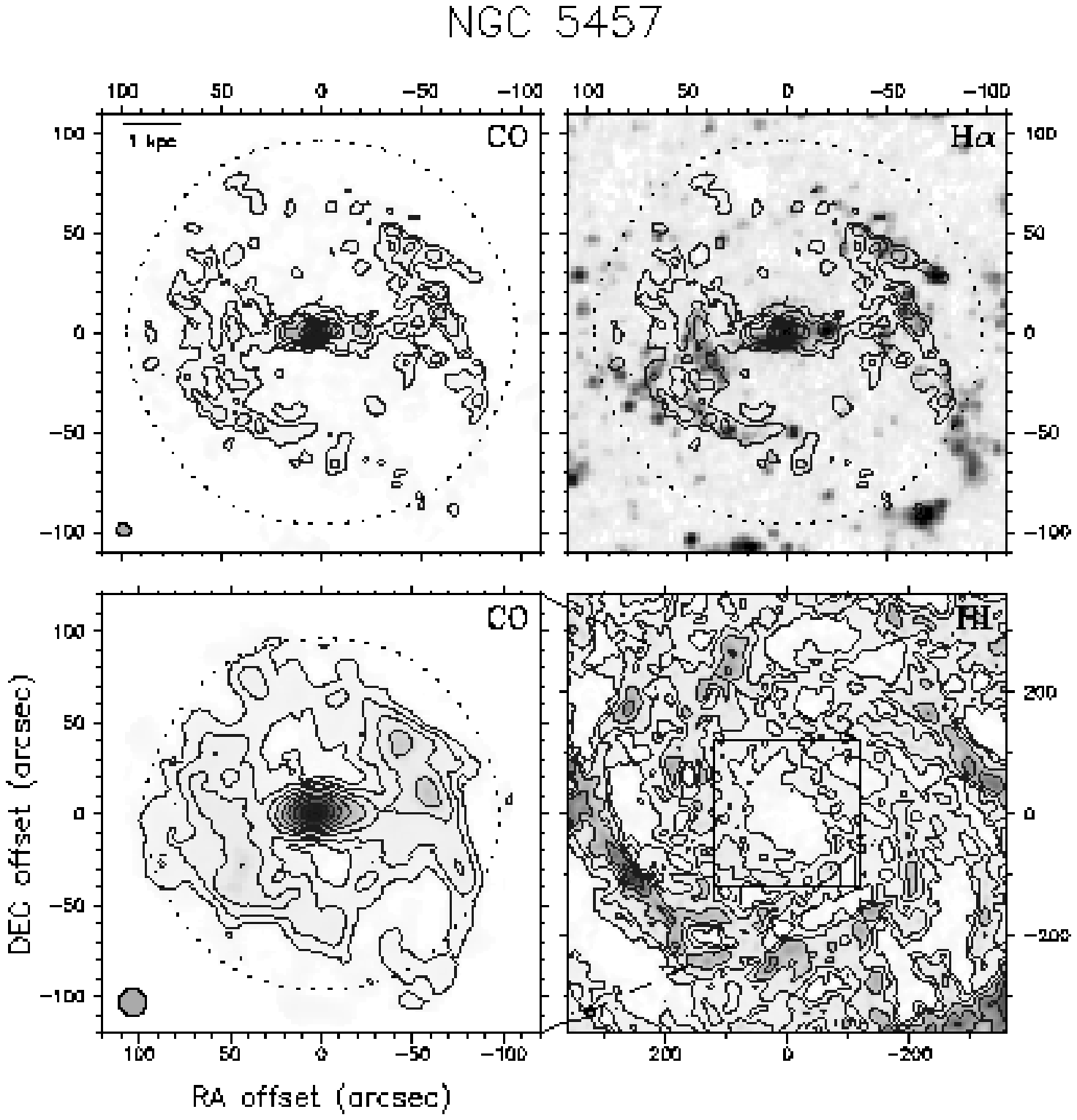}
\end{figure*}



\begin{figure*}
\epsscale{2}
\plottwo{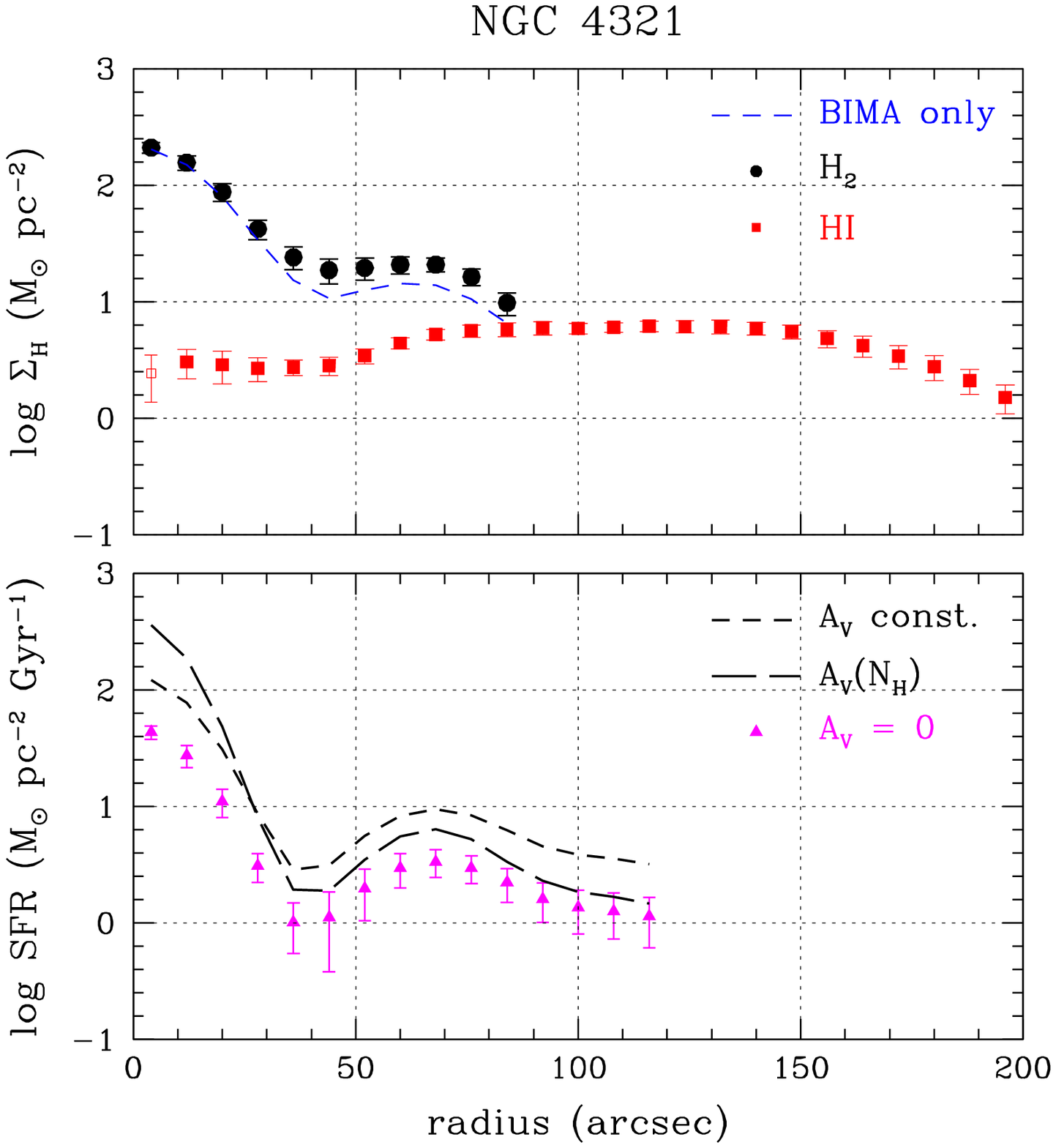}{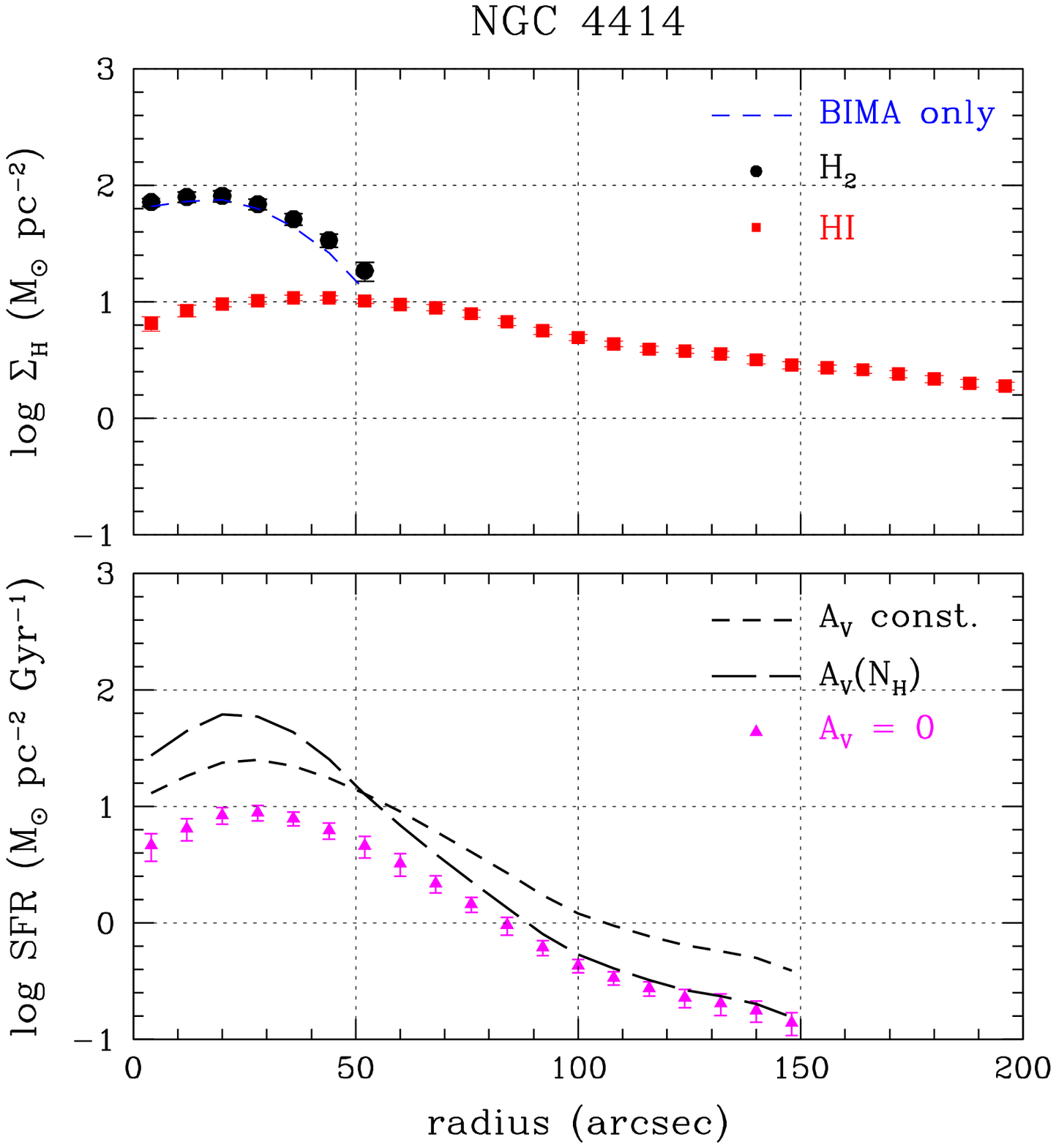}
\end{figure*}

\begin{figure*}
\plottwo{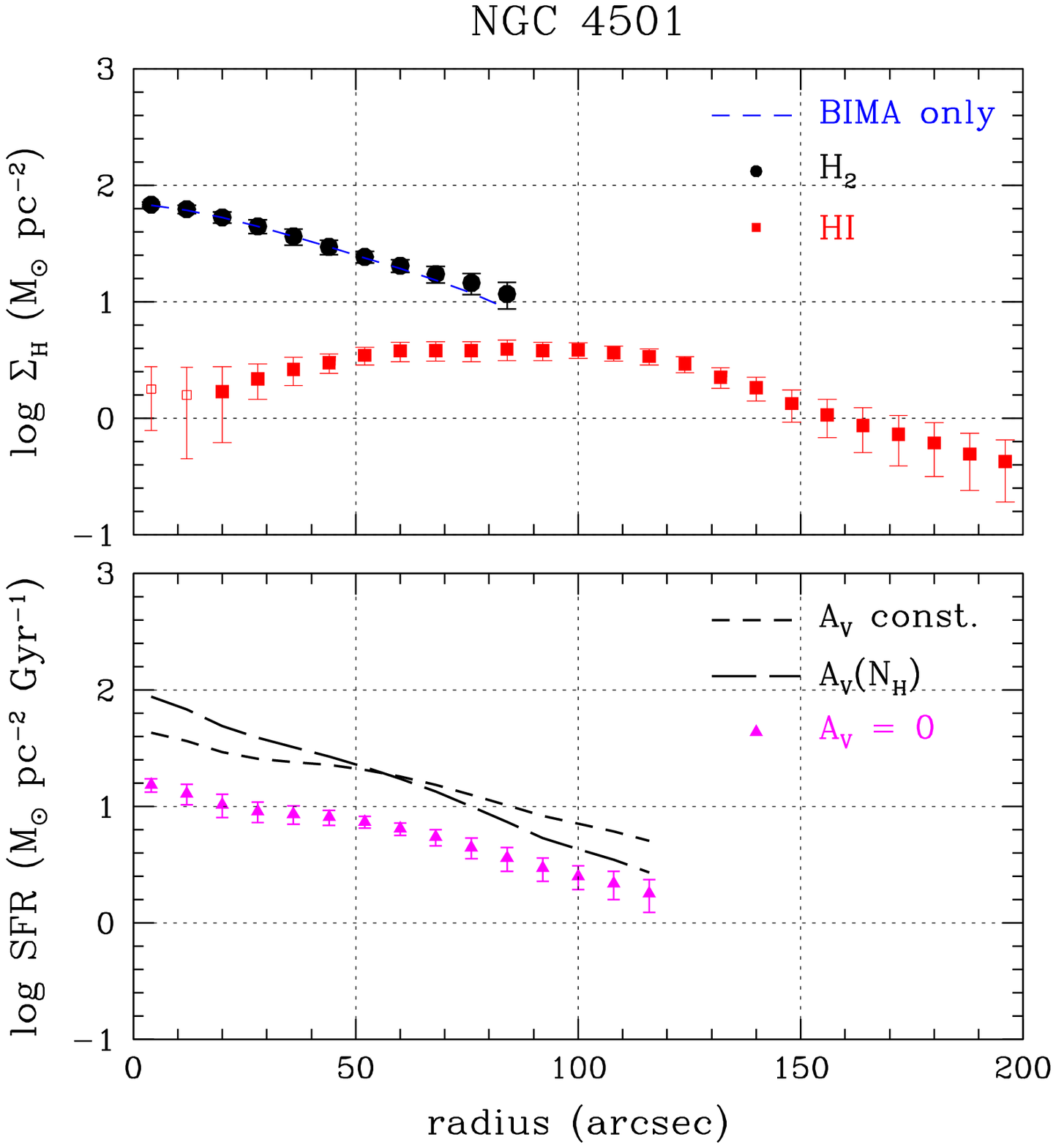}{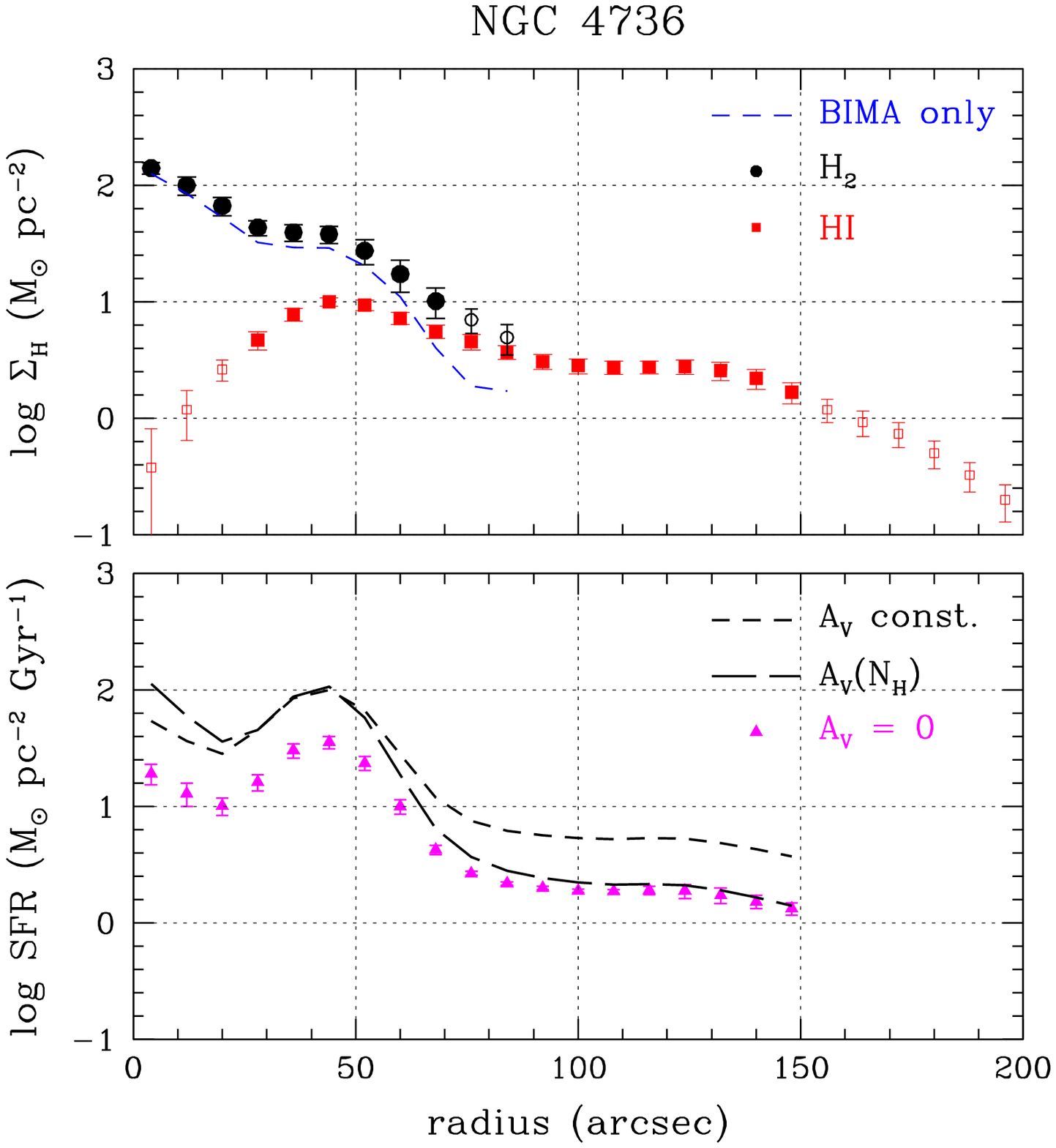}
\caption{
Radial profiles of \HI, H$_2$, and SFR surface density.  {\it Top
panels:} \HI\ and H$_2$ profiles, corrected for inclination.  The
H$_2$ profile derived from the BIMA CO data alone is shown as a dashed
line.  Unfilled plot symbols represent uncertain measurements that may
suffer from incompleteness (see text).  {\it Bottom panels:} SFR
profile (triangles), corrected for inclination but not for H$\alpha$
extinction.  The short and long dashed lines are the resulting SFR for
uniform and $N_{\rm H}$-dependent extinction corrections respectively.
\label{fig:radprof}}
\end{figure*}

\begin{figure*}
\plottwo{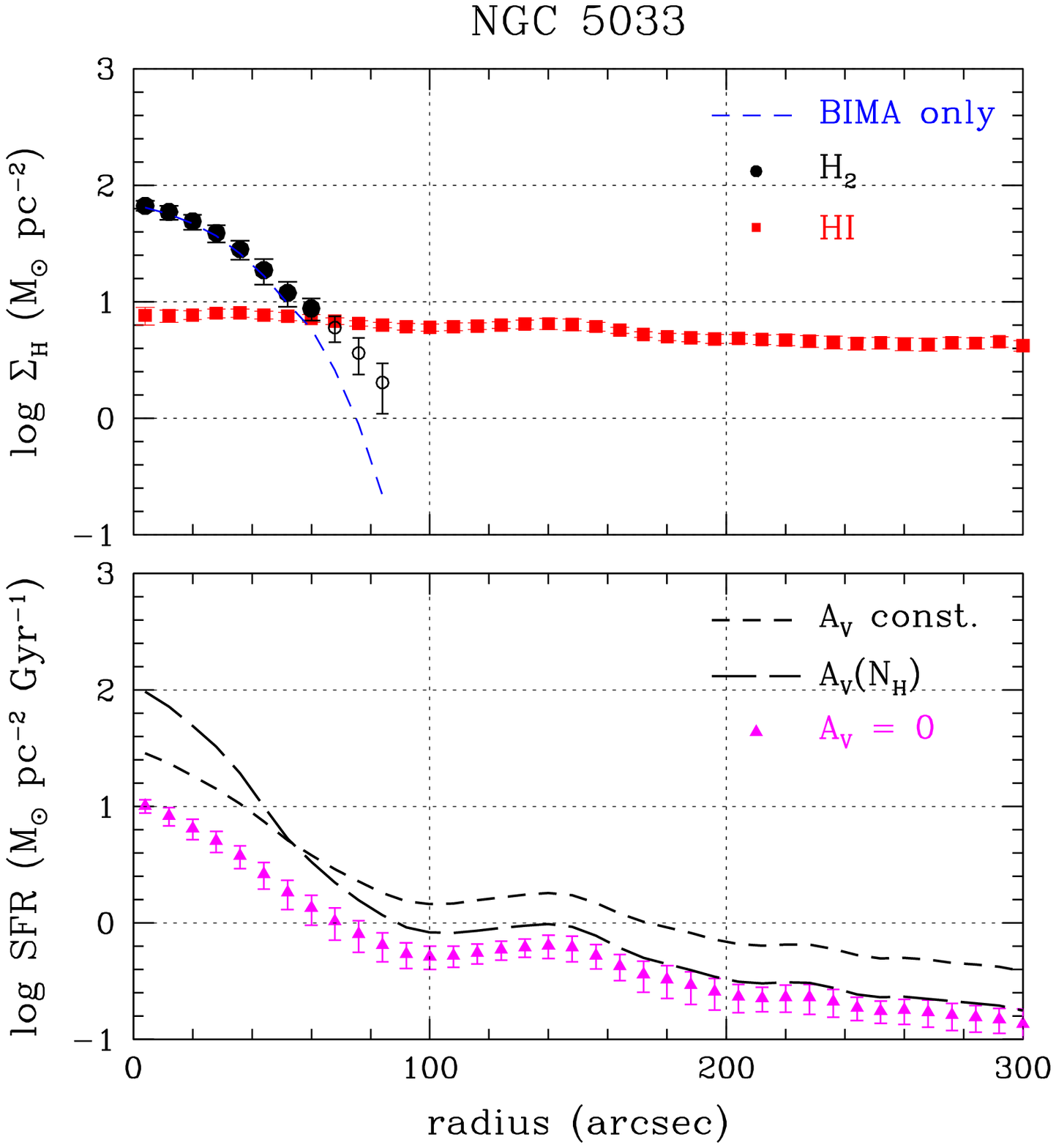}{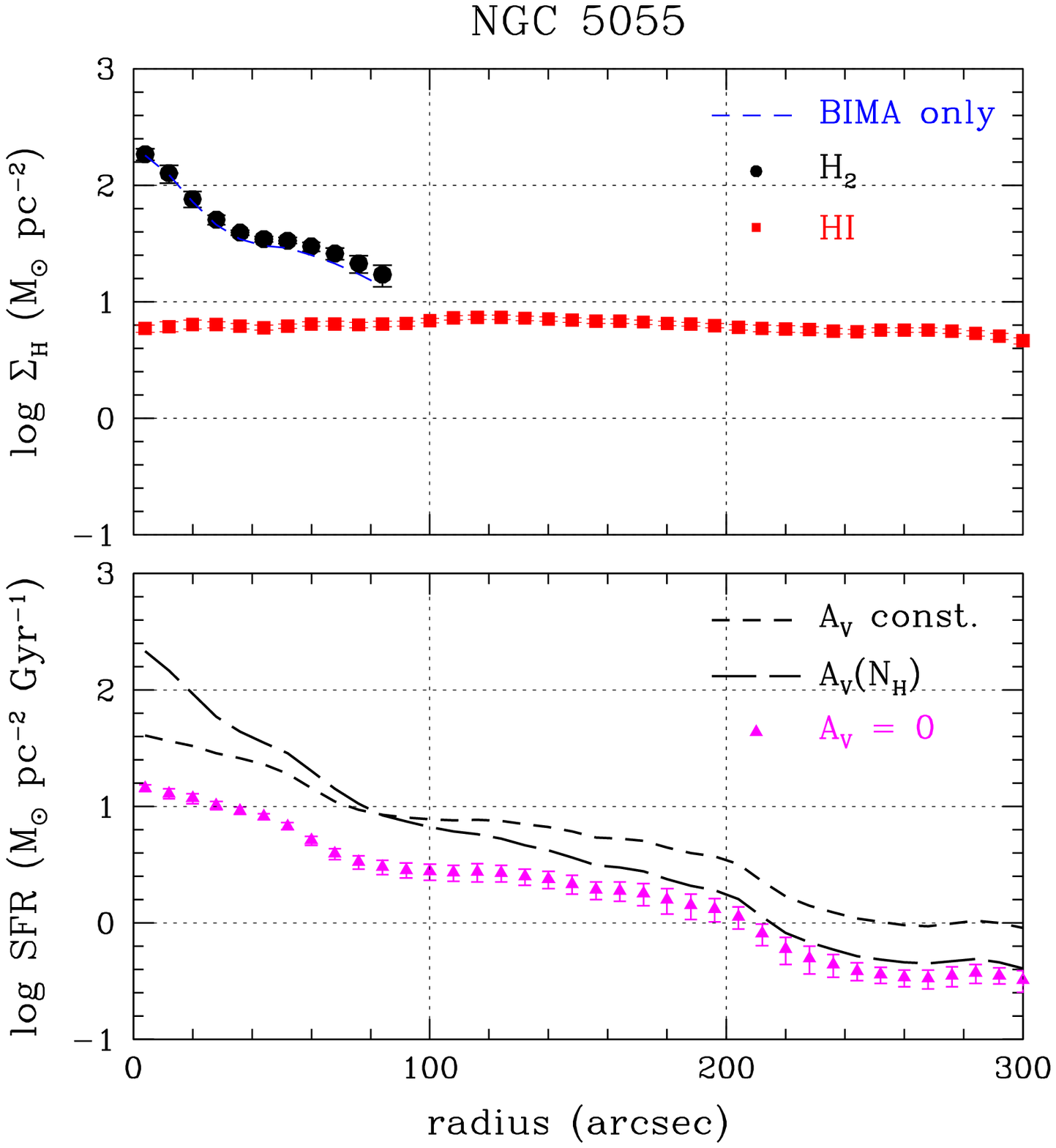}
\end{figure*}

\begin{figure*}
\epsscale{1}
\plotone{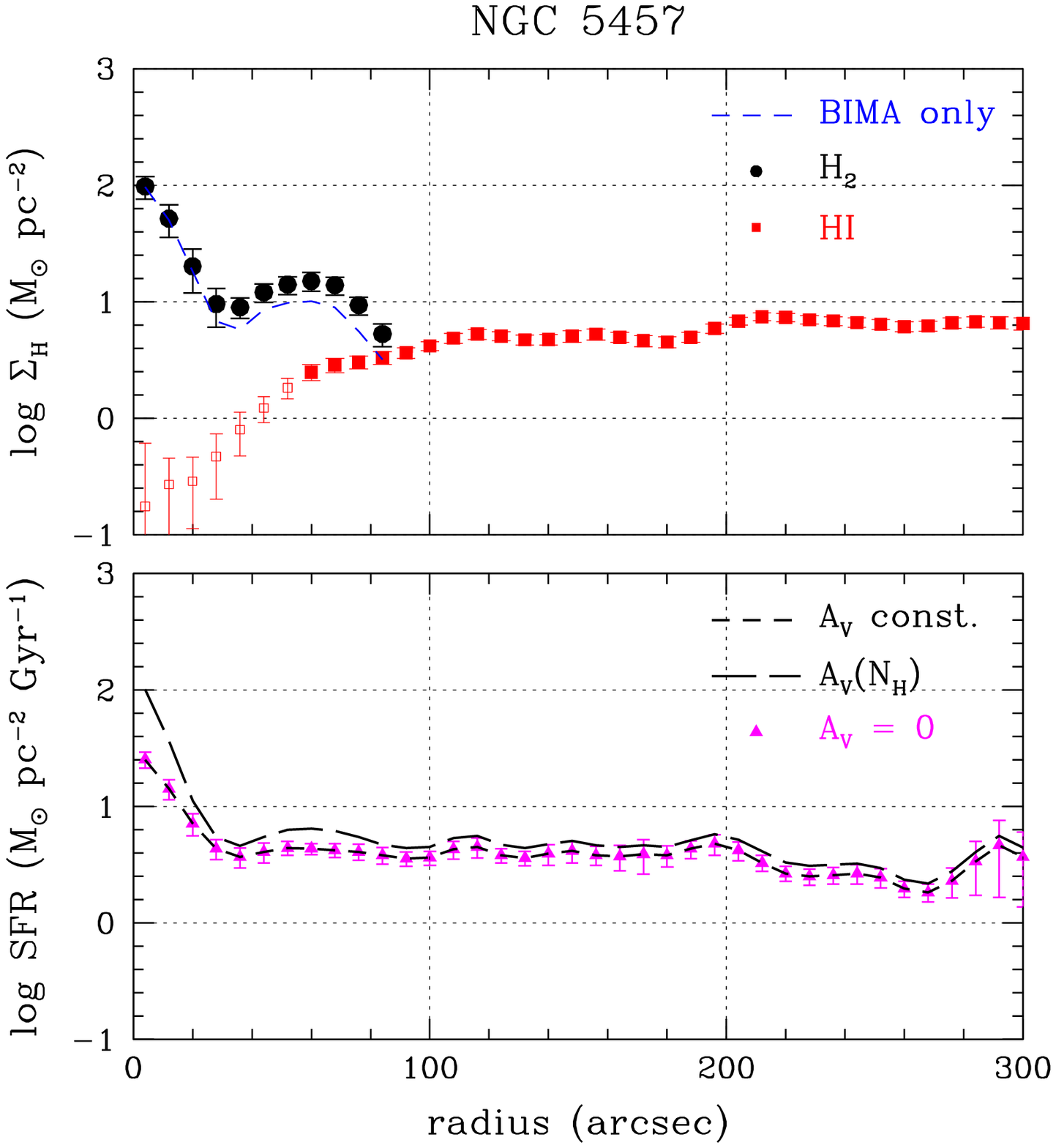}
\end{figure*}


\section{Data Analysis}\label{analysis}

\subsection{Intensity Images}\label{mom0maps}

Integrated intensity (``moment-0'') images were produced from the CO
and \HI\ datacubes using a masking technique designed to exclude
noise.  The necessity for such a procedure arises from the fact that
the emission at any given pixel is usually confined to a small range
of velocities, so a direct sum of all channels in the datacube would
result in very poor signal-to-noise \citep[see discussion
in][]{Bosma:81a,vdK:82a}.  Thus, employing a similar technique to
\citet{Braun:95}, we defined a search region in each channel map by
the 3$\sigma_{\rm smo}$ contour of a smoothed version of the map, and
included pixels within that region with intensities $|I|>\sigma_{\rm ch}$
in the sum (here $\sigma_{\rm smo}$ and $\sigma_{\rm ch}$ are the RMS
noise per channel in the smoothed and original cubes respectively).
Tests showed that while use of this masking technique does not have a
significant effect on the total flux in the datacube, it does improve
the accuracy with which weaker emission features are represented.

The resulting intensity images are displayed in
Figure~\ref{fig:mom0maps}.  For each galaxy, the top panels show
the CO intensity map at full resolution and the corresponding
H$\alpha$ image for comparison.  The field of view of the BIMA mosaic
(where the sensitivity drops to half of its peak value) is shown as a
dotted contour.  Most galaxies show a strong correspondence between CO
and H$\alpha$ emission, although the emission peaks do not always
coincide.  In this respect, NGC~4736 appears unusual since the CO and
H$\alpha$ distributions show fundamentally different symmetries, as
discussed in \citet{Wong:00}.  With the exception of NGC~4736 and
5033, the CO emission appears to extend to the edge of the observed
field, suggesting that our observations sample only the inner part of
the molecular disk.  (At a mean distance of 13 Mpc, our CO
observations extend to $R \approx 6.3$ kpc.)

The bottom panels for each galaxy show the tapered CO
intensity map alongside the VLA \HI\ intensity map.  Both maps have
the same spatial resolution (indicated by the beam in the lower left
corner of each panel).  We find that four of the galaxies (NGC 4321,
4501, 4736, 5457) show both an \HI\ deficit and CO enhancement in their
inner regions, suggesting a large-scale conversion of atomic to molecular
gas.  This phenomenon has long been noted \citep[e.g.,][]{Morris:78},
and its occurrence appears to be unrelated to the Hubble type of the
galaxy, since 4736 is classified as early-type (Sab) whereas 5457 is
classified as late-type (Scd).


\subsection{Radial Profiles}\label{prof}

Radial profiles were derived from the intensity images by averaging in
elliptical annuli spaced by roughly half the synthesized beamwidth and
centered on the position given in Table~\ref{tbl:props}.  The annuli
were assumed to be circular rings viewed at an inclination $i$ and
with a line of nodes rotated from north to east by some position angle
(PA); the adopted values of PA and $i$ for each galaxy are given in
Table~5.  As described in Paper II, the appropriate PA
for each galaxy was determined by kinematic fits to the velocity
field, while $i$ was determined from isophotal and/or kinematic fits.
The mean intensity for each annulus was assigned an uncertainty of
\begin{equation}
\sigma_{\rm mean} = \frac{\sigma_{\rm pix}}{\sqrt{N_{\rm pix}/n_{\rm
beam}}}
\end{equation}
where $\sigma_{\rm pix}$ and $N_{\rm pix}$ are the RMS pixel value and
the number of pixels in the annulus, and $n_{\rm beam}$ is the number
of pixels in each resolution element (``beam'').  The resolution of
the H$\alpha$ images was assumed to be 2\arcsec\ (Gaussian FWHM).
Note that $\sigma_{\rm mean}$ is a quantity which reflects only the
dispersion in the values within each annulus, and thus reflects a
combination of measurement errors and any intrinsic non-axisymmetry.


\begin{center}
{\sc Table 5. Adopted Orientation Parameters\label{tbl:painc}}
\vskip 0.1in
\begin{tabular*}{1.7in}{ccc}
\hline\hline
Galaxy & P.A.\tablenotemark{(a)} & $i$\\
& (deg) & (deg)\\[0.5ex]
\hline\hline
NGC 4321 & 153 & 34\\
NGC 4414 & 159 & 55\\
NGC 4501 & 140 & 63\\
NGC 4736 & 295 & 35\\
NGC 5033 & 353 & 68\\
NGC 5055 & 98  & 63\\
NGC 5457 & 42  & 21\\
\hline\hline
\end{tabular*}
\end{center}
\vskip 0.1in
{\footnotesize $^a$Position angle of receding side of line of nodes,
measured E from N.}
\vskip 0.1in


\subsubsection{CO and \HI\ Profiles}\label{cohiprof}

Since the CO and \HI\ intensity images are generated from datacubes by
a masking process, sampling effects make their noise properties
somewhat unusual.  Especially at large radii where few pixels survive
the masking, the RMS fluctuation within an annulus (given by
$\sigma_{\rm mean}$ above) can greatly underestimate the error in the
profile because masked pixels have been assigned a value of zero.  We
can derive a conservative estimate of the true uncertainty in the mean
for each annulus by considering the case in which all channels are
included in the moment map (i.e., no masking or clipping of any kind):
\begin{equation}
\sigma_{\rm est} = \frac{\sigma_{\rm ch}\,\Delta v\,\sqrt{n_{\rm ch}}}
	{\sqrt{N_{\rm pix}/n_{\rm beam}}}\;,
\label{eqn:sigest}
\end{equation}
where $\sigma_{\rm ch}$ is the RMS noise in a channel map of the
original datacube, $n_{\rm ch}$ is the number of channels, and $\Delta
v$ is the velocity width of each channel.

The derived radial gas profiles, at the resolution of the \HI\ maps,
are presented in the top panels of Figure~\ref{fig:radprof}.  Points
are separated by 8\arcsec, oversampling the profiles slightly.
Unfilled plot symbols are used to denote radii where the mean
intensity falls below 3$\sigma_{\rm est}$, since these measurements
are less secure, relying on the ability of the mask to select out the
``real'' emission.  These radii are also the most likely to suffer
from incompleteness, which can result if the mask excludes low-level
emission that still makes a substantial contribution to the sum when
integrated over a large annulus.  The profiles are given in mass surface
density units, which are derived as follows.  First the conversion
from surface brightness to column density is performed using the
CO-to-H$_2$ conversion factor given by Equation~\ref{eqn:cotoh2}, or
using the optically thin approximation for \HI:
\begin{equation}
N_{\rm HI} = 1.82 \times 10^{18} \left(\frac{I_{21}}{\rm
	K\;\kms}\right) \rm cm^{-2}\;,
\end{equation}
where $I_{21}$ is the velocity-integrated \HI\ intensity.
Column densities were then converted to face-on surface densities,
\begin{eqnarray}
\sighi & = & m_{\rm H}\, N_{\rm HI}\, \cos i\;, \\
\sightwo & = & 2m_{\rm H}\, N_{\rm H_2}\, \cos i\;,
\end{eqnarray}
excluding helium.  The BIMA-only CO profiles, converted to the
appropriate units, are also shown for comparison.  Note that the extra
flux provided by the KP data has little effect on the CO radial
profiles, aside from a constant offset which only becomes significant
at outer radii.

The assumption of a constant CO-to-H$_2$ conversion factor
($X$-factor) remains quite controversial, with some researchers
suggesting that $X$ is increased by a factor of 10 or more in the
vicinity of the Galactic Center \citep{Sodroski:95,Dahmen:98}.
Indeed, the presence of central CO condensations in many galaxies
\citep{Sakamoto:99,Regan:01} may be an indication of a change in the
$X$-factor rather than a true concentration of molecular gas.  On the
other hand, two of the seven galaxies (NGC 4501 and 5033) show
featureless, roughly exponential CO profiles.  We discuss the
possibility of a varying $X$-factor further in \S\ref{xdisc}.

\subsubsection{Total Gas Profiles}

Radial profiles of the total gas surface density (\siggas) were
derived by summing the contributions from \HI\ and H$_2$ and
multiplying a factor of 1.36 to include helium.  Since the CO emission
is only imaged within $R \lesssim 2\arcmin$, and our sensitivity to CO
emission falls off at H$_2$ column densities comparable to the peak
\HI\ column densities of $\sim$10 \Msol\ pc$^{-2}$, it was necessary
to extrapolate the CO profile to larger radii assuming an exponential
profile.  The scale length for the extrapolation was determined from a
fit to the entire radial profile, except for NGC 4414, where the fit
was made to the declining part of the CO profile, and NGC 5457, where
the scale length was determined from the radial profile of a large
(10\arcmin\ $\times$ 10\arcmin) CO map obtained at the NRAO 12m
telescope as part of BIMA SONG.\@ To take an example,
Figure~\ref{fig:totprof} shows the total gas profile for NGC 4321 and
its individual components (also corrected for helium).  The profiles
are probably most uncertain at radii just beyond $R_{\rm max}$ (refer
to Table 2), where the extrapolated H$_2$ mass is likely to be
significant.  Even so, we expect molecular gas to contribute less than
half of the gas mass at these radii, and hence the error in \siggas\
introduced by the extrapolation should be $\lesssim$50\% as long as the
CO profile continues to decline beyond $R_{\rm max}$, an assumption
supported by visual inspection of our somewhat larger OTF maps.


\vskip 0.25truein
\includegraphics[width=3.25in]{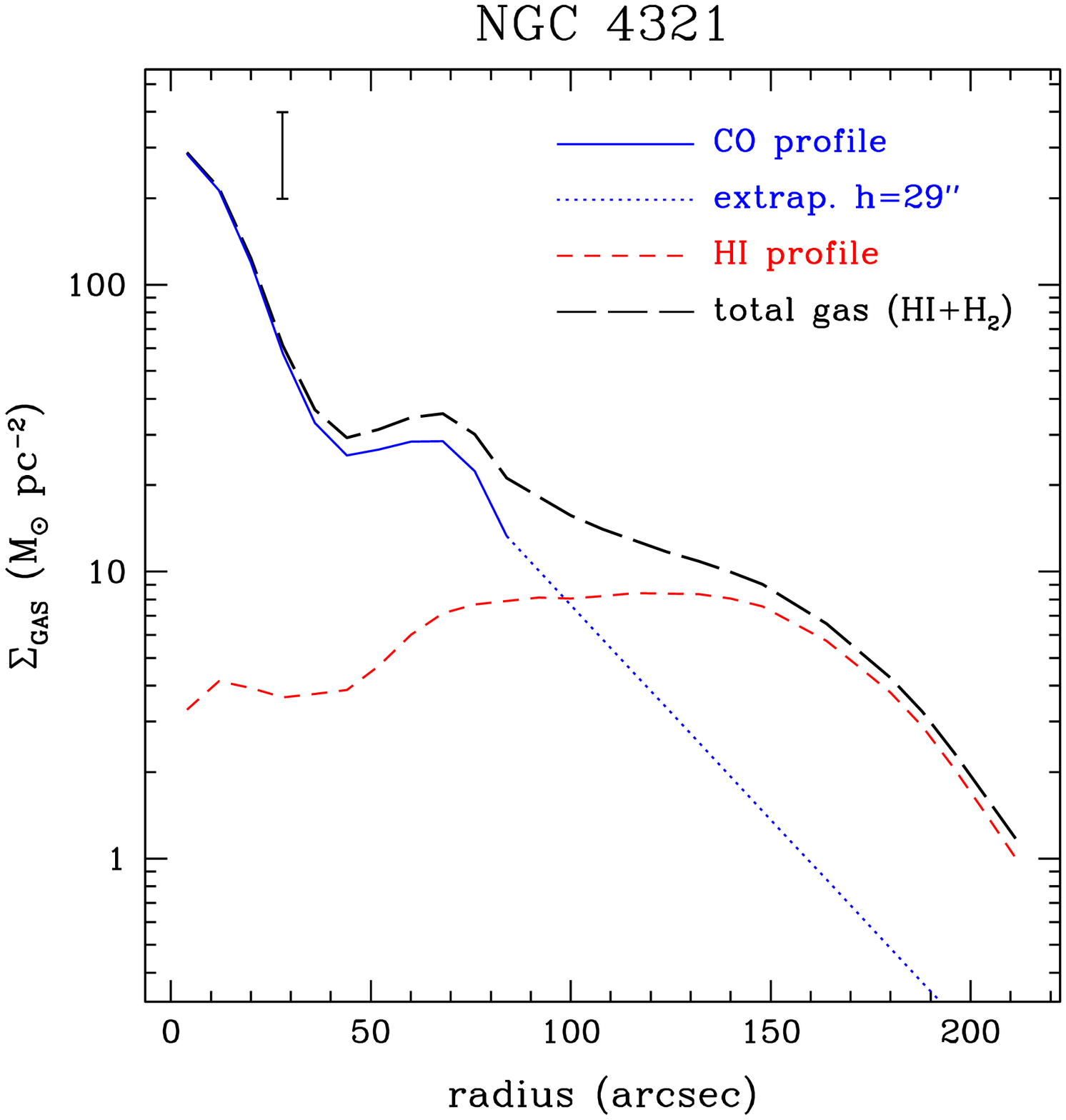}
\figcaption{
Total gas profile for NGC 4321 derived from the CO and \HI\ profiles,
with the CO profile extrapolated as shown.  The error bar in the upper
left corner represents a factor of 2 change in any of the curves.
\label{fig:totprof}}
\vskip 0.25truein



\begin{figure*}[b]
\epsscale{2}
\plottwo{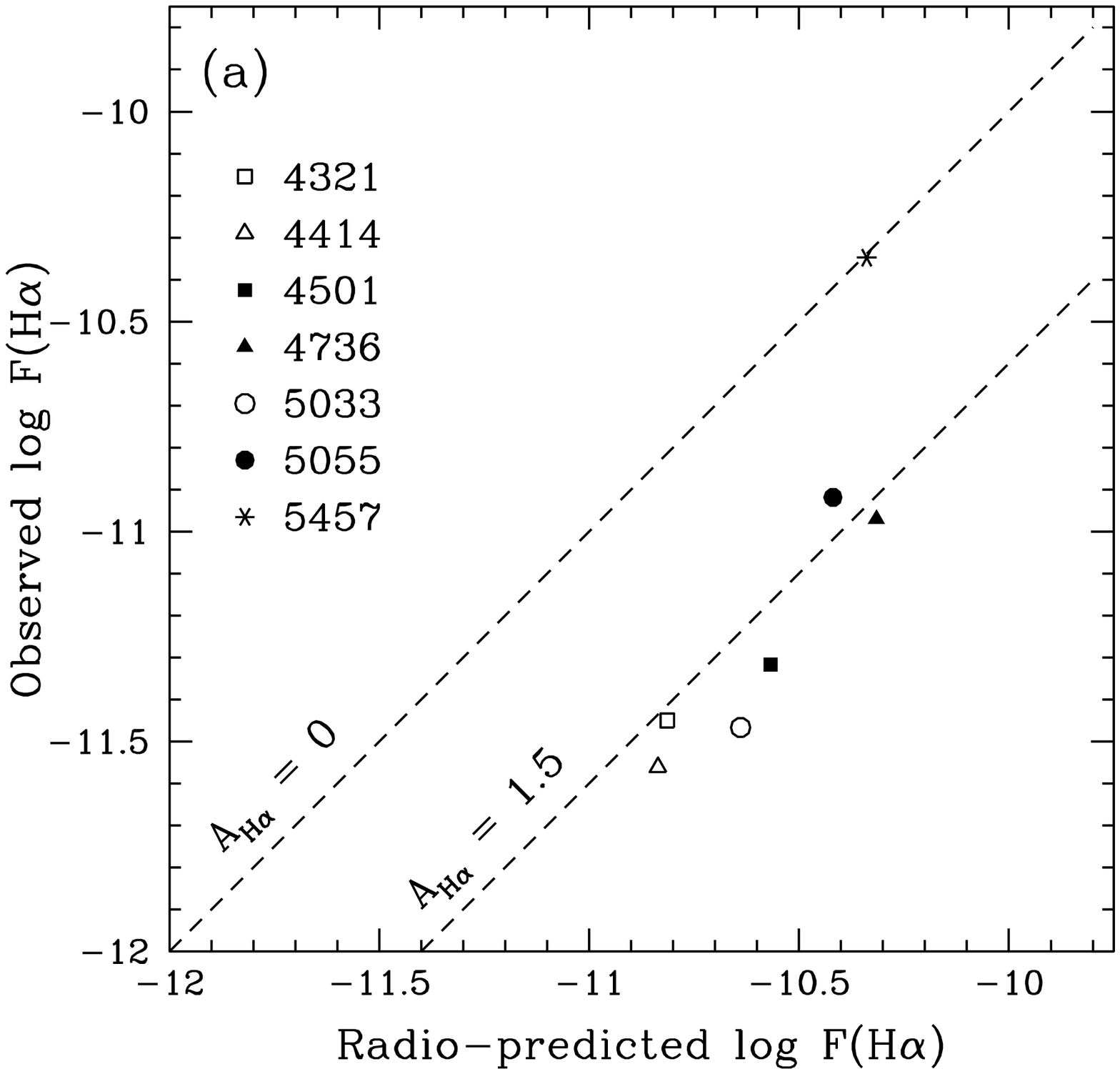}{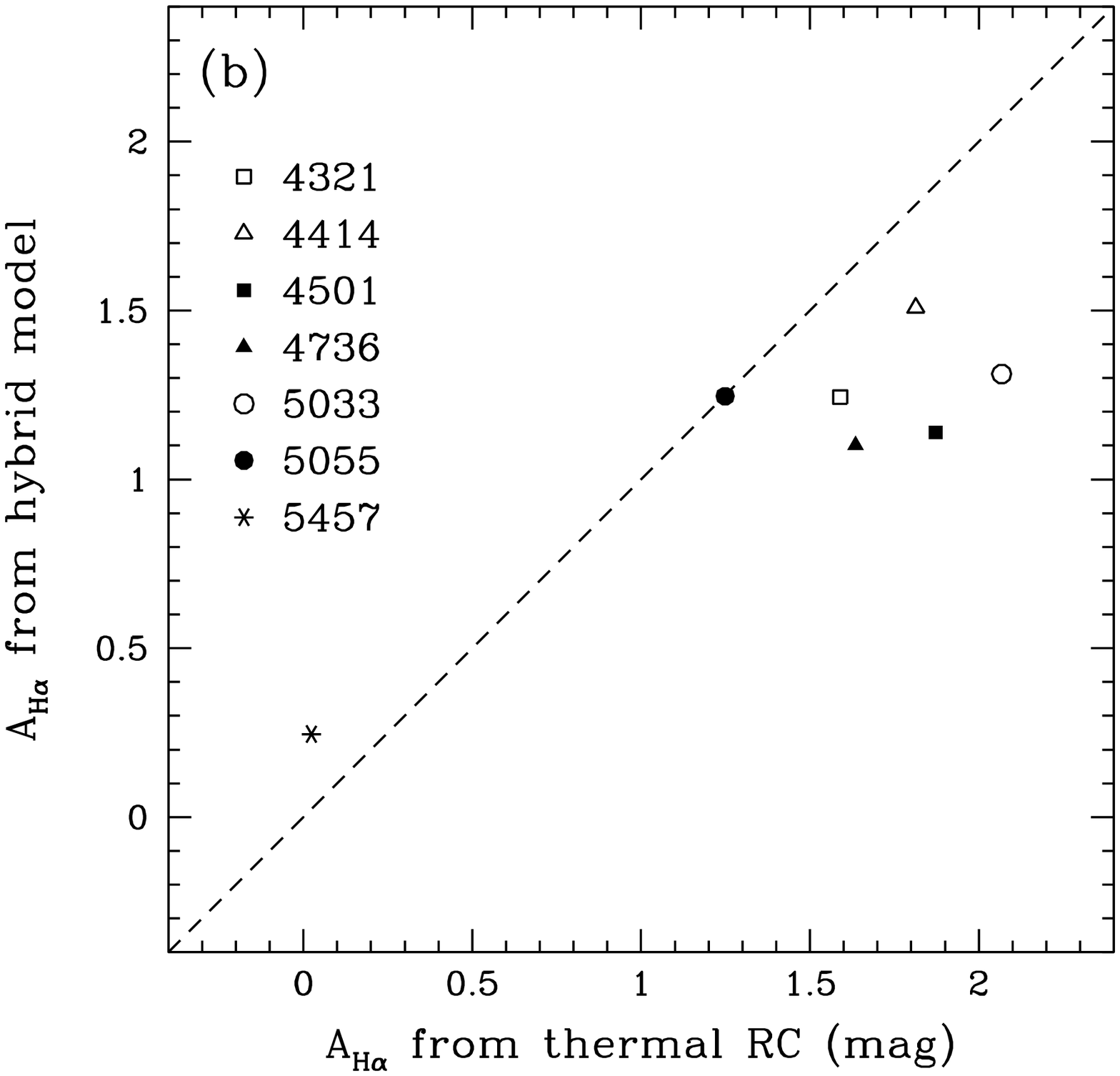}
\caption{
(a) Total H$\alpha$ flux for each galaxy (assuming \NII/H$\alpha$=0.5)
plotted against the predicted H$\alpha$ flux based on the thermal
radio continuum data.  Dashed lines correspond to extinctions of 0 and
1.5 magnitudes.  (b) Comparison of the disk-averaged extinction at
H$\alpha$ derived from comparing the radio and H$\alpha$ fluxes with
the extinction predicted by a ``hybrid'' dust model, where dust
associated with H$_2$ is well-mixed but dust associated with \HI\
constitutes a foreground screen.  The model assumes a Galactic
dust-to-gas ratio and extinction law.  The dashed line has a slope of 1.
\label{fig:radioha}}
\end{figure*}


\subsubsection{H$\alpha$ Profiles}\label{sfrprof}

The H$\alpha$ emission from spiral galaxies is dominated by young
($t<20$ Myr), massive ($>$10 \Msol) stars, and is thus a reasonable
tracer of the recent star formation rate (SFR) \citep[see the review
by][]{KC:98b}.  We expressed the H$\alpha$ brightnesses in units of
SFR density using the conversion \citep{KC:98b}:
\begin{equation}
\frac{\rm SFR}{\rm \Msol\; yr^{-1}} = 7.9 \times 10^{-42}\;
        \left(\frac{L_{\rm H\alpha}}{\rm erg\; s^{-1}}\right)\;.
\end{equation}
This conversion ignores extinction by dust and assumes a Salpeter IMF
between 0.1--100 \Msol, a constant SFR with time, and negligible
escape of ionizing radiation from the galaxy.  Since the H$\alpha$
images include a contribution from \NII, they have been multiplied by
0.67, assuming an average ratio of \NII/H$\alpha \sim 0.5$
\citep{KC:92}.  Although the use of a uniform correction is rather
simplistic, since \NII/H$\alpha$ is often enhanced in the nuclear
region \citep[e.g.,][]{Rubin:86}, it should be roughly applicable for
most of the \HII\ regions in the disk.  For comparison,
\citet{Smith:91} used long-slit spectroscopy to deduce a ratio of
H$\alpha$/(\NII+H$\alpha$) $\approx 0.62$ for the star-forming ring in
NGC 4736.

The bottom panels of Figure~\ref{fig:radprof} show the radial
H$\alpha$ profiles, converted to units of SFR density, derived from
images smoothed to the same resolution as the \HI\ images.  These can
be directly compared to the gas profiles in the top panels.  The
data points have not been corrected for extinction, whereas the
dashed lines represent the results of applying the extinction
corrections described in \S\ref{extinct}.  In \S\ref{molat} we also
compare the CO and H$\alpha$ profiles at the higher resolution of the
CO data, although we have not shown the corresponding profiles here.


\subsection{Extinction Corrections for H$\alpha$}\label{extinct}

While extinction at H$\alpha$ due to Galactic dust along the line of
sight is probably unimportant ($A_{V,Gal} < 0.13$ mag for all seven
galaxies according to \citealt{Schlegel:98}), extinction internal to
the galaxies themselves is likely to be significant, given that stars
form in regions of dense gas that have associated dust.  Thus, when
deriving quantitative estimates of the SFR from H$\alpha$ emission, it
is important to take extinction into account.

In his study of disk-averaged star formation, K98 applied a
constant extinction correction of 1.1 mag to his H$\alpha$ data, based
on an earlier comparison of integrated H$\alpha$ and radio continuum
fluxes \citep{KC:83}.  Although obviously a gross simplification, it
is difficult to do better.  The use of an H$\alpha$/H$\beta$
ratio to determine the extinction is inapplicable for galaxy-integrated
spectra, due to the difficulty of measuring the H$\beta$ flux in the
presence of stellar absorption lines \citep{KC:92}.  Even with
spectrophotometry of individual \HII\ regions, one can typically only
obtain measurements for the brightest sources, and the derived
extinctions may not be appropriate for all of the H$\alpha$ emission
\citep{Bell:01a}.  Hence, the most reliable extinction estimates
are likely to come from narrow-band imaging of NIR
recombination lines such as Br$\gamma$ or Pa$\alpha$, or from
measurements of thermal radio continuum fluxes.

We have considered four techniques for determining 
extinction corrections for our sample:

\begin{enumerate}

\item One can apply a global correction following K98, but
instead using measured radio continuum and H$\alpha$ fluxes for
individual galaxies.  \citet{Niklas:95} published radio fluxes at 10
GHz for a large set of galaxies that encompasses our sample, and
\citet{Niklas:97} have combined these with data at other
frequencies to separate the thermal and non-thermal components (the
latter often being dominant even at 10 GHz) via the spectral slope.
The thermal flux $S_{\rm th}$ can be converted into a predicted
H$\alpha$ flux using the formula \citep{Condon:92}:
\begin{equation}
F_{\rm pred} ({\rm H\alpha}) = 8.0 \times 10^{-13}
	\left(\frac{\nu}{\rm GHz}\right)^{0.1}
	\left(\frac{S_{\rm th}}{\rm mJy}\right)\,
	\rm erg\,cm^{-2}\,s^{-1}\;.
\end{equation}
This relation assumes an electron temperature of $10^4$ K, but is not
very sensitive to it ($F_{\rm pred} \propto T_e^{-0.5}$).  In
Figure~\ref{fig:radioha}(a) we compare the integrated fluxes in our
H$\alpha$ images with the radio-predicted fluxes.  While extinction
appears to be negligible for NGC 5457 (M101), a typical value of
$A_{\rm H\alpha} \sim 1.5$ mag is found for the other six galaxies.
For these galaxies, the SFR is therefore underestimated by a factor of
$\sim$4 when using the H$\alpha$ fluxes alone.

\item One can derive a radially dependent correction using the
observed radial gas profile and an assumed dust-to-gas ratio.  We
first assume a ``foreground dust screen model,'' in which half of the
observed gas column density is located in a uniform absorbing slab
between the H$\alpha$ emission and us.  We adopt standard Galactic
values for the gas-to-dust ratio, $N_{\rm H}/A_V = 2 \times 10^{21}$
cm$^{-2}$ mag$^{-1}$, and the extinction law, $A_R/A_V = 0.75$
\citep{Bohlin:78,Rieke:85}.  The inferred extinctions reach peak
values of 4--14 mag near the galaxy center, although even these
represent azimuthally averaged values at $\sim$15\arcsec\ resolution.
After correction for extinction, the global SFRs increase by factors
ranging from 1.3 (for NGC 5457) to 140 (for NGC 5055).  Comparing the
disk-averaged extinction from this model to the radio-derived
extinction, we find that for several galaxies, especially NGC 4321 and
5055, this method severely overestimates the global extinction, by up
to 4 mag.

\item A simple alternative to the foreground dust screen model is a
model in which the stars, gas, and dust are uniformly mixed.  This
model is motivated by the expectation that H$\alpha$ traces stars
which have recently formed from molecular gas, so the H$\alpha$ and
dust are likely to be physically co-extensive.  For this case we have
\citep[e.g.,][]{Lequeux:81}:
\begin{equation}
A_V = 2.5 \log\, \frac{\tau}{1-e^{-\tau}}
\label{eqn:avmix}
\end{equation}
where $\tau$ is the extinction optical depth, again derived from the
standard Galactic extinction law, $\tau = A_V/1.086 = N_{\rm H} [\rm
cm^{-2}]/(2.17 \times 10^{21})$.  In the limit of small optical depth,
this reproduces the foreground screen model, $A_V = 2.5\, \log
e^{\tau/2}$, whereas for large $\tau$, $A_V$ increases
$\propto\log\tau$ rather than $\propto\tau$.  This leads to much
smaller extinction corrections at high gas densities, and closer
agreement with the radio-derived extinctions.  However, it appears to
systematically underestimate the extinction for the galaxies with
large $A_V$ (e.g., NGC 5033).

\item Finally, we consider a ``hybrid'' model in which the H$_2$ is
mixed with the stars (Eq.~\ref{eqn:avmix}) but the \HI\ acts as a
foreground screen, motivated by the larger scale height of the \HI\
layer.  When globally averaged, this leads to reasonable agreement
(within $\sim$0.5 mag) with the radio-inferred extinctions, as shown
in Figure~\ref{fig:radioha}(b).

\end{enumerate}

For the analyses in this paper we will employ both methods (1) and
(4).  Method (1), a uniform correction across the galaxy, has the
advantage that it is based directly on measurements (of H$\alpha$ and
radio fluxes) and agrees with previous studies that have found at most
a weak trend in \HII\ region extinction with galactocentric radius in
M51 and NGC 6946 \citep[e.g.,][]{vdH:88,Belley:92,Hyman:00}.  Method
(4), which presupposes a correlation between gas column density and
extinction, agrees roughly with the radio fluxes but is not
based on actual extinction measurements.  If incorrect, it may
lead to a spurious relationship between the gas density and star
formation rate---hence the importance of applying both methods.  The
results of applying methods (1) and (4) to the radial SFR profiles are
shown by the short-dashed and long-dashed lines respectively in the
lower panels of Fig.~\ref{fig:radprof}.

A recent study by \citet{Quillen:01}, based on a comparison of
H$\alpha$ with {\sl HST} Pa$\alpha$ imaging of a number of galaxies,
including the CO-rich spirals NGC 2903, M51, and NGC 6946, lends some
support to method (4).  They find that the foreground screen model
accurately predicts $A_{\rm H\alpha}$ at low gas surface densities
but overpredicts it at high densities.  This is exactly the type
of behavior provided by our hybrid model, since the low-$\Sigma_{\rm
H}$ regions are dominated by \HI\ whereas high-$\Sigma_{\rm H}$
regions are dominated by H$_2$.  Moreover, the typical extinctions in
the central regions of their galaxies are $A_V \sim$ 2--4, consistent
with the radial extinction profiles derived using method (4).  It is
worth noting that NIR recombination-line studies seldom reveal bright
\HII\ regions suffering extinctions of $A_V>4$, even in circumnuclear
starbursts \citep[e.g.,][]{Kot:00} or in situations where much larger
extinctions might be expected based on the gas surface density
\citep{Krabbe:94}.

Our extinction corrections may still underestimate the SFR if a
substantial fraction of Lyman continuum photons escape from galaxies
or are absorbed by dust within the \HII\ regions themselves, since
such photons will not lead to observable H$\alpha$ or radio continuum
emission.  The first effect, escape from the galaxy as a whole, is
likely to be small---for example, \citet{Dove:94b} estimate an escape
fraction for the Milky Way of $\sim$14\% (but see \citealt{Beckman:00}
for a contrasting view).  While this fraction may vary with radius,
\citet{Ferguson:96} find little variation in the diffuse H$\alpha$
fraction with radius, a quantity which they argue measures leakage of
Lyman continuum photons.  As for the second effect, dust absorption,
\citet{Mathis:86} estimates a typical optical depth to Lyman continuum
photons in \HII\ regions of $\tau \sim 1.4$, in which case we have
underestimated the SFR by up to a factor of $\sim$4.  An analysis by
\citet{McKee:97}, however, indicates that such absorption is less
significant in \HII\ region envelopes than within the Str\"{o}mgren
radius; globally they find only $\sim$25\% of ionizing photons are
absorbed by dust, yielding a correction factor of 1.37.  Given the
present lack of consensus, we apply no further corrections but note
that our SFR estimates may be somewhat low.



\begin{figure*}
\plottwo{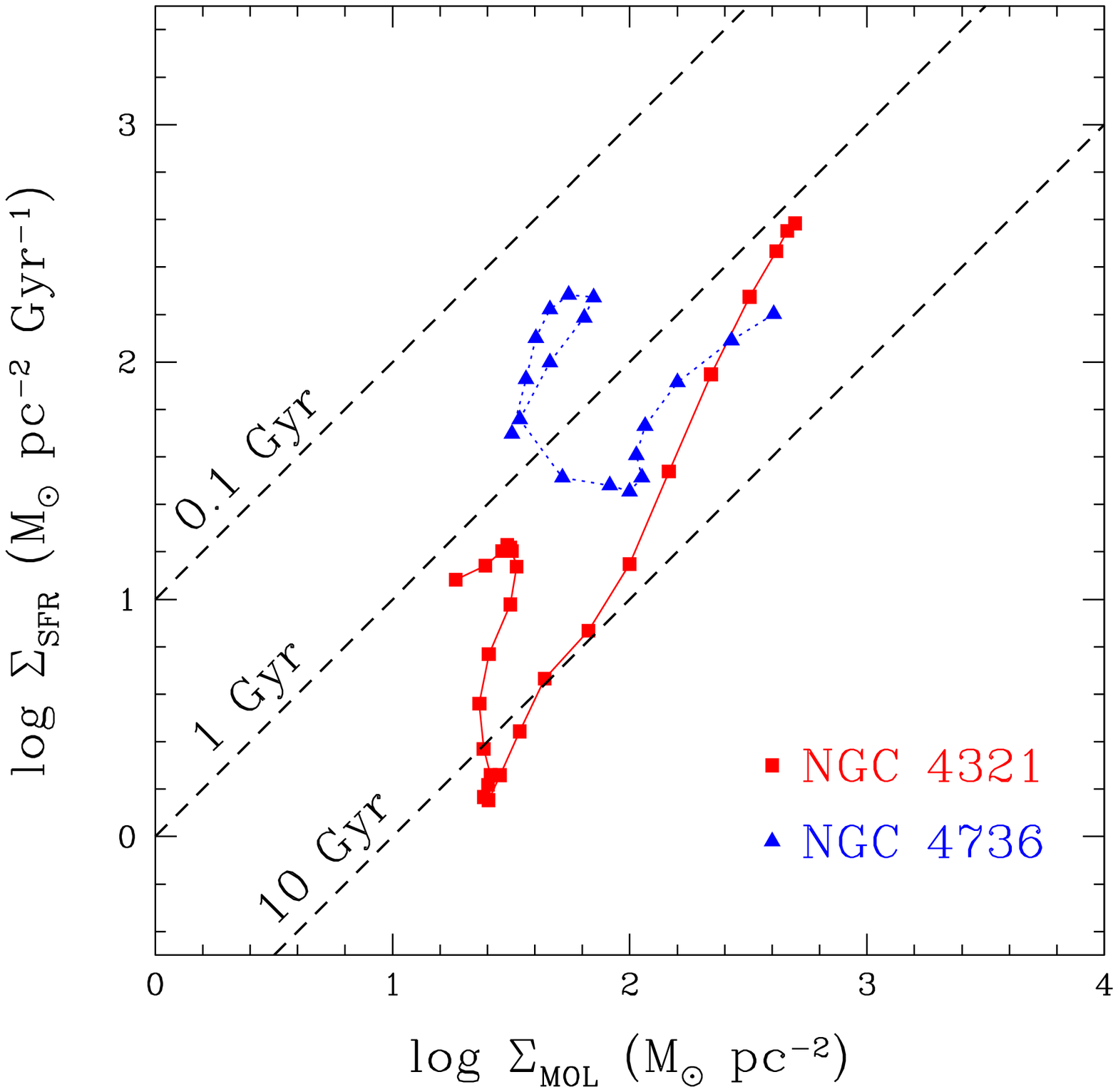}{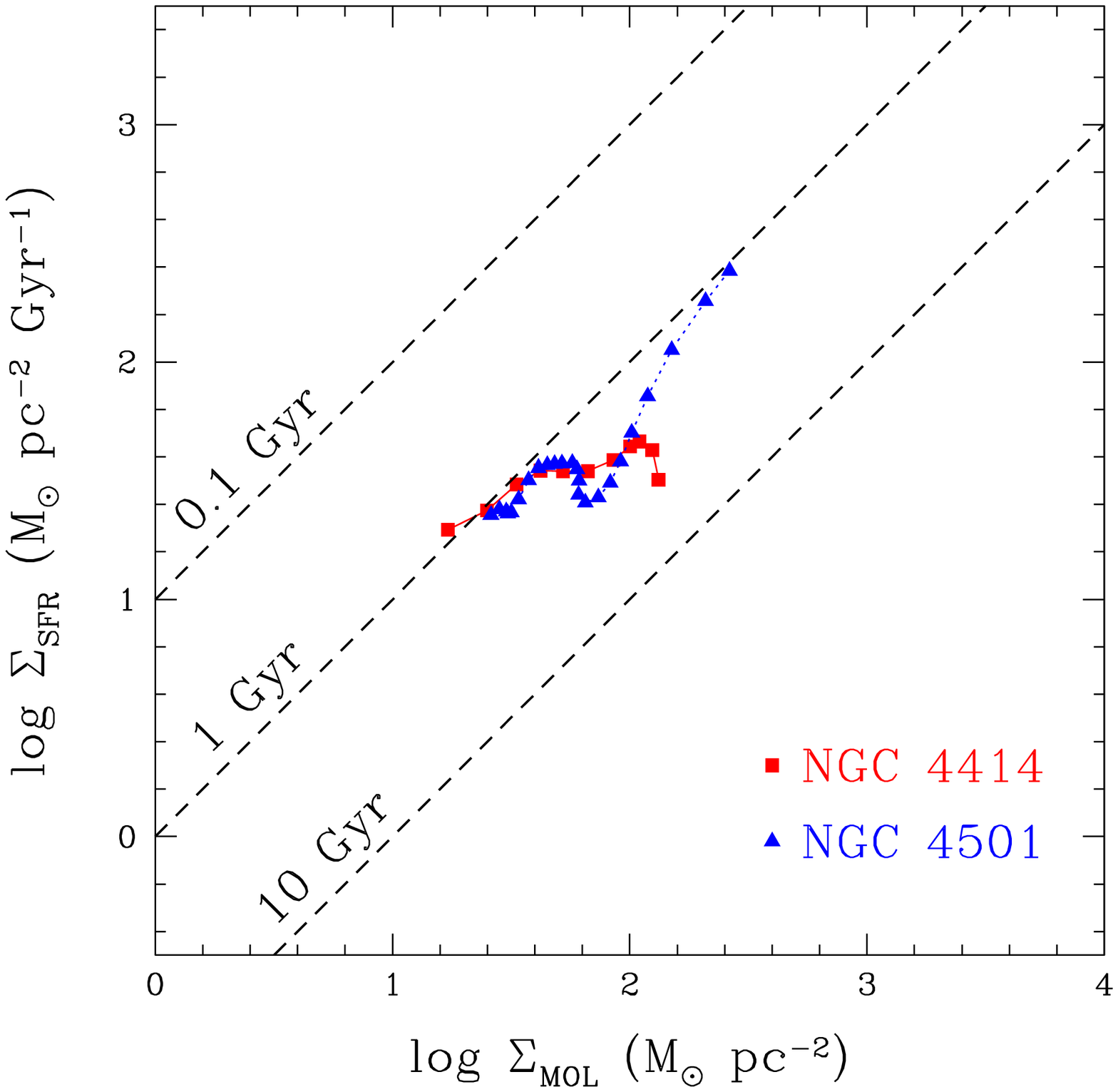}
\end{figure*}

\begin{figure*}
\plottwo{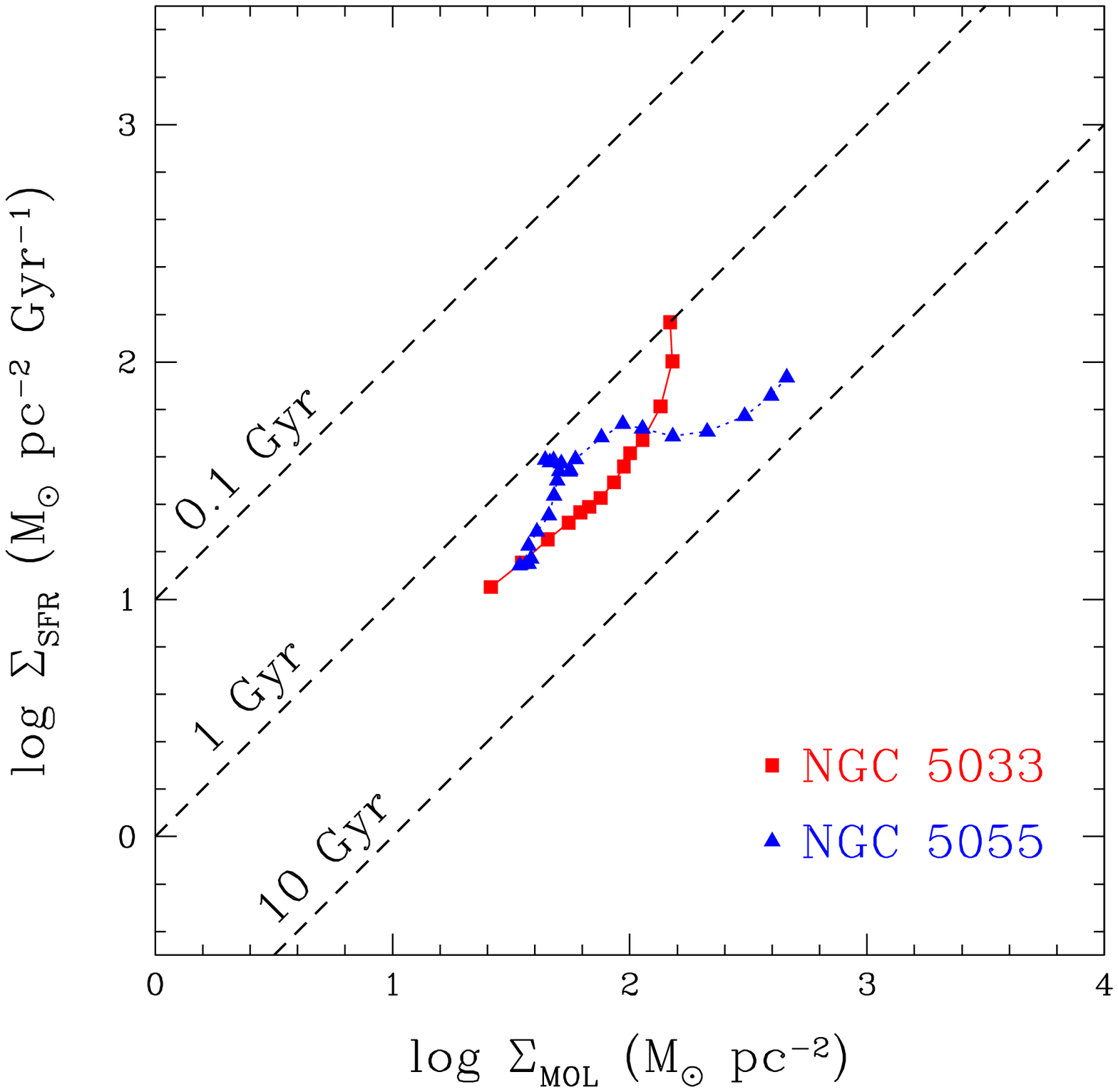}{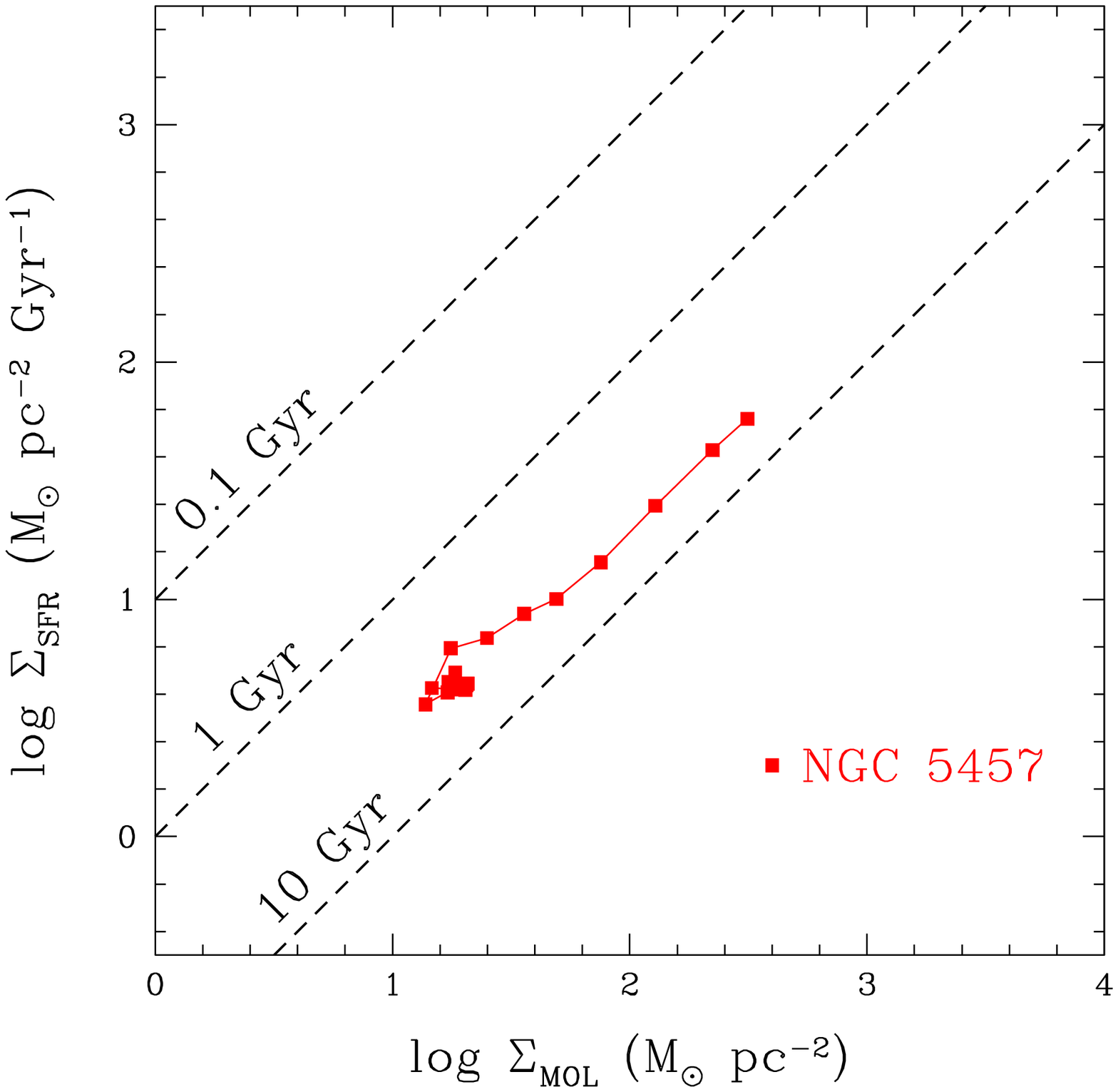}
\caption{
SFR plotted against H$_2$ gas density (including associated helium),
with both quantities azimuthally averaged, for the uniform extinction
model.  Dashed lines represent lines of constant star formation
efficiency, with the corresponding gas depletion time labeled
(assuming H$_2$ is the only fuel for star formation).  Note that we
have taken advantage of the higher resolution ($\sim$6\arcsec) of the
CO data in making the comparison.
\label{fig:sfrmol}}
\end{figure*}



\begin{figure*}
\plottwo{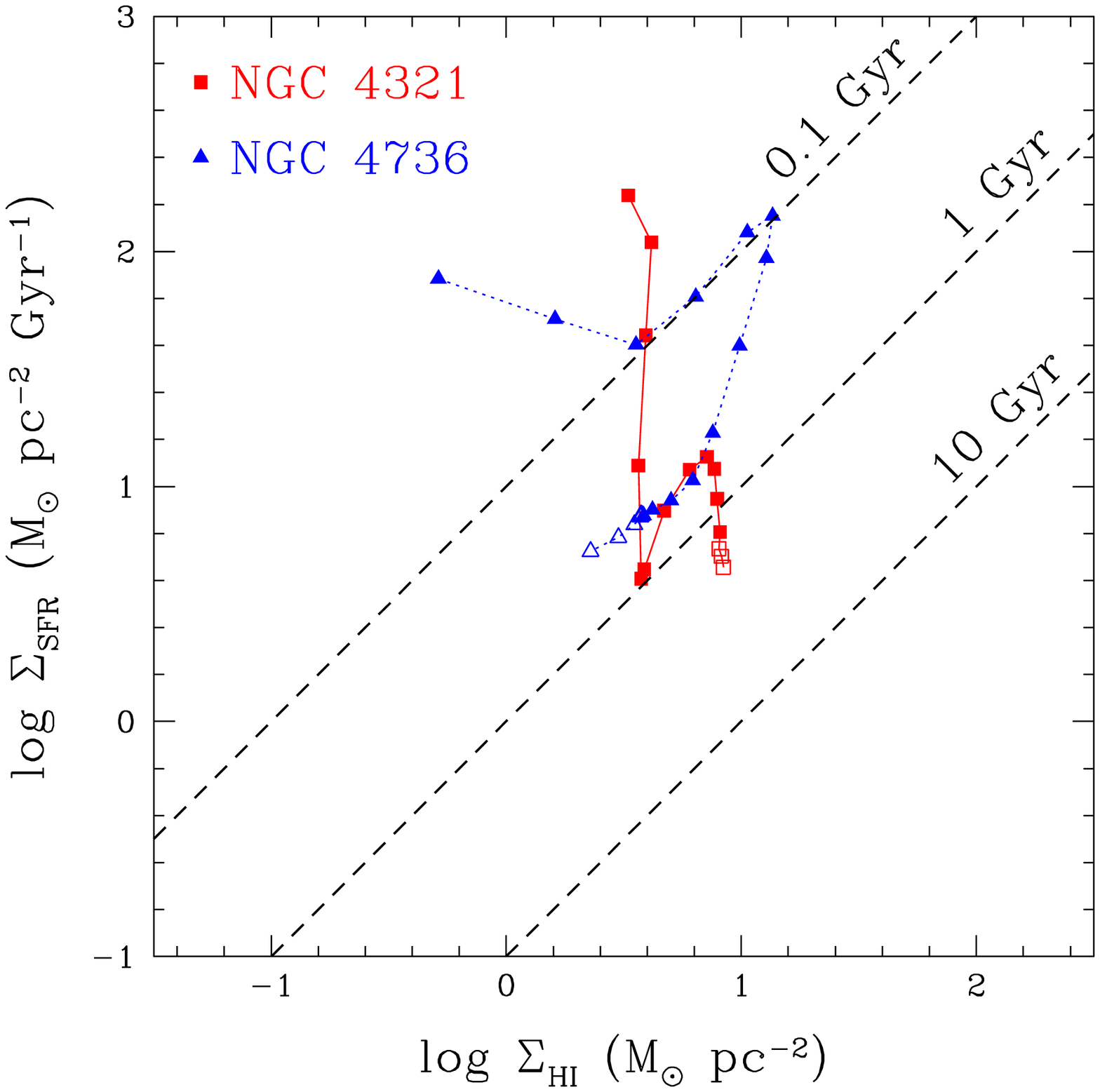}{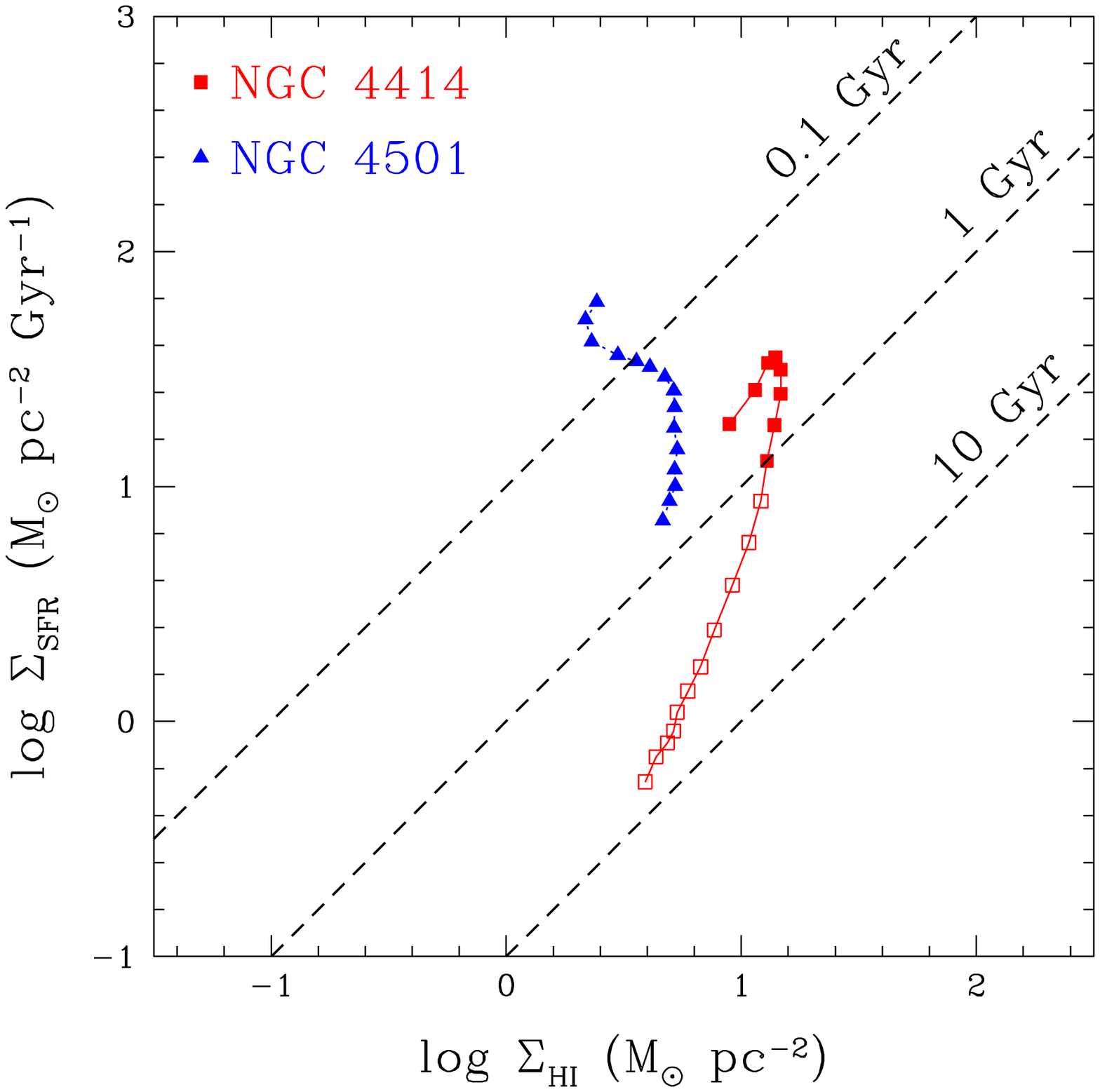}
\end{figure*}

\begin{figure*}
\plottwo{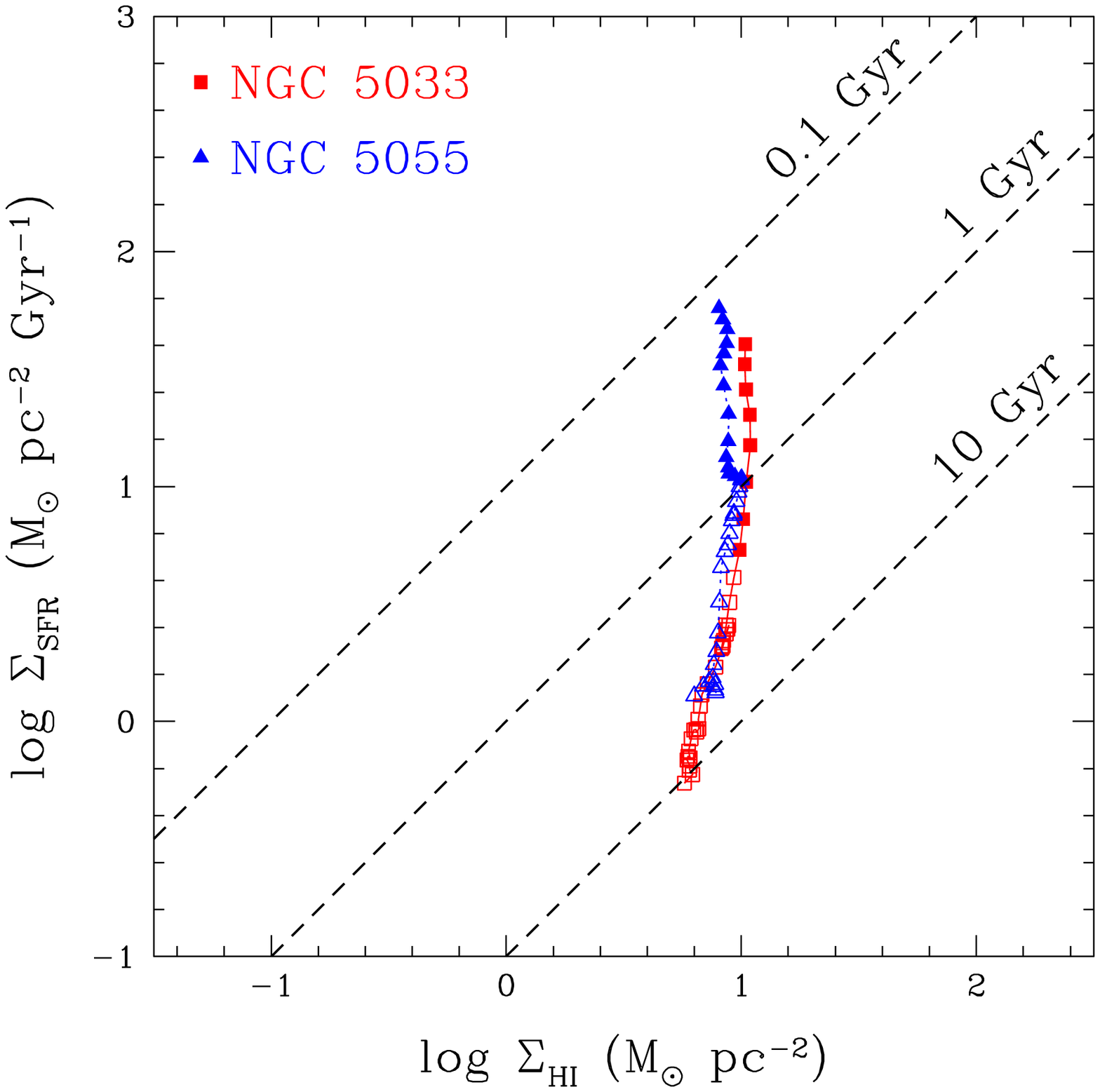}{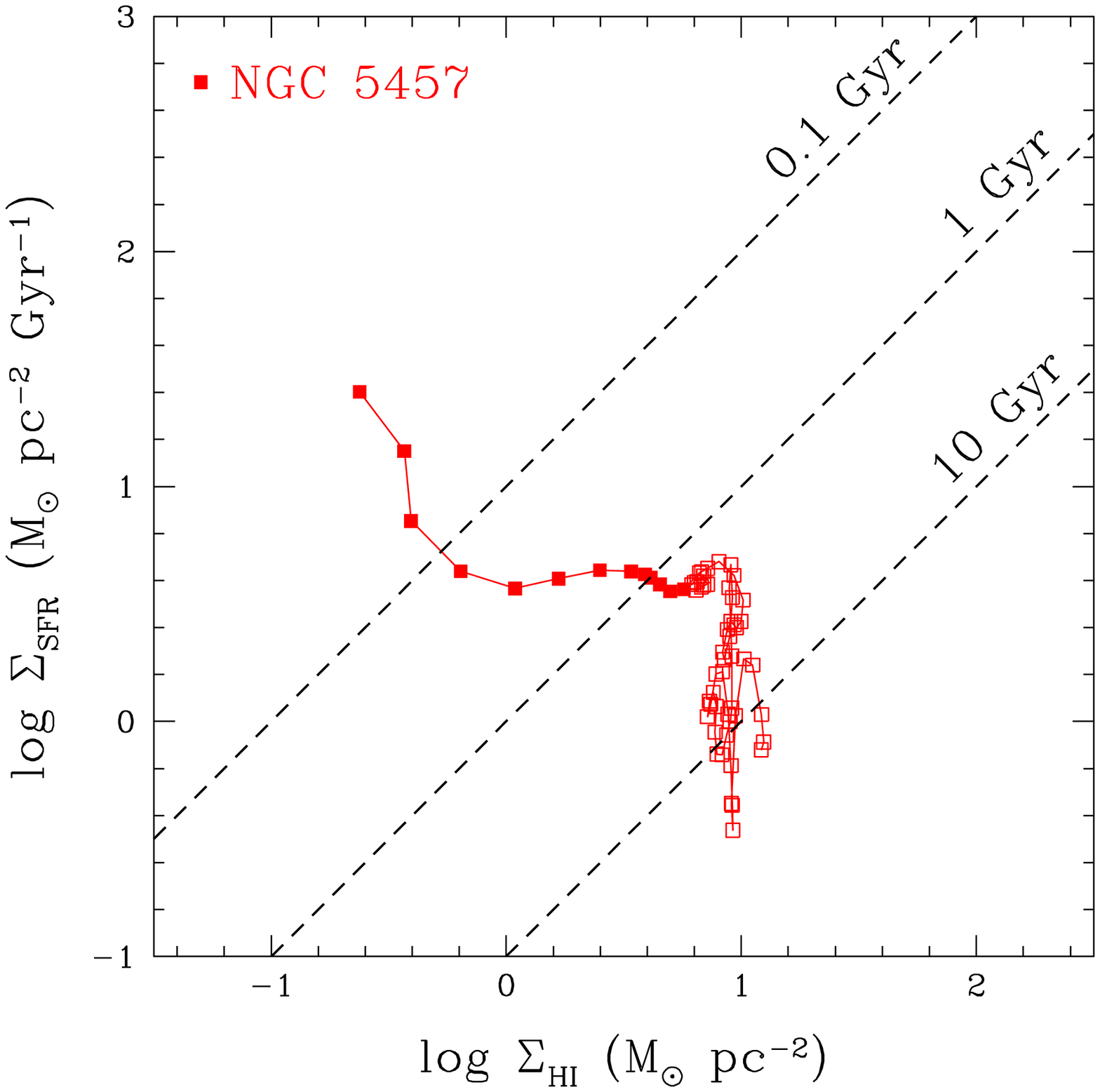}
\caption{
SFR plotted against \HI\ gas density (including associated helium),
with both quantities azimuthally averaged, for the uniform extinction
model.  Open symbols represent points where the molecular fraction is
$<$0.5.  Dashed lines represent lines of constant star formation
efficiency, with the corresponding gas depletion time labeled
(assuming \HI\ is the only fuel for star formation).
\label{fig:sfrat}}
\end{figure*}


\section{Results}\label{results}

\subsection{The Star Formation Law}\label{schmidt}

Using the radial profiles of CO, \HI, and total gas density derived in
\S\ref{prof}, we can compare the gas surface density with the observed
SFR surface density (derived from the H$\alpha$ profile) on a
ring-by-ring basis.  Such a comparison involves averaging over greater
areas for outer rings as opposed to inner rings, and thus both random
and systematic errors will vary with radius.  In particular, the lack
of independent data points may be a problem in the inner rings,
whereas errors in flat-fielding or masking may dominate for outer
rings.  To minimize the impact of our masking method, we have only
taken the radial profiles where they exceed the 3$\sigma_{\rm est}$
level defined in Equation~\ref{eqn:sigest}.  We note that despite
their limitations, radial profiles are fairly insensitive to angular
resolution and permit one to effectively average over time, which can
be useful for revealing overall trends.  A point-by-point comparison
of gas and star formation would be quite sensitive to the map
resolution and to evolutionary effects, since on small enough scales,
\HII\ regions will disrupt their natal clouds.  In a future paper, 
we will consider additional
techniques for analyzing the two-dimensional intensity maps for both
\siggas\ and \sigsfr\ provided by BIMA SONG.

\subsubsection{The SFR and Molecular vs.\ Atomic Gas}\label{molat}

We first consider the relationships between \sigsfr\ and the atomic
and molecular surface densities, \sighi\ and \sightwo, separately.
One advantage of such comparisons is that they avoid ambiguities
resulting from the extrapolation of the CO profile; moreover, the
comparison with \sightwo\ can be made at the higher resolution of the
BIMA data.  As described in \S\ref{extinct}, we consider two ways of
correcting the H$\alpha$ data for extinction: applying a uniform
correction across the disk, or a radially varying correction that
depends on the average gas column density.  While the latter method may seem
more physically motivated, evidence for radial extinction gradients is
inconclusive at present.  The results for the uniform extinction model
are shown in Figures~\ref{fig:sfrmol} and \ref{fig:sfrat}, with up to
two galaxies shown per panel.  

Most galaxies exhibit an excellent correlation between azimuthally
averaged H$\alpha$ and CO emission that can be parametrized using
a Schmidt (power) law:
\begin{equation}
n_{\rm mol} = \frac{d\log \sigsfr}{d\log \Sigma_{\rm mol}}\;,
\end{equation}
For the $N_{\rm H}$-dependent extinction corrections, the inferred
power-law indices range from $n_{\rm mol} \approx 1$ for NGC 4414 to
$n_{\rm mol} \approx 2$ for NGC 5033 (see
Table~\ref{tbl:nschmidt}).  The two exceptions are NGC 4321, where the
SFR is suppressed in the vicinity of a bar ($R$=15\arcsec--55\arcsec,
where $\log \Sigma_{\rm mol} \approx 1.4$), and NGC 4736, where the
SFR is enhanced near a resonance ring ($R$=30\arcsec--55\arcsec, where
$\log \Sigma_{\rm mol} \approx 1.6$).  These examples suggest that the
molecular content alone does not always control the SFR---dynamical
effects can also be important \citep[see e.g.,][]{Wong:00,Meier:01}.
Even in NGC 4321 and 4736, however, there are regions where a
power-law relationship is roughly obeyed.  A weighted average for all
seven galaxies gives $\bar{n}_{\rm mol}$=0.8 for the uniform
extinction model and $\bar{n}_{\rm mol}$=1.4 for the varying
extinction model.  For comparison, \citet{Rownd:99} made
point-by-point CO and H$\alpha$ comparisons for over 100 galaxies at
45\arcsec\ resolution, and found $n_{\rm mol}$=1.2--1.4 with no
extinction corrections applied.  However, they note that $n_{\rm
mol}$=1 (constant SFE for the molecular gas) is not excluded by their
data.  Assuming $A_V$ does not vary strongly with radius, our study
also finds reasonable agreement with a law in which the SFR per unit
area is proportional to the molecular gas surface density.

The relationship between \sigsfr\ and \sighi\ is more complex and
appears inconsistent with a Schmidt law.  In some cases, there is an
abrupt decline in \sighi\ even when \sigsfr\ is fairly high (NGC 4736
and 5457, and to a lesser extent NGC 4414 and 4501).  These galaxies
have central \HI\ depressions where significant star formation is
still occurring.  In other cases, the value of \sighi\ stays fairly
constant over a wide range in \sigsfr, suggesting that a maximum value
of \sighi\ has been reached.  This maximum value, roughly 10 \Msol\
pc$^{-2}$ or $N_{\rm HI} \sim 10^{21}$ atoms cm$^{-2}$, could result
from a tendency for \HI\ to convert to H$_2$ at higher column
densities due to self-shielding \citep{Federman:79,Shaya:87}.
Alternatively, the optical depth in the 21-cm line could become
substantial above this column density, especially if a cold \HI\
component dominates \citep[e.g.,][]{Braun:97}.  The opacity at the
line center is given by
\citep{Dickey:90b}:
\begin{equation}
\tau = 5.2 \times 10^{-19} \left(\frac{N_{\rm HI}}{T\Delta v}\right)\;,
\end{equation}
which produces $\tau$=1 at $N_{\rm HI}$=$10^{21}$ cm$^{-2}$ for $T$=50
K and $\Delta v$=10 \kms.  However, Fig.~\ref{fig:radprof} suggests
that the maximum deprojected $N_{\rm HI}$ is not dependent on
inclination as one would expect if it were an optical depth effect:
NGC 5033 ($\cos i$=0.37) and 5457 ($\cos i$=0.93) both exhibit
azimuthally averaged \HI\ surface densities that plateau around
$\sim$6--8 \Msol\ pc$^{-2}$.

Surprisingly, \citet{KC:89,KC:98a} has found that the disk-averaged
SFR is more strongly correlated with the disk-averaged \HI\ surface
density than with the CO surface density.  Similar conclusions were
drawn by \citet{Deharveng:94}, who traced the SFR using
far-ultraviolet (FUV) data, and \citet{Boselli:94}, who used both FUV
and H$\alpha$.  Our study, based on fully-sampled CO maps, is not
consistent with these findings.  The lack of a correlation between
\sigsfr\ and \sighi\ in our data may result from a bias in our sample
towards molecule-rich galaxies, where the \HI\ surface density tends
to saturate at a value of $\sim$10 \Msol\ pc$^{-2}$.  Galaxies with
substantially less molecular gas may show a wider range of \sighi.  It
is more difficult to understand why the correlation between global CO
flux and SFR found by Kennicutt and others should be so poor, but
K98 notes that the correlation is improved when only
metal-rich spirals (such as those in our sample) are considered, an
effect he attributes to variations in the CO-to-H$_2$ conversion
factor ($X$-factor).  Furthermore, the inhomogeneity of the CO
distribution in most galaxies suggests that incorrect H$_2$ masses may
be derived when the CO distribution is not completely mapped.


\begin{figure*}[t]
\plottwo{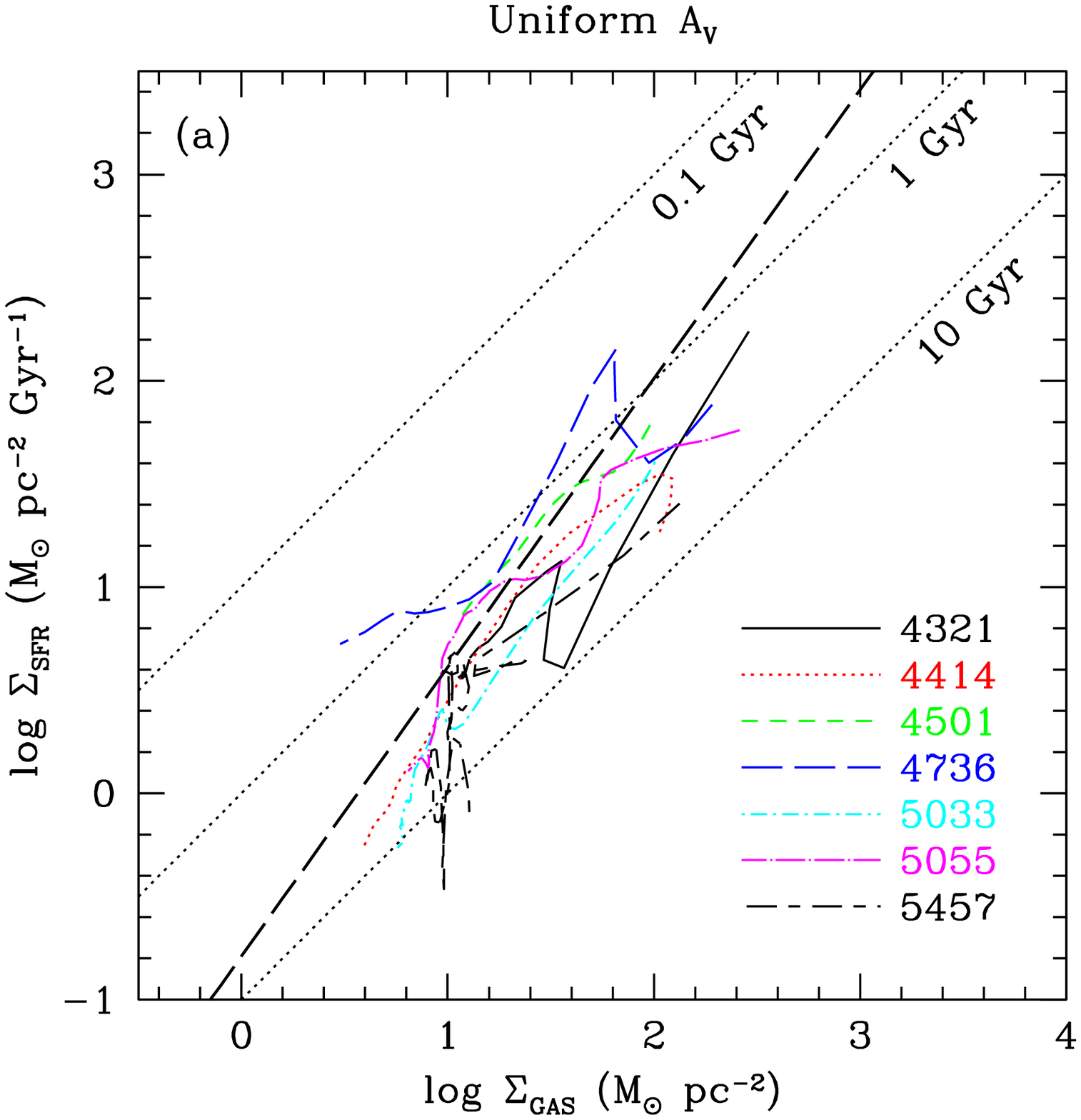}{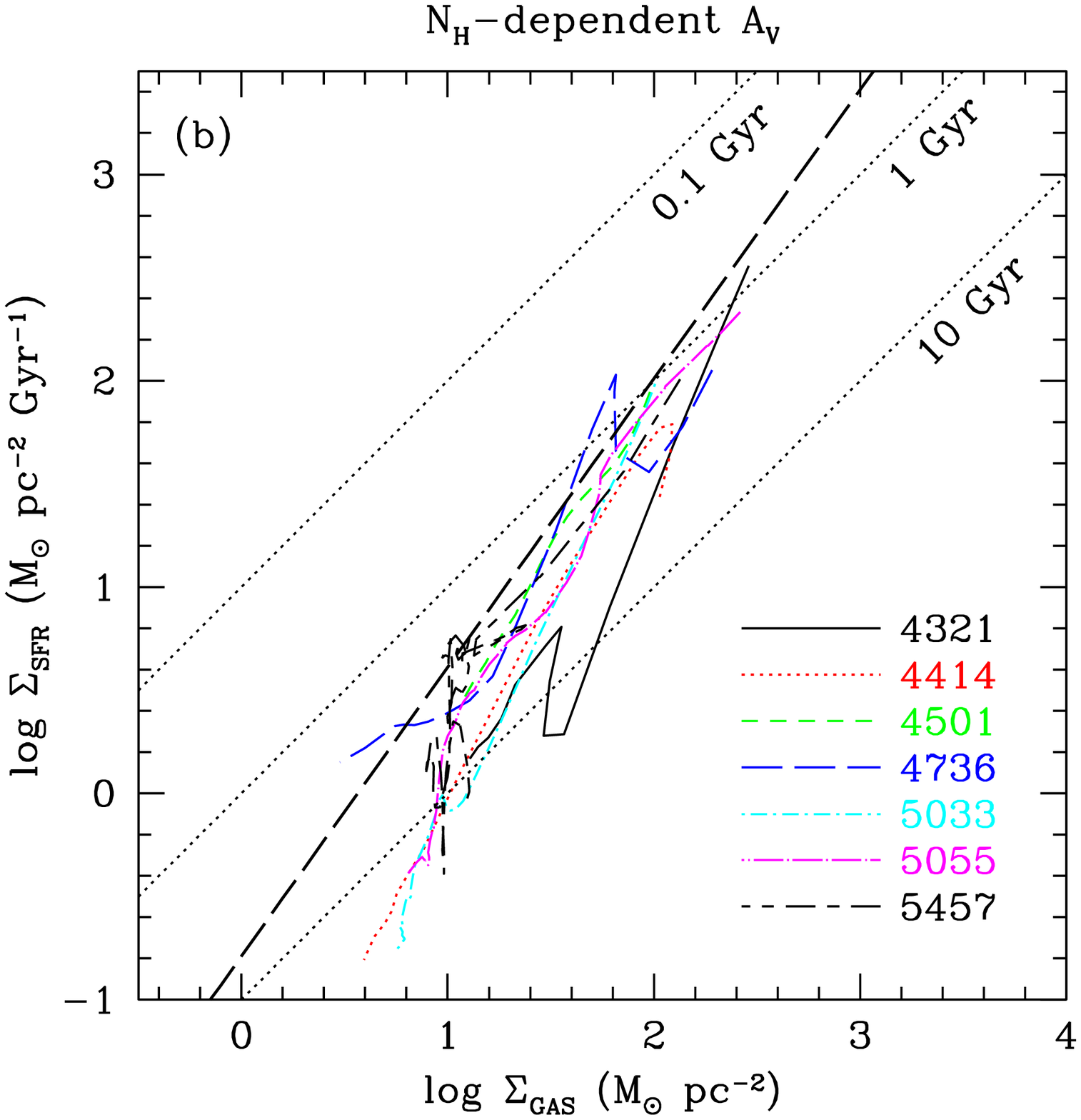}
\caption{
SFR plotted against total gas density for rings within all seven
galaxies, after applying (a) uniform extinction corrections; (b)
$N_{\rm H}$-dependent extinction corrections.  Parallel
dotted lines represent lines of constant SFR per unit gas mass,
with the corresponding gas depletion time labeled.  The heavy dashed
line is the global Schmidt law derived by K98.
\label{fig:sfrgas}}
\end{figure*}


\subsubsection{The SFR and the Total Gas Surface Density}

Combining the CO and \HI\ data (with the CO profile extrapolated to
larger radii) enables us to compare the SFR surface density with the
{\it total} gas surface density, which has been argued to be a more
relevant quantity for star formation than \sightwo\ alone
(\citealt{KC:89}; K98).  Figure~\ref{fig:sfrgas} shows the radial
profiles for all seven galaxies as plotted in the \siggas-\sigsfr\
plane for the two extinction models.  Regardless of how extinction is
treated, there is indeed a strong correlation between \siggas\ and
\sigsfr\ within galaxies, indicative of a Schmidt law
(Eq.~\ref{eqn:schmidt}).  As is clear from the results of
\S\ref{molat}, however, this correlation is driven entirely by the
molecular component for these galaxies, as \sigsfr\ shows virtually no
correlation with \sighi.  We note that the situation may be different
in galaxies where the ISM is predominantly atomic, and further studies
should be aimed at exploring the connection between star formation and
the \HI\ component in such galaxies.

Also shown in Fig.~\ref{fig:sfrgas} as a dashed line is the composite
Schmidt law derived by K98,
\begin{equation}
\Sigma_{\rm SFR} = 0.16
	\rm\left(\frac{\Sigma_{gas}}{1\,\Msol\,pc^{-2}}
	\right)^{1.4}\;\Msol\,Gyr^{-1}\,pc^{-2}\;.
\end{equation}
(Note that the numerical coefficient has been adjusted to include
helium in \siggas.)  Kennicutt's parameterization is based on averages
over the star-forming disks of 61 normal galaxies and the inner
$\sim$1 kpc disks of 36 starbursts.  Two features are evident from
Fig.~\ref{fig:sfrgas}.  First, the use of an $N_{\rm H}$-dependent
extinction correction produces a steeper dependence of \siggas\ on
\sigsfr\ than a uniform correction, as would be expected
\citep[e.g.,][]{Buat:89}.  Second, most of the observed data points
lie to the right of Kennicutt's relation.  Note, however, that many of
the points at larger \siggas\ in Fig.~\ref{fig:sfrgas} occur at inner
radii, which are weighted less when taking a global average.
Averaging our data across the entire observed disk region leads to
much better agreement with Kennicutt's results (Figure~\ref{fig:rck}),
even though our outer radius is generally smaller than the optical
radius adopted by K98.  Indeed, we find that the offsets between our
disk-averaged values and Kennicutt's points for the same galaxies in
Fig.~\ref{fig:rck} can largely be attributed to a change in the
normalizing radius, which shifts points along a line of unit slope for
the same total SFR and $M_{\rm gas}$.  We conclude that within the
uncertainty of $\pm$0.2 dex adopted by K98, our results are consistent
with his.

Table~\ref{tbl:nschmidt} gives the average Schmidt law index,
\begin{equation}
n = \frac{d\log \sigsfr}{d\log \siggas}\;,
\end{equation}
within each galaxy as derived from an unweighted least-squares fit to
the curves in Fig.~\ref{fig:sfrgas}.  An average of the indices for
all seven galaxies, weighted by the inverse of their variances, gives
$\bar{n} = 1.1 \pm 0.2$ for the uniform extinction model and $\bar{n} =
1.7 \pm 0.3$ for the $N_{\rm H}$-dependent extinction model.  We
examined the effect of changing the radial CO profile in the region
where it had been extrapolated: in the extreme cases of a sharp
truncation or complete flattening of the CO profile, we obtained
values of $\bar{n} =$ 0.8 and 1.45 respectively for the uniform
extinction model.  However, such extreme profiles seem unlikely, and
we consider the 1$\sigma$ errors on $\bar{n}$ stated above to be
realistic.  Given the uncertainties, our derived indices are roughly
consistent with the Schmidt law {\it among} galaxies ($n \approx 1.4$)
found by K98.  Note that there would not necessarily have been a
correspondence between the azimuthally averaged and global Schmidt
laws if star formation in galaxies depended on some quantity other
than \siggas, but for which a disk-averaged $\left<\siggas\right>$ was
a convenient proxy.  Thus, the validity of the Schmidt law on both
local and global scales probably reflects an underlying physical link
between \siggas\ and \sigsfr.


\vskip 0.25truein
\includegraphics[width=3.25in]{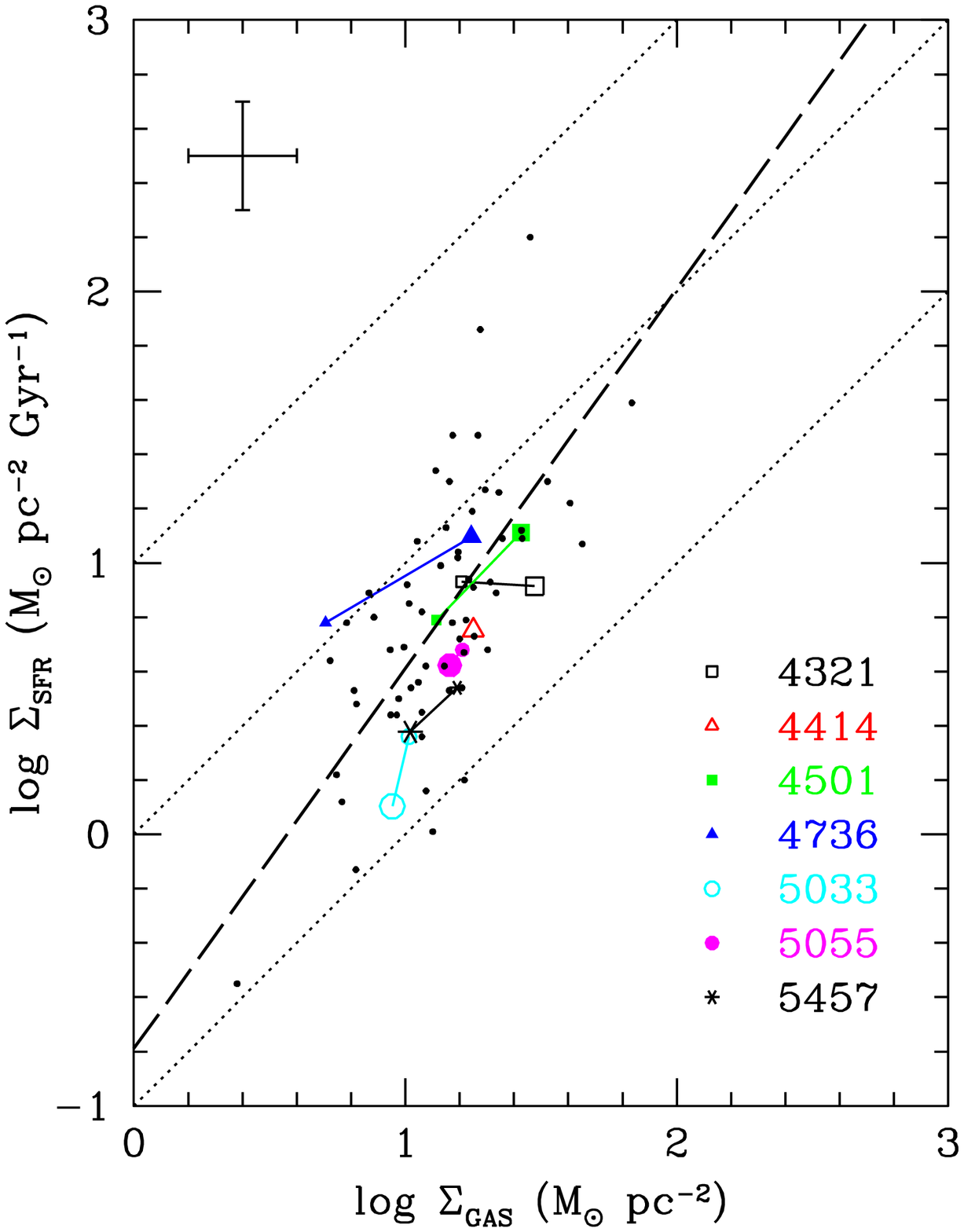}
\figcaption{
Disk-averaged SFRs plotted against total gas density.  Large plot
symbols are averages derived from radial profiles given in this paper,
with radially varying extinction corrections applied.  When connected
by a line to a smaller symbol, the small symbol represents the
corresponding point for the same galaxy as given by K98.
The small dots are the other disk averages calculated by K98, 
the heavy dashed line is his global Schmidt law, and the error
bars in the upper left represent his estimate of the uncertainties.
\label{fig:rck}}
\vskip 0.25truein



\begin{table*}[t]
\begin{center}
\caption{Observed Schmidt Law Indices for Total Gas and H$_2$ Only
\label{tbl:nschmidt}}
\bigskip
\begin{tabular}{ccccc}
\hline\hline
& \multicolumn{2}{c}{total gas} & \multicolumn{2}{c}{molecular gas}\\
Galaxy & $A_V$ const. & $A_V(N_{\rm H})$ & 
$A_V$ const. & $A_V(N_{\rm H})$\\[0.5ex]
\hline\hline
NGC 4321 & 1.13$\pm$.12 & 1.74$\pm$.14 & 1.41$\pm$.16 & 2.08$\pm$.16\\
NGC 4414 & 1.15$\pm$.07 & 1.71$\pm$.05 & 0.32$\pm$.06 & 0.98$\pm$.06\\
NGC 4501 & 0.95$\pm$.04 & 1.60$\pm$.03 & 0.90$\pm$.10 & 1.63$\pm$.10\\
NGC 4736 & 0.82$\pm$.09 & 1.23$\pm$.09 & 0.04$\pm$.24 & 0.74$\pm$.24\\
NGC 5033 & 1.38$\pm$.04 & 2.06$\pm$.04 & 1.27$\pm$.13 & 2.01$\pm$.13\\
NGC 5055 & 1.14$\pm$.07 & 1.79$\pm$.06 & 0.55$\pm$.08 & 1.30$\pm$.08\\
NGC 5457 & 1.18$\pm$.17 & 1.56$\pm$.17 & 0.88$\pm$.03 & 1.45$\pm$.04\\
Weighted~Avg. & 1.12$\pm$.22 & 1.75$\pm$.25 & 0.78$\pm$.34 & 1.36$\pm$.35\\
\hline\hline
\end{tabular}
\end{center}
\end{table*}



\begin{figure*}[b]
\plottwo{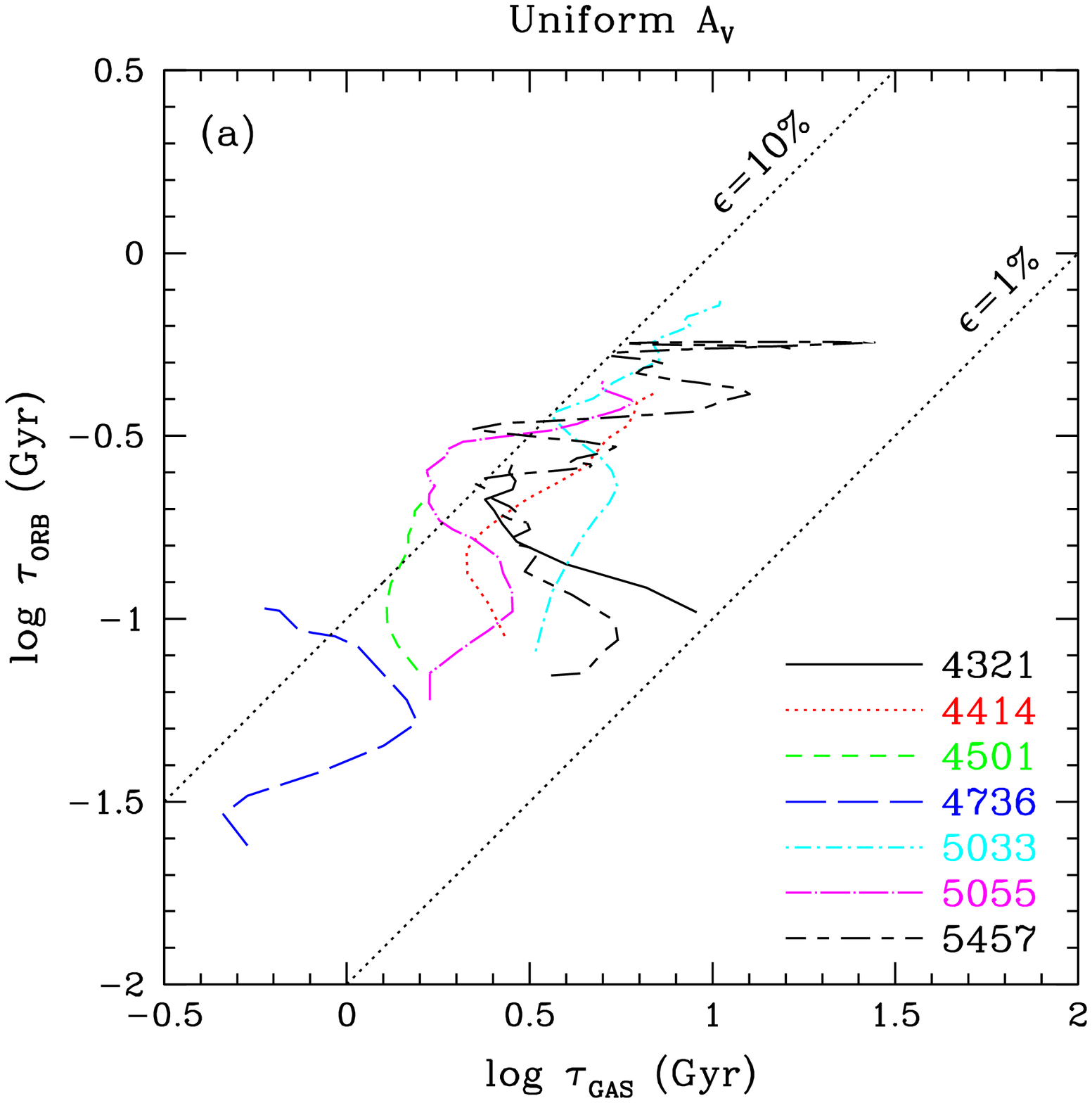}{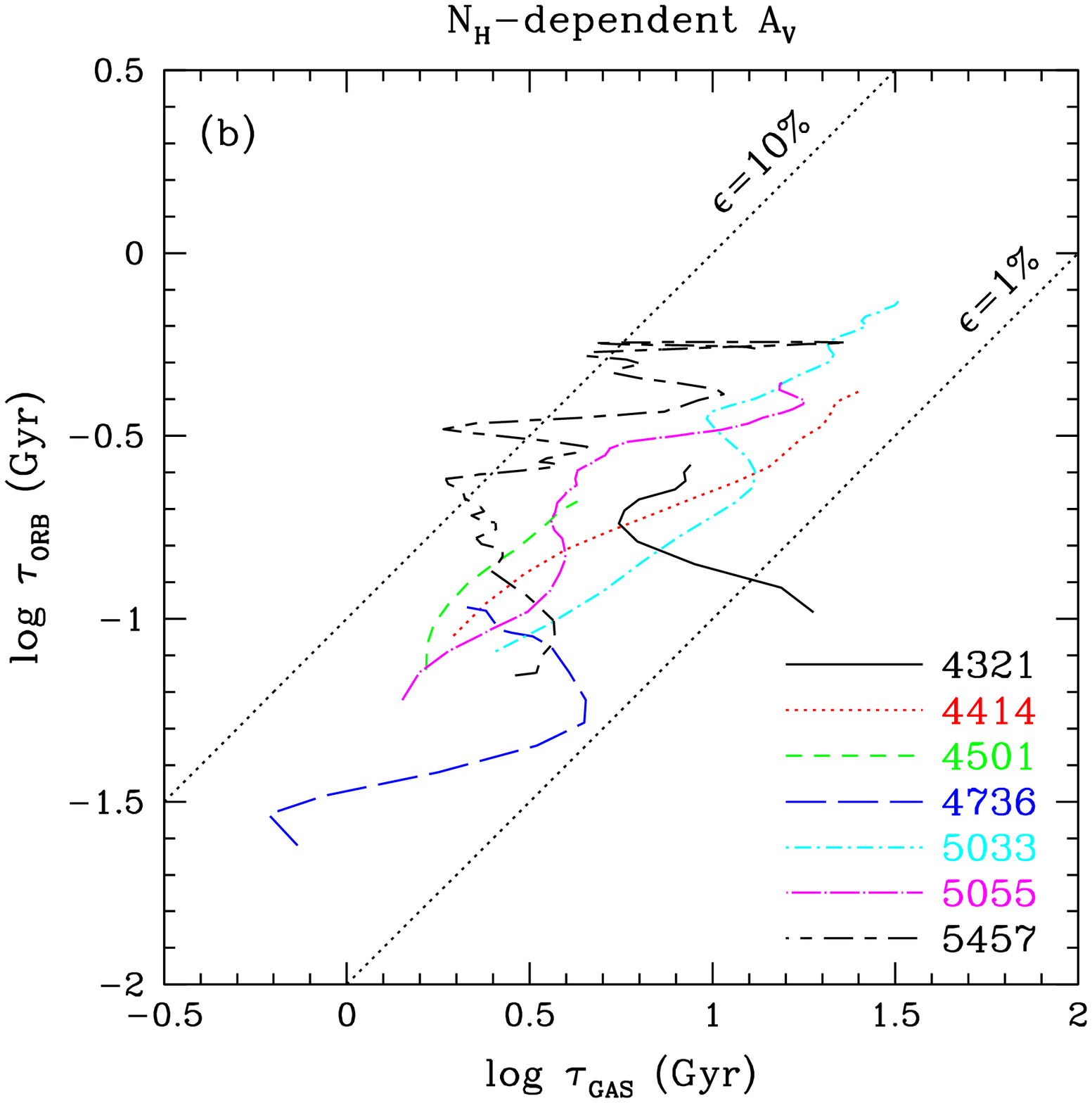}
\caption{
Orbital period $\tau_{\rm orb} = 2\pi/\Omega$ plotted against
gas depletion time \tgas\ for annuli within each of the seven 
galaxies, after applying (a) uniform extinction corrections; (b)
$N_{\rm H}$-dependent extinction corrections.
Parallel dotted lines denote points where 1\% or 10\% of the
available gas is consumed by star formation during each orbit.
\label{fig:tautau}}
\end{figure*}


\subsubsection{Gas Depletion vs.\ Orbital Timescales}\label{torb}

We have also considered an alternative description of the star formation 
law given by
\begin{equation}
\sigsfr \propto \siggas \Omega\;,
\end{equation}
where $\Omega$ is the orbital frequency
\citep[e.g.,][]{Larson:88,Silk:97}.  This law can be expressed as
\begin{equation}
\tau_{\rm orb} = \epsilon \tgas\,,
\label{eqn:silk}
\end{equation}
where $\tgas \equiv \siggas/\sigsfr$ is the timescale for gas depletion by star
formation, $\tau_{\rm orb}$ is the orbital period and $\epsilon$ is an
efficiency factor that represents the fraction of available gas that
is consumed in each orbit.  K98 finds that Equation~\ref{eqn:silk},
with $\epsilon \sim 0.1$, provides an equally good fit to the
disk-averaged data in comparison to the Schmidt law.  However, this
prescription can only be properly tested using azimuthally averaged
data {\it within} galaxies, since a disk-averaged $\left<\tau_{\rm
orb}\right>$ may simply be a reflection of the galaxy's dynamical
mass.

Figure~\ref{fig:tautau} shows the relation between $\tau_{\rm orb}$
and \tgas\ for the two extinction models we are considering.
$\tau_{\rm orb}$ was determined from rotation curves fitted to the CO
and \HI\ velocity fields, which had been derived from the datacubes by
an automated Gaussian fitting routine (details will be given in Paper
II).  Only points where $R>30\arcsec$ have been plotted, since inside
of that region the adopted rotation curve has a higher spatial
resolution than the total gas profile.  Applying a uniform extinction,
we find that although a roughly linear trend {\it among} galaxies is
apparent, as found by K98, no strong linear relation exists {\it
within} galaxies.  This is not surprising, since \tgas\ tends to be
constant with radius when uniform extinction corrections were applied,
whereas Eq.~\ref{eqn:silk} predicts $\tgas \propto R$ for a flat
rotation curve.  The correlation improves somewhat for the varying
extinction model [Fig.~\ref{fig:tautau}(b)], and indeed the scatter in
$\log [\sigsfr/(\Sigma_{\rm gas}\Omega)]$ is comparable to the scatter
in $\log(\sigsfr/\Sigma_{\rm gas}^{1.4})$ for this model.  Thus a star
formation law where $\tgas \propto \tau_{\rm orb}$ appears to be as
valid a prescription for the SFR as the Schmidt law, but only if the
extinction increases towards the center to allow for a significant
gradient in \tgas.  To truly distinguish between the two prescriptions
will require better constraints on the extinction, or the use of star
formation tracers impervious to extinction.



\begin{figure*}
\plottwo{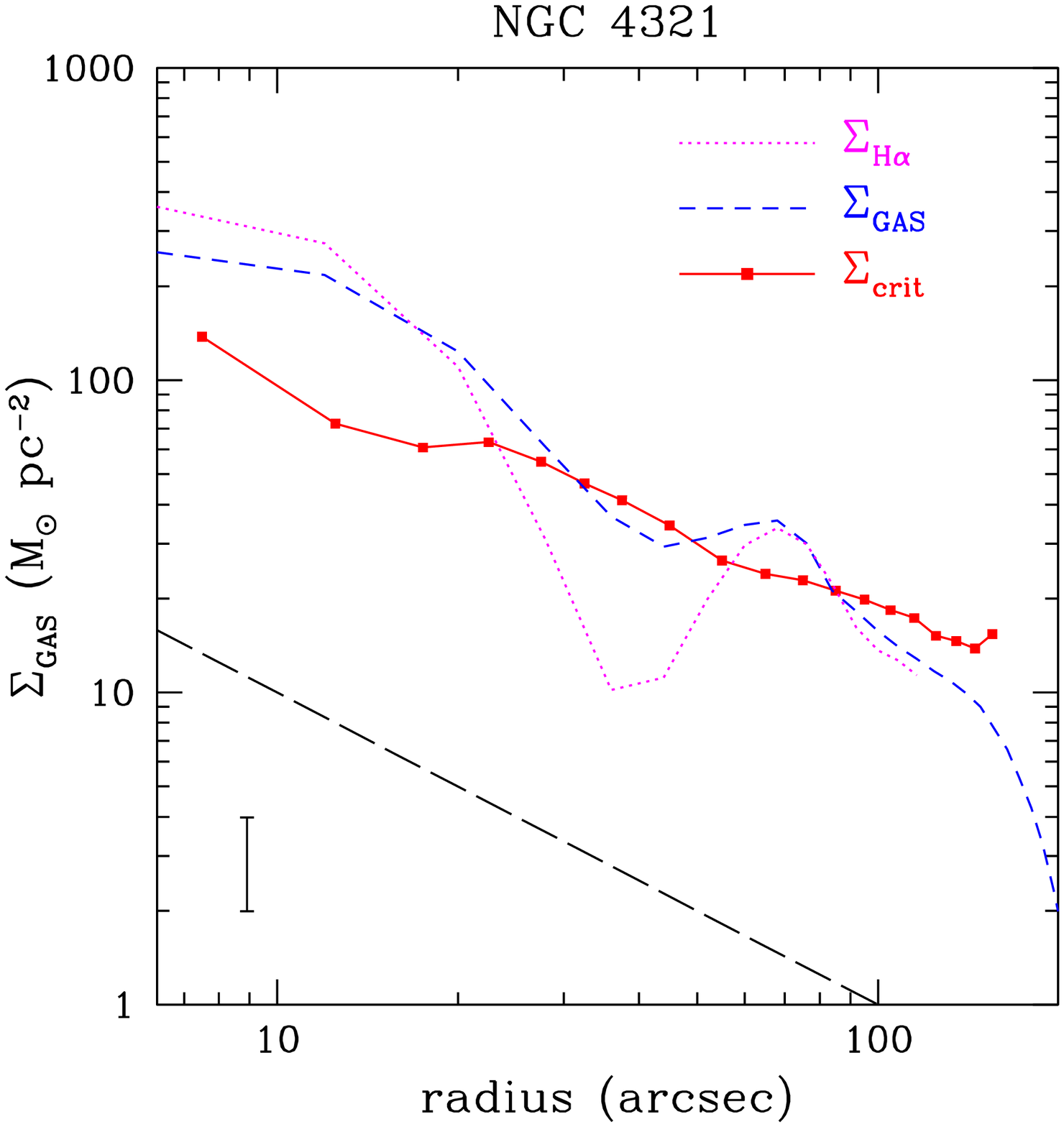}{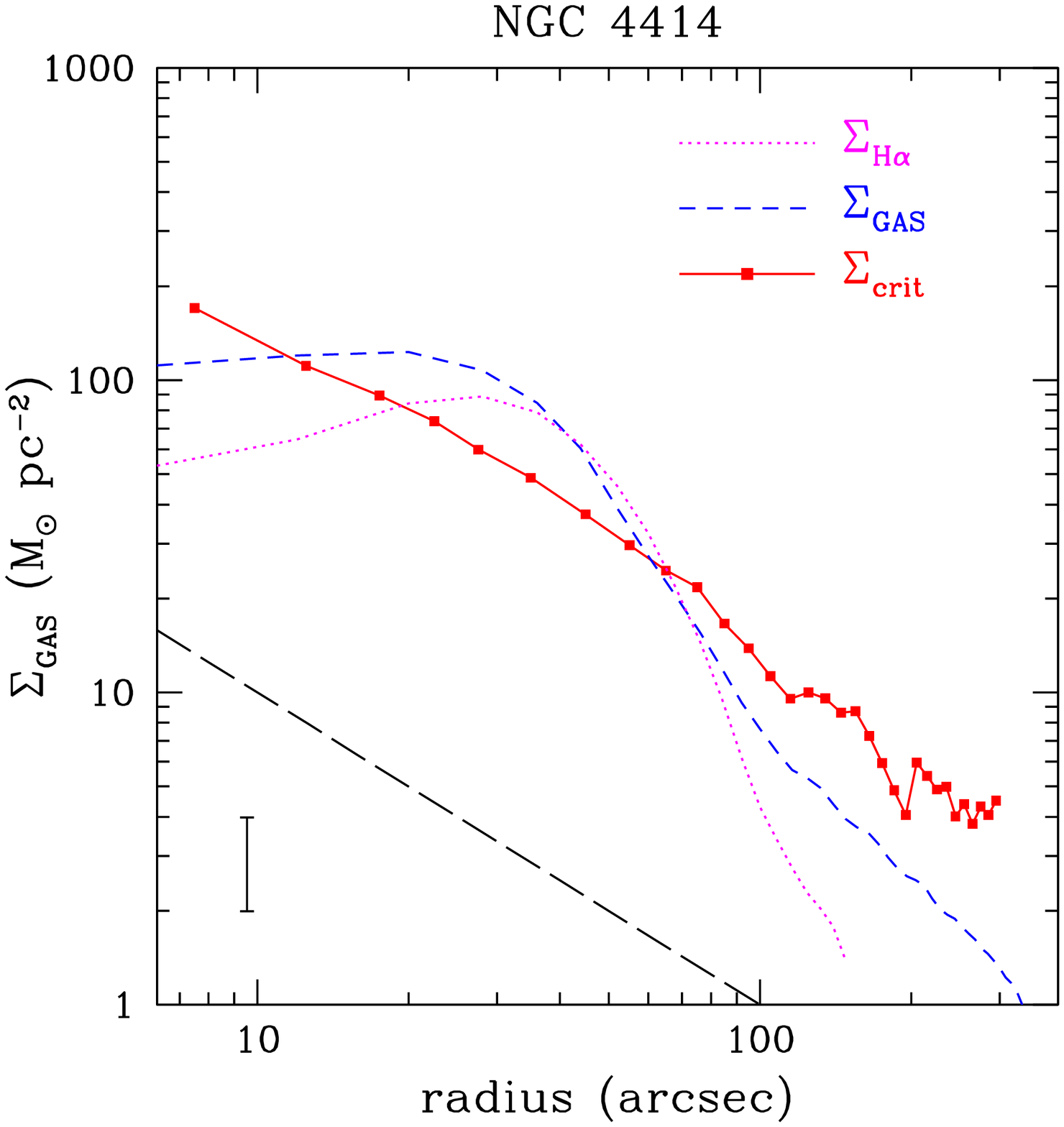}
\end{figure*}

\begin{figure*}
\plottwo{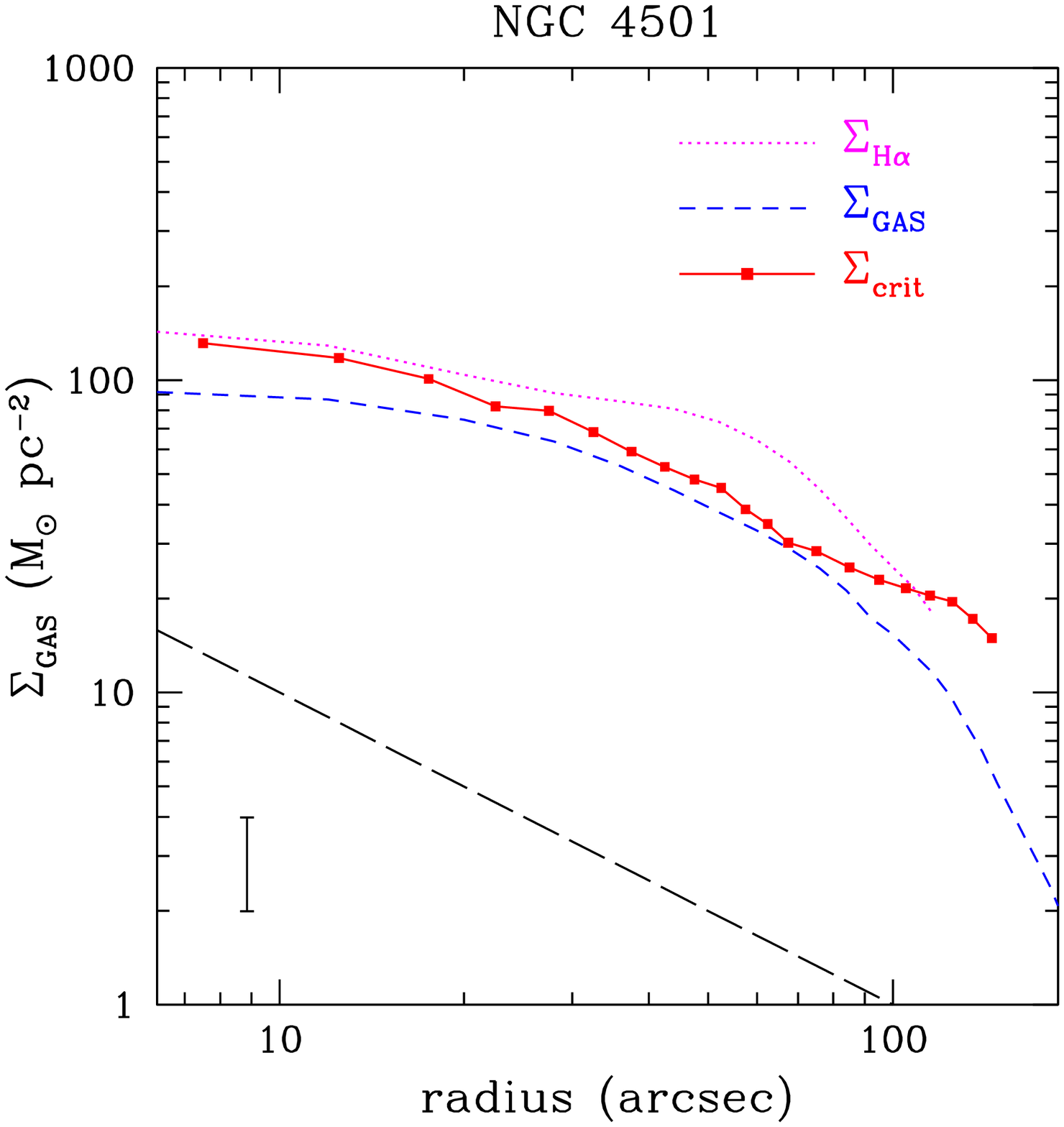}{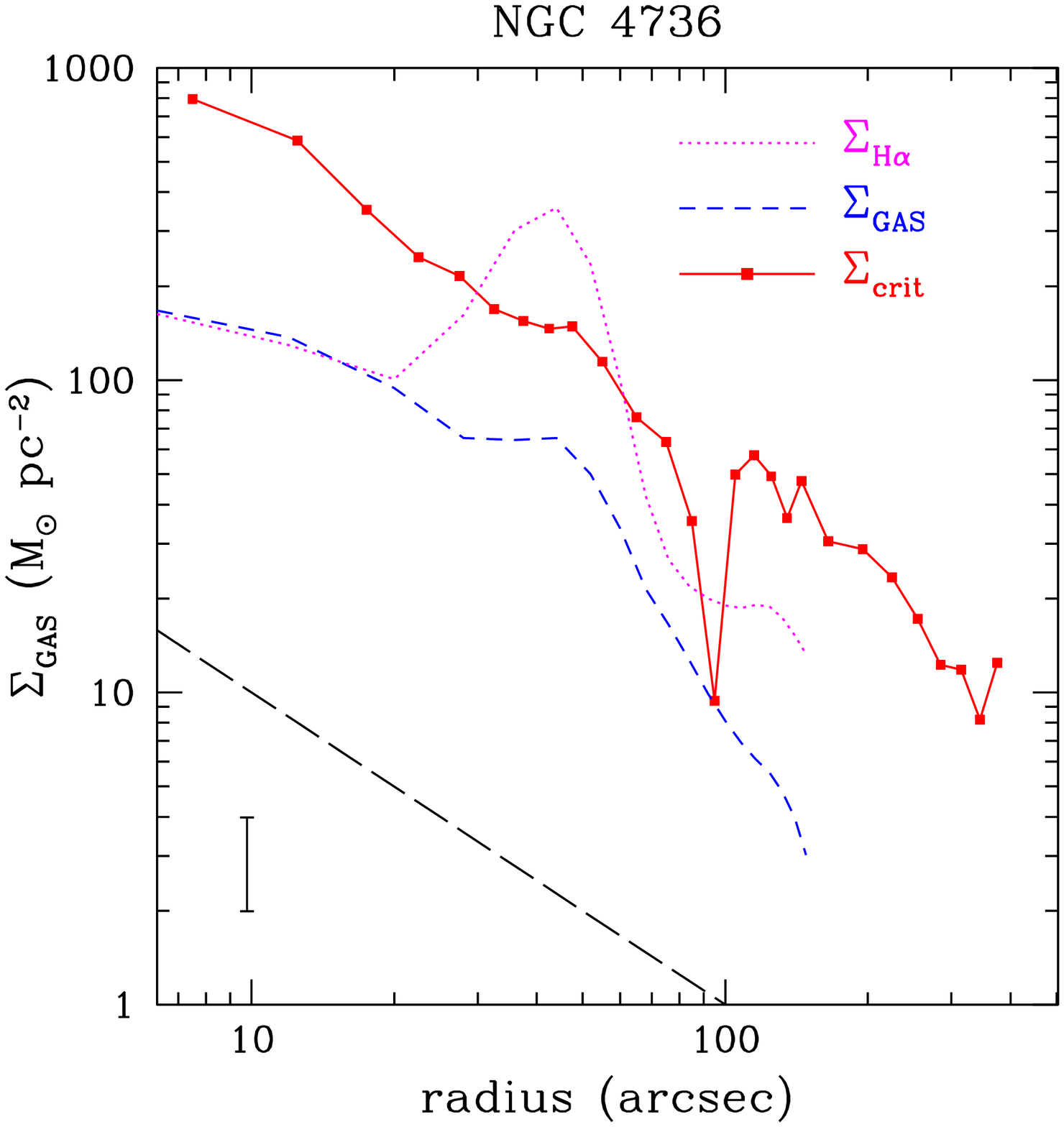}
\caption{
Comparison of total gas density with the critical density for
gravitational instability for NGC 4321, 4414, 4501, and 4736.  The
H$\alpha$ profile, uncorrected for extinction, is also shown for
comparison ({\it dotted line}).  The error bar at the lower left
represents a factor of 2 change, and the heavy dashed line represents
an $R^{-1}$ profile, as expected for $\Sigma_{\rm crit}$ with a flat
rotation curve.
\label{fig:sigcrit1}}
\end{figure*}



\begin{figure*}
\plottwo{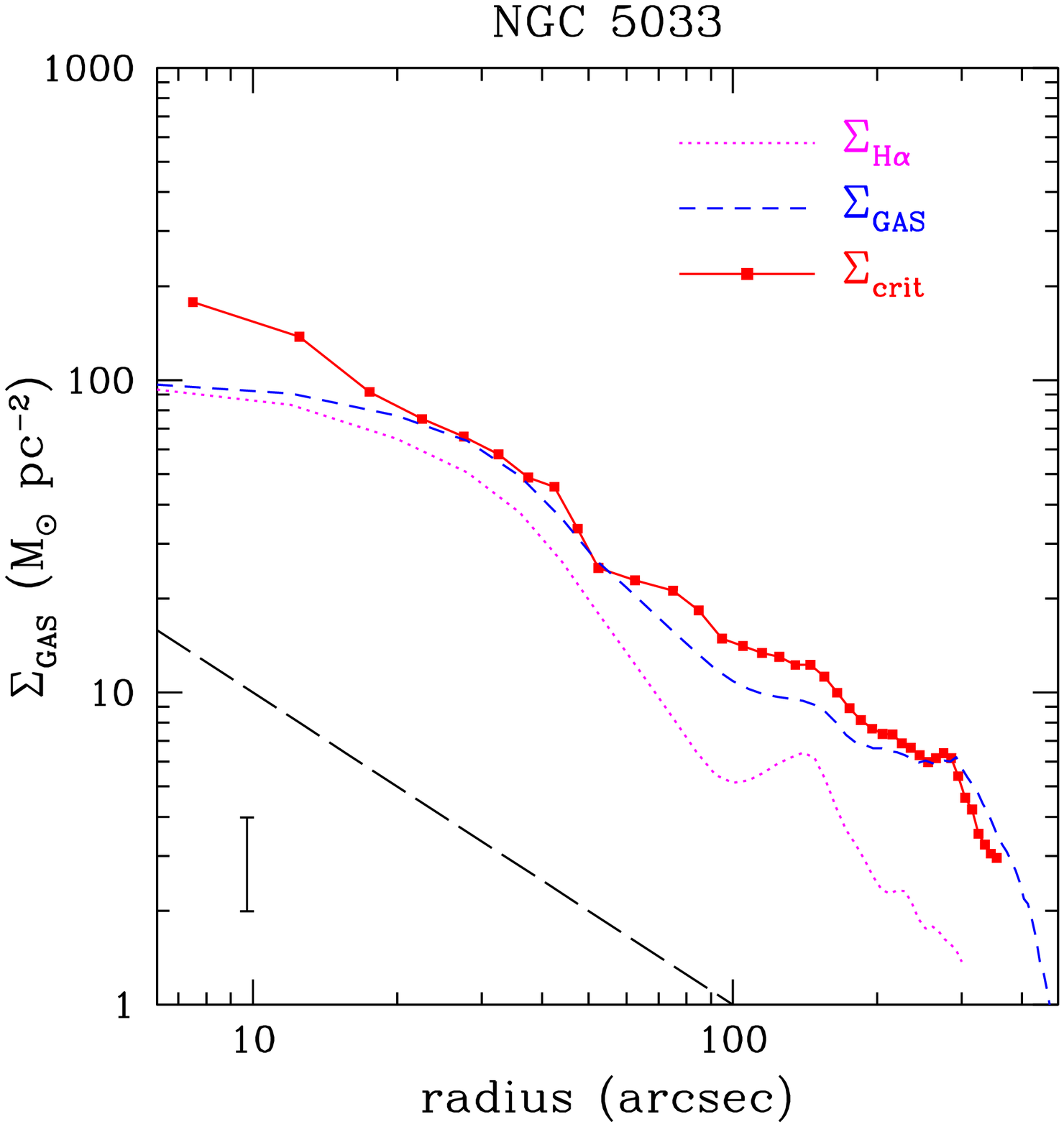}{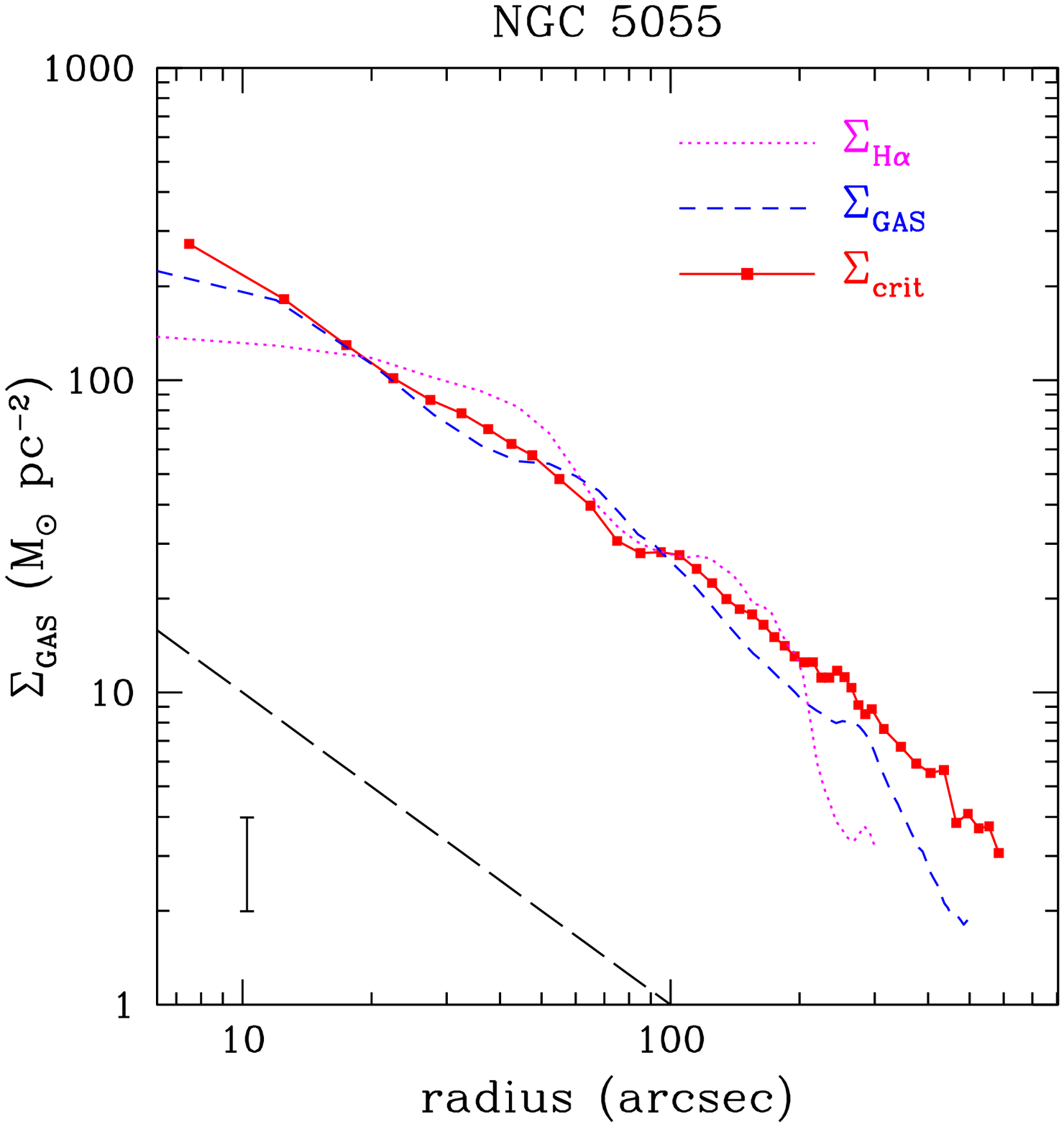}
\end{figure*}

\begin{figure*}
\plottwo{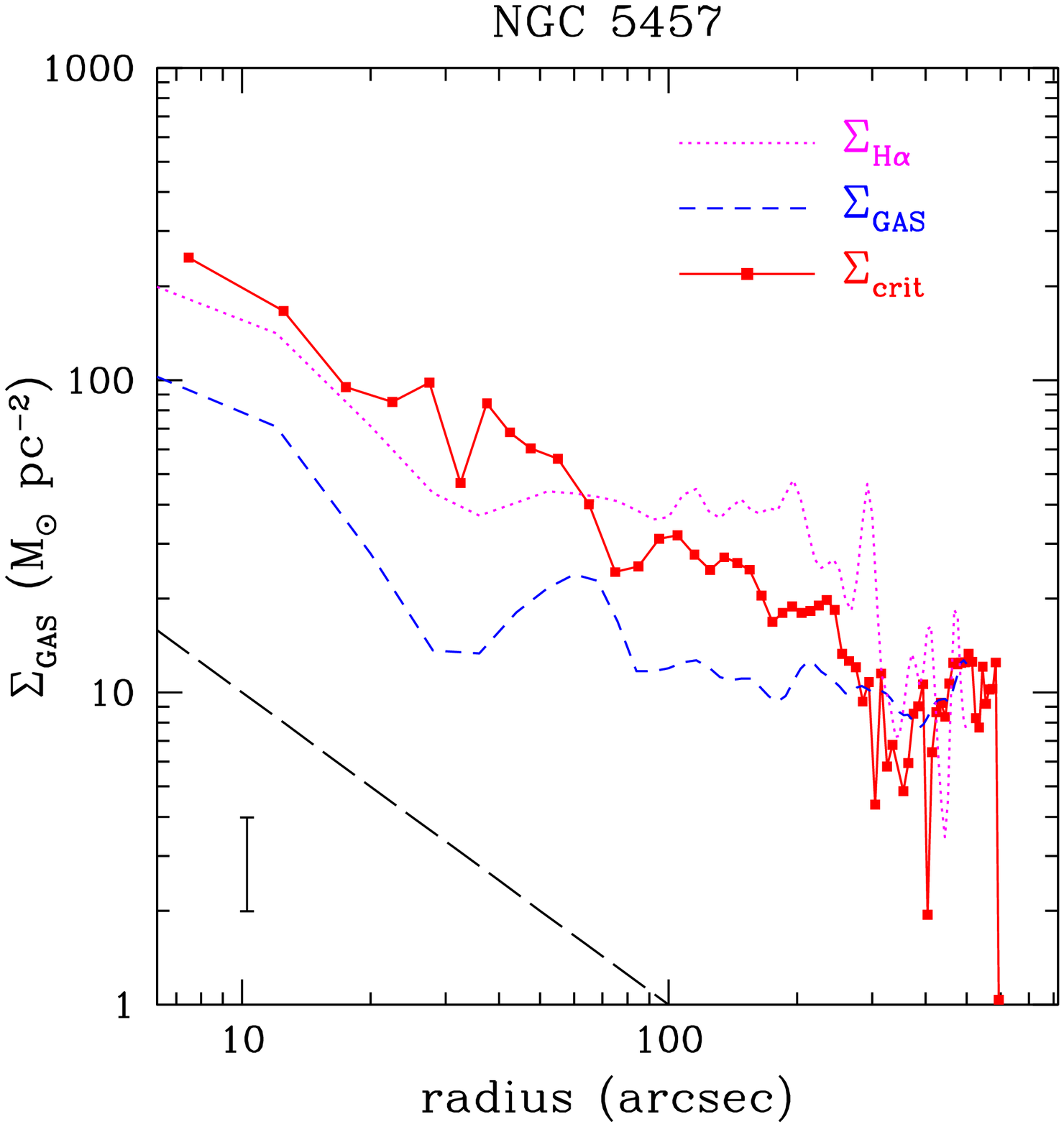}{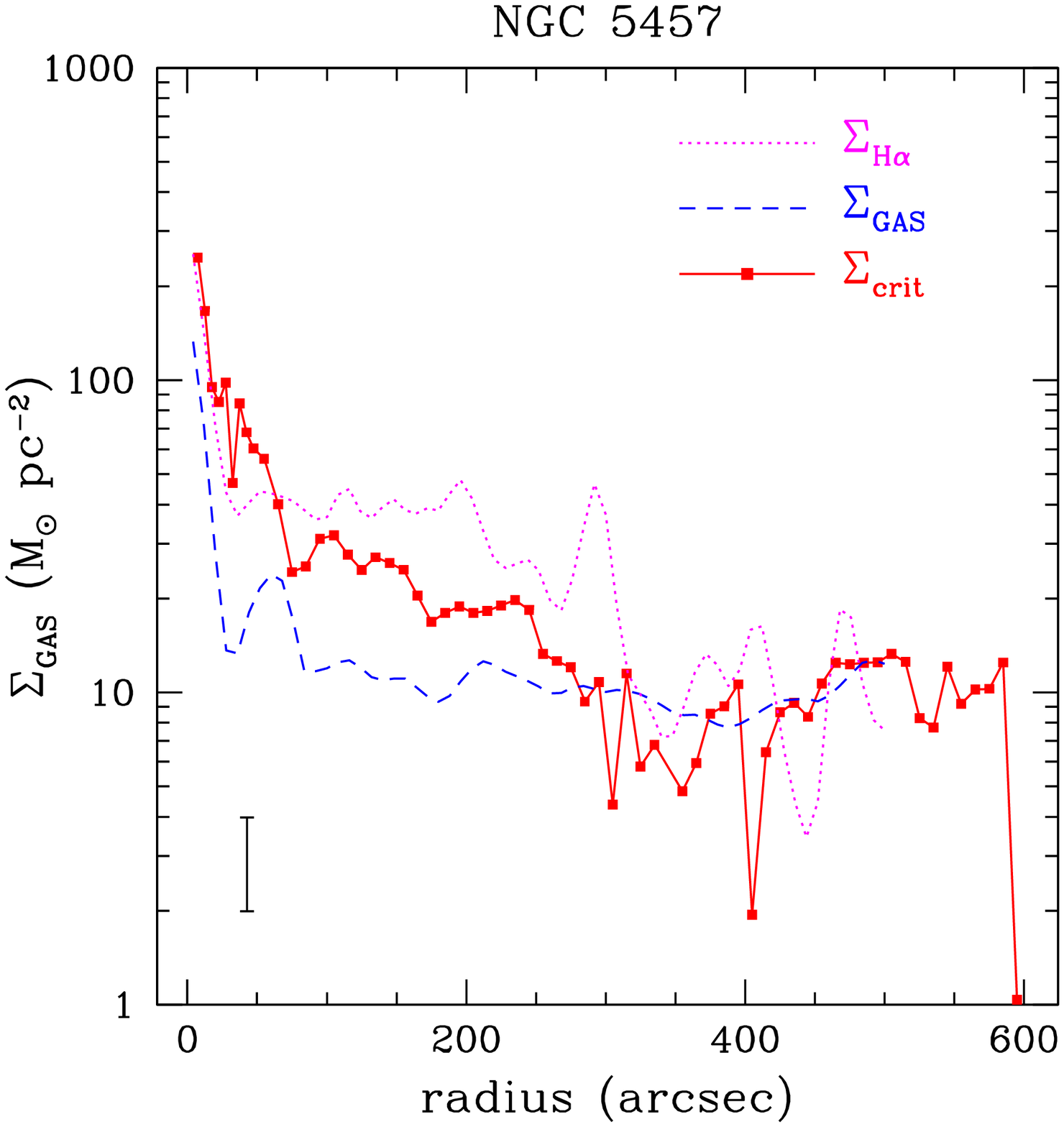}
\caption{
Same as Figure~\ref{fig:sigcrit1}, for NGC 5033, 5055, and 5457.
For NGC 5457 the horizontal axis is also shown on a linear scale
({\it lower right}).
\label{fig:sigcrit2}}
\end{figure*}


\subsection{Star Formation Thresholds}\label{crit}

Is star formation suppressed below a threshold or ``critical'' gas
density?  The observation that \HI\ in spiral galaxies often extends
well beyond the optical disk has led to widespread support for the
notion of a star formation threshold that is constant or depends on
local conditions.  As proposed by \citet{Quirk:72}, such a threshold
may be related to the parameter $Q_g$ for gravitational instability in
the gas layer \citep{Safranov:60,Toomre:64}:
\begin{equation}
Q_g \equiv \frac{\Sigma_{\rm crit}}{\Sigma_{\rm gas}}\;,
\quad
{\rm where}
\quad
\Sigma_{\rm crit} = \frac{\kappa c_g}{\pi G}\;.
\end{equation}
Here $\kappa$ is the epicyclic frequency and $c_g$ is the gas velocity
dispersion.  For a thin gaseous disk, values of $Q_g>1$ correspond to
stability and $Q_g<1$ to instability.  \citet{KC:89} found that in a
sample of 15 mostly late-type galaxies, massive star formation was
limited to regions of the disk where $Q_g \lesssim 1.6$, suggesting
that gas beyond some threshold radius is no longer susceptible to
gravitational instability.  Within the star-forming part of the disk,
he also found that $Q_g$ did not vary much as function of radius,
suggesting that star formation is self-regulated so that $Q_g \sim 1$.

We constructed radial profiles for $\Sigma_{\rm crit}$ using the
adopted rotation curves presented in Paper II.  The epicyclic
frequency $\kappa$ is given by
\begin{equation}
\kappa^2 = \frac{2v_c}{R}\left(\frac{dv_c}{dR}+\frac{v_c}{R}\right)\;.
\end{equation}
Note that for a flat rotation curve, $\kappa^2=2\Omega^2$.  The
derivative of the rotation curve was determined by simple two-point
interpolation.  Following \citet{KC:89}, we assumed a value of 6 \kms\
for the velocity dispersion $c_g$, which is consistent with detailed
observations of \HI\ in the outer regions of NGC 1058
\citep{vdK:84,Dickey:90a} and of CO in NGC 628 \citep{Combes:97}.  The
resulting profiles of $\Sigma_{\rm crit}$ are shown in
Figures~\ref{fig:sigcrit1}--\ref{fig:sigcrit2}, along with profiles
for $\Sigma_{\rm gas}$ and $\Sigma_{\rm SFR}$.  Note that $\Sigma_{\rm
crit} \propto R^{-1}$ for most galaxies, a reflection of the
relatively flat rotation curves ($\Omega \propto R^{-1}$).  In a few
cases (e.g., NGC 4736 and 5457), large spikes in $\Sigma_{\rm crit}$
are seen; these are likely due to errors in determining $dv_c/dR$ via
interpolation.

At first glance, it is impressive how well the relation $\Sigma_{\rm
gas} \approx \Sigma_{\rm crit}$ holds for some of the galaxies (NGC
4501, 5033, and 5055).  On the other hand, there are two galaxies (NGC
4321 and 4414) where $\Sigma_{\rm gas} > \Sigma_{\rm crit}$ over a
significant range of radius, while in NGC 4736 and 5457, $\Sigma_{\rm
gas} < \Sigma_{\rm crit}$ across most of the disk.  The
discrepancies are generally within a factor of 2, which is certainly
within the plausible range of error for either $\Sigma_{\rm gas}$
(where a constant $X$-factor has been used and the CO profile
extrapolated) or $\Sigma_{\rm crit}$ (which depends on the assumed
velocity dispersion, inclination, and distance).  Note, however, that
the greatest uncertainties are probably associated with the $X$-factor
and $c_g$, and given our choices for these parameters we have
probably {\it overestimated} $\Sigma_{\rm gas}$ and {\it
underestimated} $\Sigma_{\rm crit}$.  Thus, the large inferred values
for $Q_g=\Sigma_{\rm crit}/\Sigma_{\rm gas}$ are almost certainly
significant for NGC 4736 and 5457, yet star formation is hardly
suppressed in these galaxies.  (Surprisingly, these two galaxies are
among those used by Kennicutt to argue for a $Q$ threshold.)

Indeed, we find no clear evidence that $\Sigma_{\rm crit}$ is relevant
to $\Sigma_{\rm SFR}$ in the inner disk, since in most cases the
$\Sigma_{\rm crit}$ profile is relatively featureless, a reflection of
the relatively featureless rotation curve.  For instance, in NGC 4321
it appears to be the case that star formation is enhanced when
$\Sigma_{\rm gas} > \Sigma_{\rm crit}$ and vice versa.  Yet this may
simply be due to the SFR profile roughly tracking the gas density
profile (see also Fig.~\ref{fig:radprof}).  Similarly, the starburst
ring in NGC 4736 is associated with an increase in $\Sigma_{\rm
gas}$, not a dip in $\Sigma_{\rm crit}$.  Thus one should carefully
examine claims that star formation occurs in a certain region because
the gas density exceeds the critical density.  The fact that the gas
surface density is higher is probably relevant, but the critical density
itself may not be.

While the $Q$ model was primarily intended to explain the outer disk
cutoff of star formation \citep{KC:89}, our data do not provide strong
constraints in this regime.  Deep, wide-field H$\alpha$ imaging is
required, ideally with spatially resolved measurements of the gas
velocity dispersion.  Such a study has been undertaken by
\citet{Ferguson:98b}, who found that a threshold radius could not be
defined for NGC 6946 when using measured values for $c_g$ from
\citet{Kamphuis:th}, although qualitative agreement with the $Q$ model
could be achieved if $c_g$ were assumed to be constant with radius.
Since $c_g$ is very difficult to measure in galaxies that are not
face-on, it is unclear whether this result indicates a failure of the
$Q$ model or reflects inaccuracies in the measurement of $c_g$.


\subsection{A Relationship Between Atomic and Molecular Gas}\label{hitoh2}

The results of \S\ref{molat}, which showed that the SFR traces the
radial CO distribution and not the \HI, imply that the SFR on large
scales will be strongly influenced by the fraction of interstellar
matter in molecular form.  In a steady-state configuration, this
fraction is determined by a balance between molecular cloud formation
and destruction processes.  The timescales of these processes and the
amount of material involved can only be determined empirically when
there are organized features, such as spiral arms, with which one can
trace the \HI/H$_2$ ratio azimuthally.  In the inner disk of M51, for
example, virtually all of the \HI\ is believed to result from
dissociation of H$_2$ in the spiral arms
\citep{Vogel:88,Tilanus:89,Rand:92}.  Conversely, \citet{Heyer:98}
find that the gas entering the Perseus arm in the outer Galaxy is
primarily atomic; in this case the \HI\ must be a precursor to the
H$_2$ in the arm.  We will defer discussion of time-dependent
processes to a future paper that includes azimuthal comparisons, and
will focus instead on what determines the time-averaged \HI/H$_2$
ratio.  Thus we ignore the question of whether \HI\ is a product of
H$_2$ or vice versa, since in a time-averaged sense the distinction
has little meaning.

In comparing the azimuthally averaged values of \sighi\ and \sightwo,
we are restricted to a fairly narrow region in the disk:
$R>20\arcsec$, in order to minimize the effect of beam smearing on the
profile, and $R \lesssim 80\arcsec$, the extent of our CO
observations.  Furthermore, we have rejected points where $\sighi < 1$
\Msol\ pc$^{-2}$, because the \HI\ profile is generally uncertain at
these radii (this was only an issue for $R < 36\arcsec$ in
NGC 5457).  In practice, this means that we only probe a region
where the molecular fraction $f_{\rm mol}$ falls from $\sim$0.9--1 to
$\sim$0.6--0.7, emphasizing the need for high sensitivity, large-field
CO data to fully analyze the transition region.
Figure~\ref{fig:molat1} compares the azimuthally averaged \HI\ and
H$_2$ surface densities for all seven galaxies over the region
satisfying the above criteria.  The two types of behavior seen in the
radial gas profiles in Fig.~\ref{fig:radprof} are reflected in whether
\sighi\ is constant with increasing \sightwo\ or begins to decrease at
some point.  In NGC 4414, 5033, and 5055, the \HI\ profile remains
relatively flat across the inner disk, while in NGC 4321, 4501, 4736,
and 5457, the \HI\ profile declines toward the center.


\vskip 0.25truein
\includegraphics[width=3.25in]{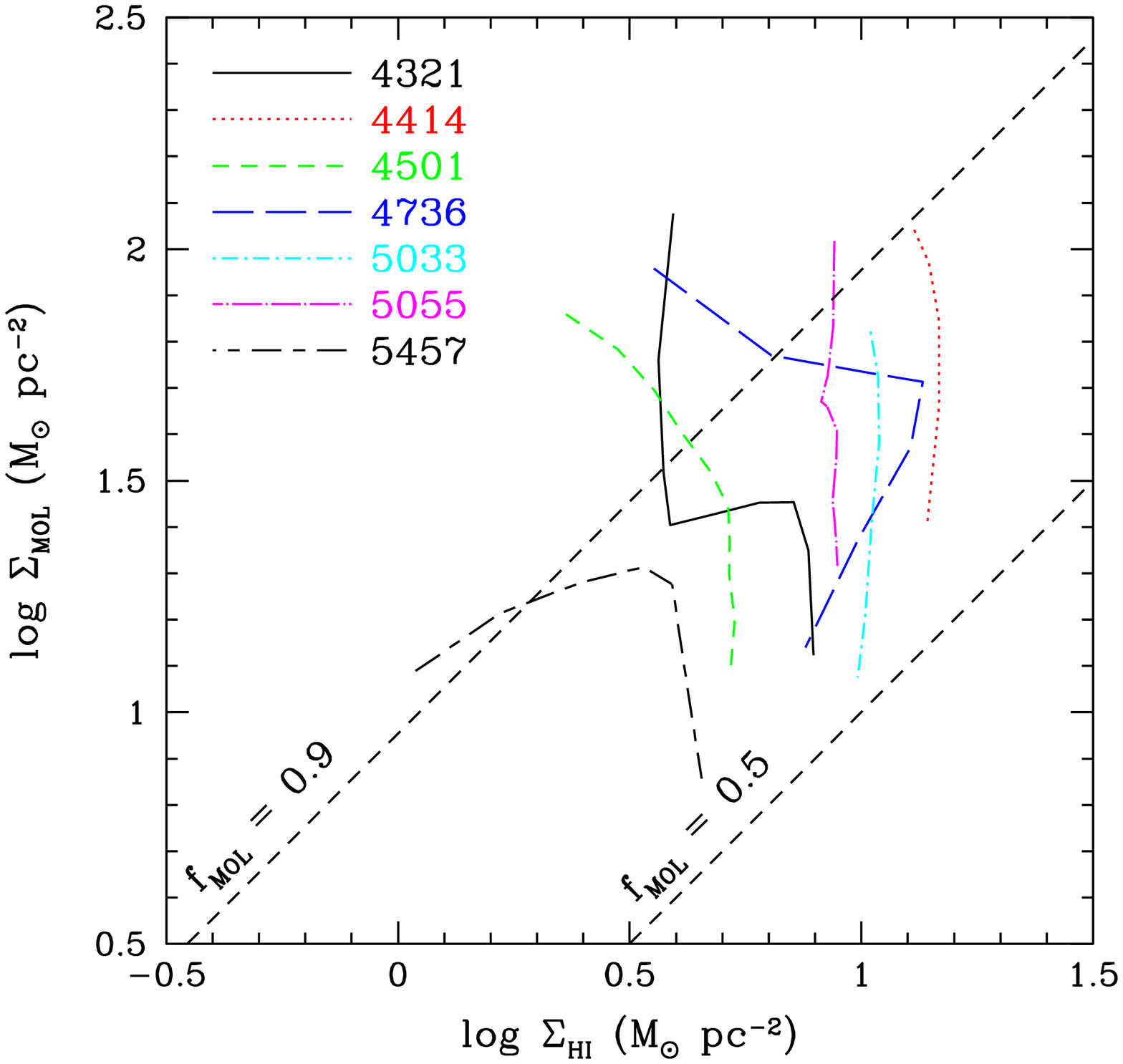}
\figcaption{
Relation between molecular and atomic gas surface density for all
seven galaxies in the region where observations overlap 
($R_{\rm min}$=20\arcsec, $R_{\rm max} \sim 80$\arcsec) and
where $\sighi > 1$ \Msol\ pc$^{-2}$.
\label{fig:molat1}}
\vskip 0.25truein


As shown in Figure~\ref{fig:molat2}, the ratio of \sighi\ to \sightwo\
appears to depend on radius, increasing as roughly
\begin{equation}
\frac{\sighi}{\sightwo} = K R^{1.5}
\label{eqn:h1h2}
\end{equation}
over a factor of $\sim$2 (0.3 dex) in $R$.  The fact that the curves
in Fig.~\ref{fig:molat2} are parallel but are spread out over nearly
an order of magnitude in $R$ indicates that there is a second
parameter $K$ that is roughly constant {\it within} a galaxy but
varies from galaxy to galaxy.  If $K$ is taken to be
$(R_{25})^{-1.5}$, where $R_{25}$ is the optical radius given in
Table~\ref{tbl:props}, then the residual variation in $\log R$,
although still substantial, is reduced
to $\sim$0.5 dex.  The most striking aspect of the relation given by
Equation~(\ref{eqn:h1h2}) is that it is independent of whether the
\HI\ profile is flat or declining towards the center: note especially
how it ``straightens out'' the lines for NGC 4321, 4736, and 5457 when
compared to Fig.~\ref{fig:molat1}.  Thus, {\it the ratio of \sighi\ to
\sightwo\ appears to depend on galactocentric radius, normalized by
$R_{25}$, and little else.}  In galaxies where the inner \HI\ profile
is flat, \sightwo\ declines as roughly $R^{-1.5}$, whereas the
galaxies with \HI\ decreasing towards the center have a
correspondingly flatter H$_2$ profile across the transition region.
We discuss a possible interpretation of these trends in
\S\ref{atmoldisc}.


\vskip 0.25truein
\includegraphics[width=3.25in]{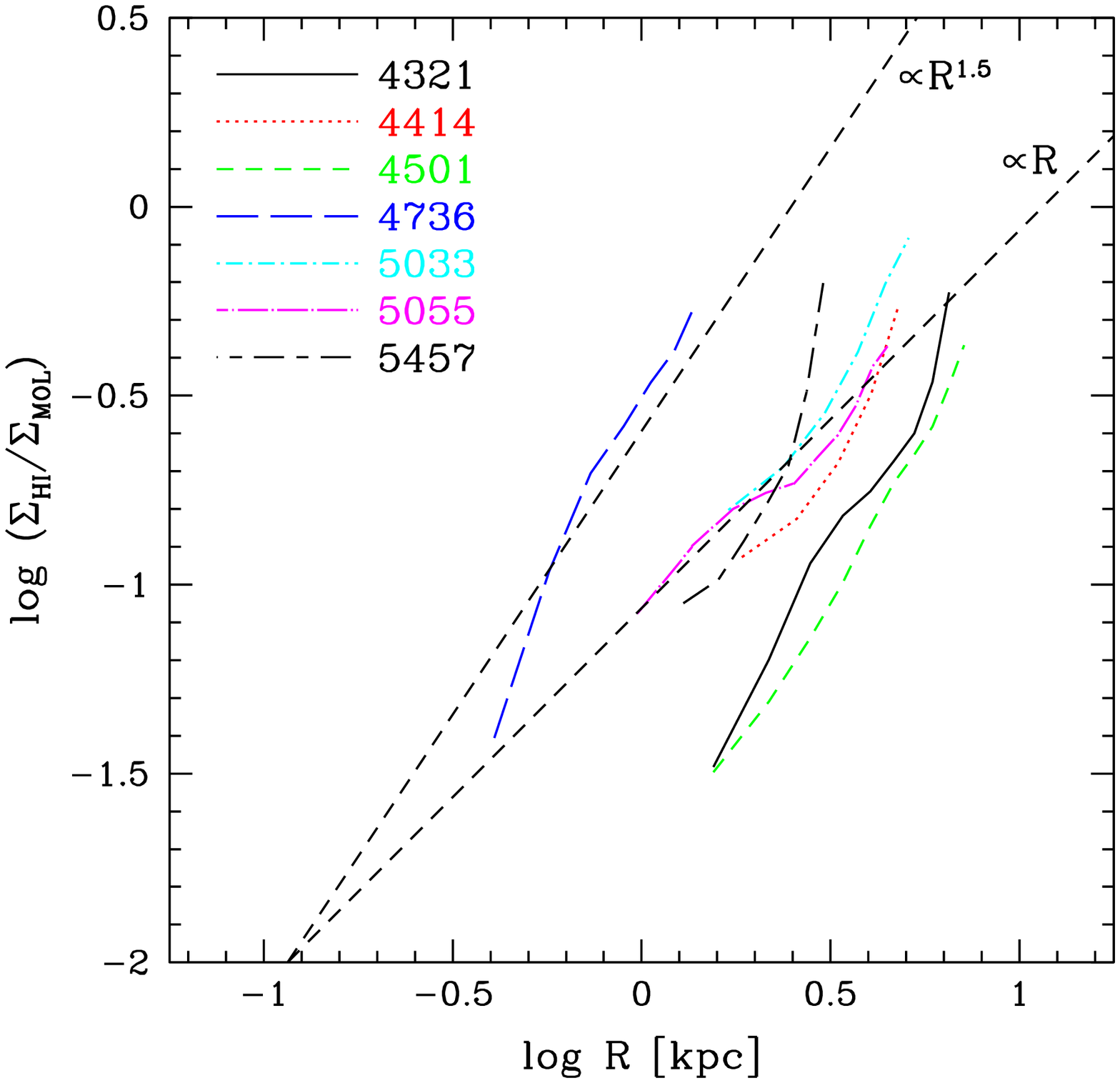}
\figcaption{
Ratio of \HI\ to H$_2$ surface density plotted against radius
for all seven galaxies.  Two power-law parameterizations are shown,
one where the ratio scales with $R$ and another where the ratio
scales with $R^{1.5}$.  The data are the same as shown in 
Figure~\ref{fig:molat1}.
\label{fig:molat2}}
\vskip 0.25truein



\section{Discussion}\label{disc}

\subsection{Does CO Trace Molecular Mass?}\label{xdisc}

We have interpreted the linearity between azimuthally averaged CO and
H$\alpha$ emission (\S\ref{molat}) as indicative of a constant star
formation efficiency, but it could conceivably be a result of CO
excitation effects.  That is, if energy input from young stars is
responsible for exciting the CO emission, then CO may trace star
formation directly rather than the H$_2$ mass
\citep[e.g.,][]{Dopita:94}.  Evidence in favor of this viewpoint comes
from observations of the [\ion{C}{2}] 158 $\mu$m line, a tracer of
photodissociation regions, which is found to scale linearly with the
CO(1--0) flux in infrared-luminous galaxies \citep{Crawford:85}.
However, relatively stronger CO emission (compared to [\ion{C}{2}]) is
found in spirals with more typical SFRs \citep{Stacey:91}, and further
studies with higher-resolution instruments such as SOFIA will be
essential.

In any case, there are at least three difficulties with the suggestion
that we have significantly overestimated the H$_2$ mass in regions of
active star formation.  First, the gas depletion times are already
quite short in these regions, $\sim$1--3 Gyr, and become even shorter
if there is less H$_2$; it seems unlikely that we would happen to
witness the final stages of star formation in these regions.  Second,
observations of cold dust in nearby galaxies with the SCUBA
submillimeter camera show a strong correlation between CO and 850
$\mu$m emission \citep[e.g.,][]{Alton:01}, indicating that variations
in the $X$-factor with radius cannot significantly exceed variations
in the gas-to-dust ratio.  Finally, the spatial correspondence between
CO and young massive stars (which would dominate the radiative
excitation) does not appear to be close when observations of
sufficient angular resolution are performed.  Note for instance the
different azimuthal symmetries of CO and H$\alpha$ in NGC 4736
\citep{Wong:00}, or the spatial offset between the brightest CO
emission and the most populous young clusters in the LMC
\citep{Cohen:88}.


\subsection{Interpretation of the Schmidt Law}\label{evoldisc}

Previous evidence for a power-law relationship between \siggas\ and
\sigsfr, with $n \approx 1.5$, has been discussed by \citet{KC:97} and
K98.  As shown in \S\ref{schmidt}, this relation appears to hold
within galaxies as well.  The most straightforward interpretation of
such a Schmidt law is that the SFR on large scales is controlled by
the self-gravity of the gas.  As a result, the rate of star formation
is proportional to the gas mass divided by the timescale for
gravitational collapse:
\begin{equation}
\rho_{\rm SFR} = \frac{\epsilon_* \rho_{\rm gas}}
	{(G\rho_{\rm gas})^{-0.5}} \propto \rho_{\rm gas}^{1.5}\;.
\end{equation}
The problem with this explanation, however, is that the efficiency
factor $\epsilon_*$ is observed to be much less than 1.  In other
words, the gas depletion time on large scales ($\tgas \sim 10^9$ yr),
while often much less than a Hubble time, is still significantly longer than
$\tau_{\rm ff} = (G\rho_{\rm gas})^{-0.5} \approx 10^7$--$10^8$ yr for
$n_{\rm H} \sim 1$--100 cm$^{-3}$.  This leads to two competing
interpretations: (1) the star formation timescale really is $\tau_{\rm
ff}$, but only a small fraction $\epsilon_*$ of the gas participates
in star formation \citep{Elm:00}, or (2) the star formation timescale
is $\tau_{\rm ff}$ multiplied by a factor of $\epsilon_*^{-1}$, due to
the inhibiting effects of stellar feedback (i.e., winds, radiation,
and supernovae) or magnetic support on star formation
\citep{Larson:88}.  Neither description, however, offers a clear
prediction for the value of $\epsilon_*$ or explains why it should be
constant.  Replacing $\tau_{\rm ff}$ with the orbital timescale
$\tau_{\rm orb}$, as discussed in \S\ref{torb}, yields a similar
problem: why is a small but fixed fraction of the gas consumed per orbit?

Given that the theoretical basis of the Schmidt law remains unclear,
it is worthwhile to consider an alternative explanation, motivated by
our finding in \S\ref{molat} that \sigsfr\ scales roughly linearly
with \sightwo---a result that is also consistent with previous
extragalactic \citep{Rownd:99} and Galactic \citep{McKee:89} studies.
In this picture, stars form only in molecular gas, and the timescale
for star formation is controlled by stellar feedback, magnetic
diffusion, or other processes which do not vary strongly with location
in a galaxy, so that $\sigsfr \propto \sightwo$ ($n_{\rm mol}$=1).  As
suggested by \citet{Tacconi:86} and \citet{Wang:90}, the Schmidt law
between \sigsfr\ and the total gas surface density \siggas\ follows
from an amalgamation of two ``laws,'' the first relation governing the
formation of H$_2$ from \HI, and the second (linear) relation
governing the formation of stars from H$_2$.  The composite Schmidt
law index is given by
\begin{eqnarray}
n & = & \frac{d\ln \sigsfr}{d\ln \siggas}\\ 
& = & \left(\frac{d\ln\sigsfr}{d\ln \sightwo}\right)
	\left(\frac{d\ln \sightwo}{d\ln \siggas}\right)\\ 
& = & \left(\frac{d\ln\sigsfr}{d\ln \sightwo}\right)
	\left(\frac{d\ln \siggas + d\ln f_{\rm mol}}{d\ln \siggas}\right)\\ 
& = & n_{\rm mol} \left(1+\frac{d\ln f_{\rm mol}}{d\ln \siggas}\right)\;.
\end{eqnarray}

For $n_{\rm mol} \approx 1$ and $f_{\rm mol}$ increasing with \siggas\
(since $f_{\rm mol}$ increases with interstellar pressure which
scales with \siggas---see \S\ref{atmoldisc} below), one naturally
arrives at $n \gtrsim 1$.  Physically this also makes more sense than
a star formation law based on the {\it total} gas surface density,
since when one looks in the third dimension,
one finds that the vertical distribution of star formation corresponds
to that of H$_2$, not \HI.  The sharp increase in the Schmidt law
index $n$ seen at low \siggas, which \citet{KC:89} attributes to a
$Q$-dependent threshold for star formation, could instead be due to a
sudden decrease in $f_{\rm mol}$ at low \siggas\ (or low $\rho_{\rm
gas}$, if the volume density is more relevant due to flaring of the HI
layer, \citealt{Ferguson:98b}), although the two thresholds may turn
out to be equivalent if molecular cloud complexes are formed by
large-scale gravitational instabilities.

This ``two-step'' model is not without its difficulties---it still
does not predict the SFE in molecular clouds quantitatively, nor does
it explain the enhanced SFE's found in merging galaxies
\citep{Young:96} or starburst rings \citep{Wong:00}, which could be
related to cloud collisions \citep[e.g.,][]{Tan:00}.  However, it does
provide an attractive explanation for the relatively constant star
formation histories of disks \citep{KC:94}, namely that the
requirement of converting \HI\ into H$_2$ could limit the SFR at early
times, despite a greater supply of gas.  \citet{Rana:86} have
presented such a model, in which the H$_2$ formation rate at early
epochs is limited by the lack of metals, since metals serve as
important coolants in the ISM and constitute the dust grains on which
most of the H$_2$ forms.  Even today, the confinement of H$_2$ to a
thinner disk than the \HI\ \citep[see e.g., studies of NGC 891
by][]{Scoville:93,Swaters:97} suggests that the formation of H$_2$ is
inhibited outside of the midplane, perhaps because of insufficient
pressure or dust abundance (see also the discussion in
\S\ref{atmoldisc} below).


\subsection{Is Q Relevant to Star Formation?}\label{critdisc}

We have found that the $Q_g$ parameter, defined using a constant
velocity dispersion of 6 \kms, appears to be remarkably close to 1 in some
galaxies, but is significantly greater than 1 in NGC 4736 and 5457,
which are still actively forming stars.  Our observations thus lend
support to earlier studies which have failed to show a clear
relationship between $Q_g$ and the presence of massive star formation
\citep{Thornley:95,Meurer:96,Ferguson:98b,Hunter:98,Pisano:00}.  This
is not altogether surprising; \citet{Elm:99} raises the question of
``why $Q<1$ has anything at all to do with star formation in a
supersonically turbulent medium filled with supernovae, magnetic
fields, multiple gas phases and other seemingly important details.''
Yet the problem remains of explaining why $Q_g \sim 1$ is often
observed in galaxy disks, a fact that has long been taken to indicate
that star formation on large scales is regulated by gravitational
instability \citep{Quirk:72,KC:89,Silk:97}.  A possible clue is that
$Q_g$ is significantly greater than 1 in the two galaxies with the
lowest gas fractions (NGC 4736 and 5457), as inferred from a
comparison of \siggas\ with the stellar surface density $\Sigma_*$
derived by assuming a maximum disk (see Paper II).  This suggests 
that $Q_g$ is actually a measure of the gas fraction in the disk.

A simplified analysis suggests that this is indeed the case.
For a \citet{Mestel:63} disk, with $\Sigma_{\rm tot} \propto R^{-1}$, a 
flat rotation curve is obtained such that
\begin{equation}
v_c^2 = 2 \pi G \Sigma_{\rm tot} R\;,
\end{equation}
\begin{equation}
\Sigma_{\rm tot} = \frac{v_c\Omega}{2\pi G}\;.
\end{equation}
[See \citet{Binney:87}, p. 76.]  But for a flat rotation curve, the
critical density is given by 
\begin{equation}
\Sigma_{\rm crit} = \frac{\sqrt{2}\Omega c_g}{\pi G}\;.
\end{equation}
Defining the gas fraction as
$\mu=\Sigma_{\rm gas}/\Sigma_{\rm tot}$, we have
\begin{equation}
\mu Q_g = \frac{\Sigma_{\rm crit}}{\Sigma_{\rm tot}} = 
	\frac{2\sqrt{2}c_g}{v_c}\;,
\end{equation}
\begin{equation}
Q_g \approx \frac{2.8c_g}{v_c\mu}\;.
\end{equation}
Substituting typical values for spiral galaxies ($c_g$=6 \kms,
$v_c$=200 \kms, and $\mu$=0.1) we find that $Q$=0.84.  Thus, {\it the
fact that $Q_g$ is of order unity in many spiral galaxies reflects the
fact that the present-day gas fraction is roughly 0.1}.  Note that we
have implicitly assumed a maximum disk model with negligible bulge or
halo, but this analysis will be roughly appropriate if $\Sigma_{\rm
tot}$ is considered to be whatever mass distribution dominates the
rotation curve.  For instance, dwarf galaxies exhibit large $Q_g$
values \citep[e.g.,][]{Meurer:96,vanZee:97} because they rotate slowly
and have gas surface densities that are small compared to the dark
matter surface density.  We therefore propose that $Q_g \sim 1$ is
primarily a reflection of observational selection favoring luminous,
gas-rich galaxies for this kind of analysis; galaxies with smaller gas
fractions should show $Q_g>1$.  We note that this explains the larger
$Q_g$ values found in early-type galaxies by \citet{Martin:01}.
Still, the lack of galaxies with $Q_g \ll 1$ suggests that {\it
strongly unstable} gas disks are precluded.

That the $Q_g$ parameter is most sensitive to the gas fraction and
thus cannot be the same for all galaxies was recognized many years ago
by \citet{Larson:88}.  He suggested that an increase in $c_g$ would
partially offset the larger $\Sigma_{\rm gas}$ at early times,
maintaining $Q_g \sim 1$.  It is unlikely, however, that the velocity
dispersion of the gas, when averaged over large scales, can continue
to decrease much below a value of $\sim$6 \kms, since measurements in
the face-on galaxy NGC 1058 indicate that it does not fall below this
value even well beyond the optical radius \citep{vdK:84,Dickey:90a}.
\citet{Sellwood:99} attribute this minimum velocity dispersion to
magnetohydrodynamic turbulence that is ultimately fed by the galaxy's
differential rotation.  Based on these arguments, a secular increase
in $Q_g$ seems inevitable as gas consumption continues, unless new gas
is constantly accreted to maintain $\mu$ at a constant level.

In spite of the apparent difficulties in using $Q_g$ as a predictor of
star formation activity, it is still remarkable how well $Q_g \approx
1$ is satisfied in certain galaxies, such as NGC 5033 and 5055
(Fig.~\ref{fig:sigcrit2}).  A possible explanation is that
gravitational instability {\it does} regulate star formation, but the
appropriate $Q$ parameter must take into account both the gas and
stellar disks.  \citet{Elm:95} and \citet{Jog:96}, extending the work
of \citet{Jog:84}, have shown that the gravitational interaction of
these two components leads to greater instability in the combined
system than in either component individually.  They derive an
``effective'' $Q$ parameter $Q_{\rm eff}$ which is $\approx Q_g$ for
gas-dominated disks, but can be close to $Q_*$ when the gas fraction
is small.  For $Q_*$ to be regulated at a value of $\sim$1 requires
only that the stellar velocity dispersion $c_*$ increase in proportion
to the stellar surface density, and is in good agreement with
observations if one assumes a constant $M/L$ ratio in the disk
\citep{Bottema:93}.  Thus, even galaxies with low gas fractions may be
susceptible to mild gravitational instabilities as a result of $Q_*
\sim 1$, and gravitational instability may be able to account for the
presence of continued star formation in these galaxies after all.

As noted by \citet{Skillman:96b}, an alternative to a $Q$-based star
formation threshold is a constant column density threshold of $N_{\rm
H} \sim 10^{21}$ cm$^{-2}$ associated with the formation of molecular
clouds.  In this picture, star formation cuts off at a radius where
there is insufficient self-shielding from UV radiation to allow
molecular clouds to form.  In practice, the column density threshold
may vary with metallicity, pressure, and the interstellar radiation
field, all of which affect the transition from \HI\ to H$_2$
\citep{Elm:93}.  A similar criterion, in which the minimum column
density is defined by the requirement of sufficient pressure to allow
a two-phase ISM, has been discussed by \citet{Elm:94c}.  Supporting
this hypothesis, \citet{Braun:97} has presented evidence that cold
\HI\ clouds are not observed outside the optical radius of galaxies;
the inability to form such clouds may be responsible for star
formation thresholds.  A comparison of sensitive CO, \HI, and
H$\alpha$ data near the edges of galaxy disks would be needed to test
these ideas.


\subsection{The Atomic-to-Molecular Gas Ratio}\label{atmoldisc}

In \S\ref{hitoh2} we found that the ratio of \sighi\ to \sightwo\
varies within galaxies as roughly $R^{1.5}$ in the region of overlap.
However, the interpretation of this dependence is ambiguous, since
many galaxy properties vary with radius.  Moreover, the limited range
in radius over which the ratio is calculated may allow for a variety
of functional forms, of which this relation is only the simplest.
For purposes of discussion, we consider here two possible models for
how \sighi\ and \sightwo\ should be related, and compare them with
our results.

First, \citet{Wyse:86} has suggested that the formation of GMCs is
tied to collisions between atomic clouds as they pass through spiral
arms.  Assuming a negligible pattern speed (i.e., well inside the
corotation radius), the rate of spiral arm crossings at radius $R$
is $\propto \Omega(R)$, and one expects
\begin{equation}
\sightwo \propto (\sighi)^2\Omega\;.
\end{equation}
This naturally introduces an $R$ dependence into the ratio
\sighi/\sightwo, qualitatively similar to the $R^{1.5}$ dependence we
derived above.  However, Figure~\ref{fig:wyse} shows that while the
majority of the galaxies show {\it some} correlation between \sightwo\
and $(\Sigma_{\rm HI})^2\Omega$, the expected relation (given by the
dashed line) is a much poorer description of the data than the simple
$R^{1.5}$ relation shown in Fig.~\ref{fig:molat2}.


\vskip 0.25truein
\includegraphics[width=3.15in]{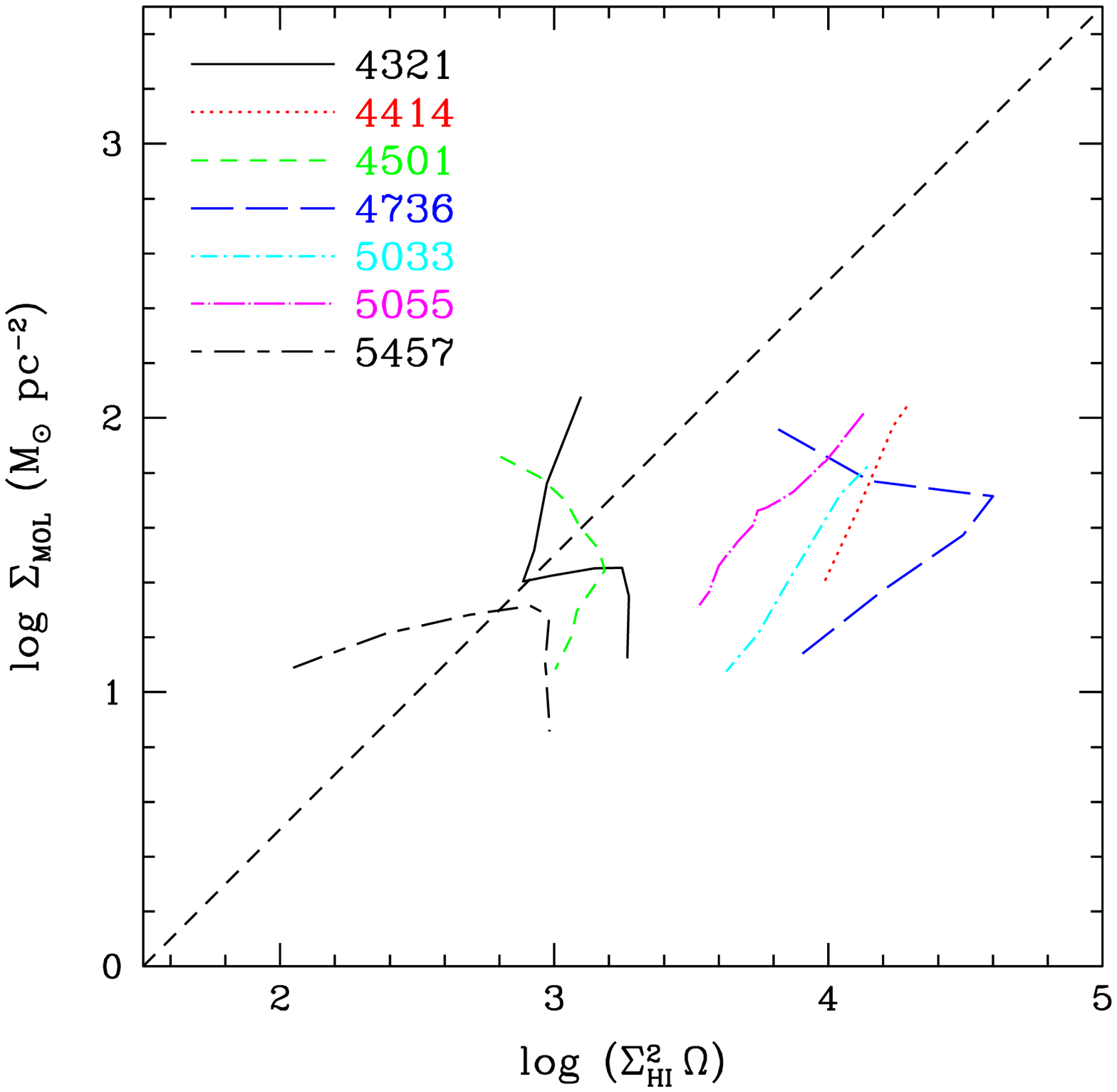}
\figcaption{
Surface density of molecular gas vs.\ the prediction from
\citet{Wyse:86}, where $\sightwo \propto \sighi^2\Omega$.
The dashed line represents a slope of 1, in agreement with
the prediction.
\label{fig:wyse}}
\vskip 0.2truein



\begin{figure*}[t]
\plottwo{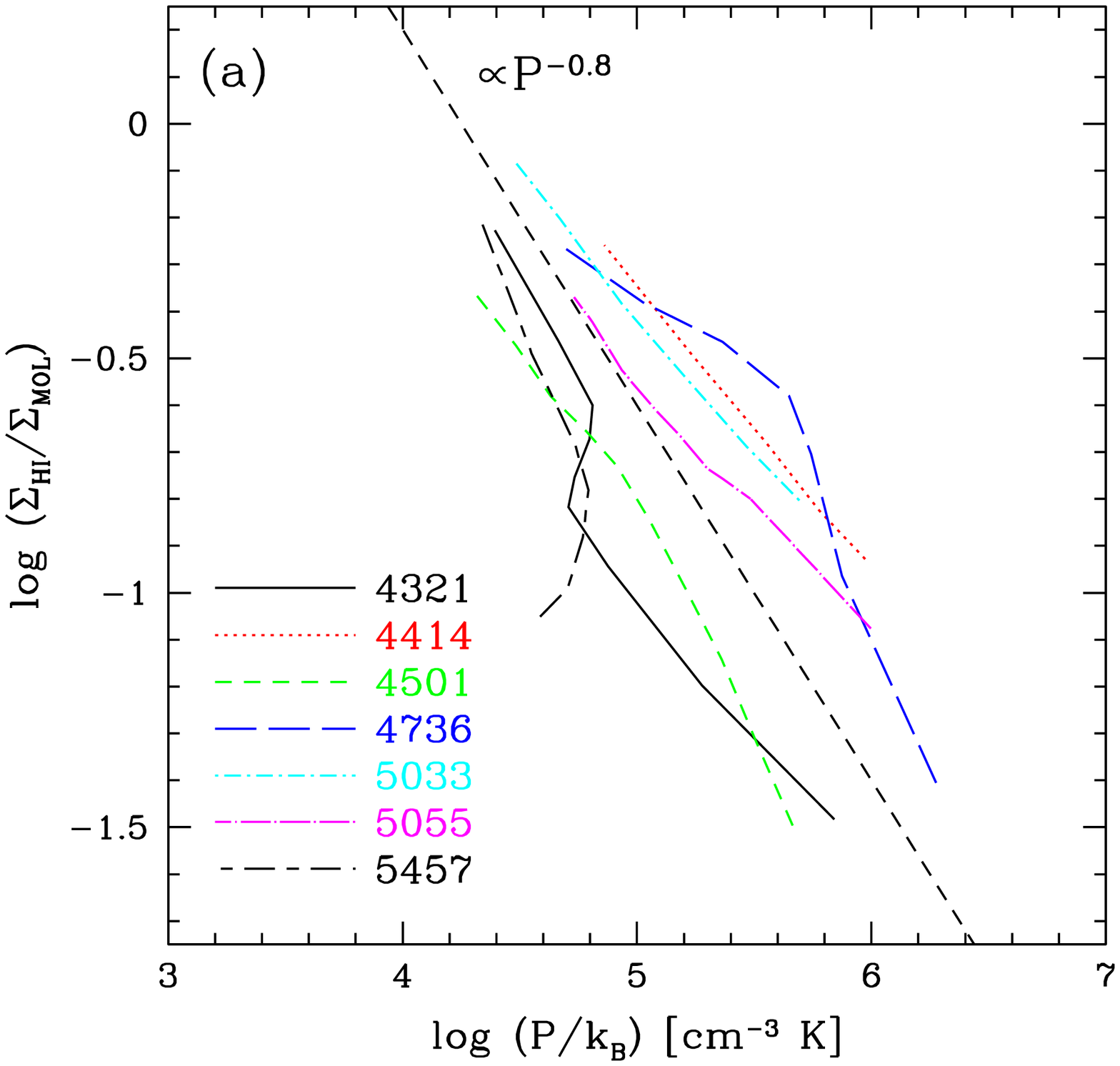}{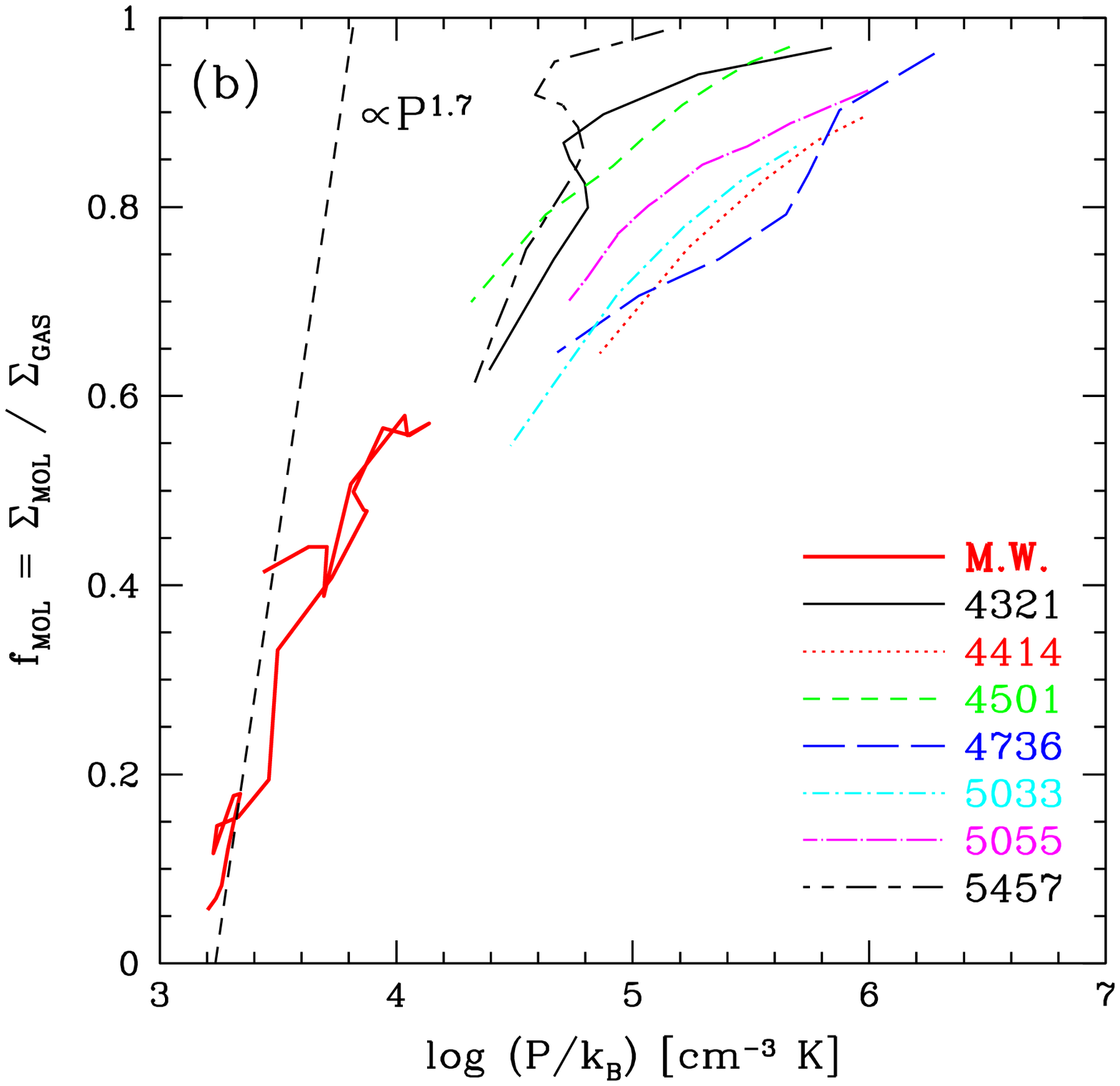}
\caption{
(a) Ratio of \HI\ to H$_2$ surface density plotted against the
hydrostatic disk pressure at the midplane, as given by \citet{Elm:89}.
The straight dashed line represents the mean slope of $-0.8$.  (b)
Molecular fraction plotted against midplane pressure.  The straight
dashed line represents the slope of 1.7 predicted by \citet{Elm:93}.
Data for the Milky Way is based on \citet{Dame:93a}, adopted a stellar
scalelength of 3 kpc \citep{Sackett:97} and $\Sigma_* = 35$ \Msol\
pc$^{-2}$ at the solar circle \citep{Gilmore:89}.
\label{fig:elm}}
\end{figure*}


Second, \citet{Elm:93} argues that the degree to which a cloud is
molecular is determined in part by the ambient pressure $P$, since the
molecule formation rate depends on the volume density $n$, which
scales with $P$ for both diffuse and self-gravitating clouds.  The
pressure at the midplane of a two-component disk in hydrostatic
equilibrium is given by \citep{Elm:89}:
\begin{equation}
P \approx \frac{\pi}{2}G\siggas\left(\siggas+\frac{c_g}{c_*}
	\Sigma_*\right)\,,
\end{equation}
where $c_g$ and $c_*$ are the velocity dispersions of the gaseous and
stellar disk respectively.  The observations of \citet{Bottema:93}
typically give $c_* \sim 60$ \kms, so we assume a value for $c_g/c_*$
of 0.1.  (This value for $c_*$ is consistent with the velocity
dispersion of an isothermal disk, $c_*=\sqrt{\pi G\Sigma_* z_0}$, for
$\Sigma_* \sim 300$ \Msol\ pc$^{-2}$ and $z_0 \sim 1$ kpc.)  We have
derived radial profiles of $\Sigma_*$ using $K$ or $I$-band profiles
with a mass-to-light ($M/L$) ratio chosen to provide a maximal disk
contribution to the rotation curve, as discussed further in Paper II.
As shown in Figure~\ref{fig:elm}(a), a relation of the form
\begin{equation}
\sighi/\sightwo \propto P^{-0.8}
\label{presfit}
\end{equation}
appears to match the data reasonably well.  For comparison,
\citet{Elm:93} predicts a very sensitive dependence of the molecular
fraction $f_{\rm mol}$ on pressure: $f_{\rm mol} \propto
P^{2.2}j^{-1}$, where $j$ is the volume emissivity of dissociating
radiation.  For $j \propto \Sigma_* \propto \siggas$, this leads to
$f_{\rm mol} \propto P^{1.7}$, which is clearly steeper than what is
observed within our galaxies [Fig.~\ref{fig:elm}(b)].  However, his
parametrization is intended for use in the low $f_{\rm mol}$ regime,
which is more appropriate for the Milky Way than for H$_2$-dominated
galaxies; indeed the match to the Galactic data of \citet{Dame:93b} is
reasonably good [Fig.~\ref{fig:elm}(b)].

Residual offsets between galaxies in Fig.~\ref{fig:elm}(a)
may result from uncertainties in the adopted $M/L$ ratio, or could be
related to additional physical parameters, such as metallicity.  
Specifically, we find that galaxies with oxygen abundances
significantly above solar in their inner ($R<200\arcsec$) disks (NGC
4321, 4501, 5055, and 5457) display smaller values of \sighi/\sightwo.
Higher metallicity is expected to favor the formation of H$_2$, since
dust attenuates the dissociating UV radiation field and provides a
surface on which to form H$_2$.  The importance of metallicity in the
\HI-H$_2$ balance is discussed further by \citet{Rana:86} and
\citet{Honma:95}.  However, we caution that the use of CO to trace
H$_2$ may also introduce a metallicity dependence in our derived 
value of \sighi/\sightwo.

Thus, while the range in $f_{\rm mol}$ probed by our observations is
fairly small, the trend in the \HI/H$_2$ ratio with radius suggests
that pressure is playing the dominant role, with metallicity as a
secondary factor.  This is supported by observations indicating that
the global $M_{\rm HI}/M_{\rm H_2}$ ratio increases from early-type to
late-type galaxies \citep{Young:89a,Casoli:98}, since early-type
galaxies generally have higher mass surface densities (and
metallicities).  Moreover, the tendency for recent star formation to
be correlated with the distribution of stellar light, noted for
example by \citet{Ryder:94} and \citet{Hunter:98}, can be explained if
stars currently dominate the pressure which controls H$_2$ and star
formation.  Further observations spanning a wider range of $f_{\rm
mol}$ would be useful to verify the importance of pressure in
determining the \HI/H$_2$ balance.


\section{Summary and Conclusions}\label{conc}

We have investigated the relationship between gas content and star
formation rate, commonly termed the star formation law, in seven
nearby galaxies using CO, HI, and H$\alpha$ images at resolutions of
$\sim$20\arcsec\ or higher.  An important advance is the recent
availability of CO data that combines the resolution of an
interferometer with the sensitivity to large-scale structure afforded
by a single-dish telescope.  Although our sample is biased toward
luminous, molecule-rich galaxies, none of the galaxies display
signs of strong interactions.

We find that the correlation of the azimuthally averaged SFR surface
density \sigsfr\ with \sighi\ is much poorer than with \sightwo,
contrary to the results of studies based on integrated fluxes.
Whereas a roughly linear relation exists between \sigsfr\ and
\sightwo, consistent with a constant star formation efficiency for the
molecular gas, \sighi\ tends to reach a maximum value of $\sim$10
\Msol\ pc$^{-2}$, or even decline in regions of high SFR.  Thus the
star-forming gas in these galaxies exists predominantly in molecular form,
although the situation may be different in low-mass galaxies.

Despite the poor correlation of \sigsfr\ with \sighi, the dominance of
H$_2$ in these galaxies leads to a strong correlation between \sigsfr\
and total gas surface density \siggas\ when azimuthally averaged, in
rough agreement with the global Schmidt law derived by K98.  The index
of the power law is given by $n \approx 1.1$ if the extinction at
H$\alpha$ is assumed to be uniform with radius, but steepens to $n
\approx 1.7$ if the mean extinction is assumed to vary according to
the gas surface density.  A star formation law of the form $\sigsfr
\propto \siggas\Omega$ is also consistent with the data, but only if
the $N_{\rm H}$-dependent extinction corrections are applied.  We
suggest that the observed Schmidt law results from an effective
dependence of the molecular fraction on \siggas\ by way of the
interstellar pressure, coupled with a roughly linear relation between
\sigsfr\ and \sightwo.

The gravitational instability parameter $Q_g$ is found to be $\sim 1$
in several galaxies but is clearly $>1$ in the two galaxies with the
smallest gas fractions, NGC 4736 and 5457.  We suggest that $Q_g$ is
best considered a measure of the gas fraction, and that when
discussing a range of galaxy types, the contribution of the stellar
component to the disk instability should not be neglected.  While a
combined (gaseous and stellar) instability may ultimately provide a
better description of star formation thresholds, alternative
explanations that attribute such thresholds to the inability to form
cold, dense clouds remain attractive as well.

Finally, over the limited range in radius where our CO and \HI\ data
overlap, we find that the relation $\sighi/\sightwo \propto R^{1.5}$
holds remarkably well within galaxies, although the proportionality
constant varies among galaxies.  We show that such a relation supports
the view that the interstellar pressure plays the dominant role in
determining the \HI-H$_2$ balance.

These results have demonstrated the importance of including the
molecular component in studying the star formation law.  Moreover,
considering just the total gas density rather than the \HI\ and H$_2$
surface densities separately may obscure underlying physical processes
that are essential to star formation.  In a future study, we will
develop these ideas using the much larger CO database provided by the
BIMA SONG project, supplemented by archival and recently obtained \HI\
synthesis imaging.  We also emphasize that significant advances will
be made possible by improved measurements of extinction within
galaxies to enable more accurate estimates of the SFR, and by studies of
the vertical structure of edge-on galaxies, which will provide further
insight into the \HI-H$_2$ transition and its relationship to the ISM
pressure.

\acknowledgements

We thank our collaborators on BIMA SONG, particularly D. Bock,
T. Helfer, M. Regan, K. Sheth, M. Thornley, and S. Vogel, for their
role in conducting many of the observations used in this study, and
for extensive discussions and suggestions.  C. McKee provided useful
comments to an earlier draft of this paper, and suggested a more
careful treatment of extinction.  Constructive comments by W. J. G. de
Blok, B. Elmegreen, S. Ryder, and the referree, C. Wilson, are also
appreciated.  We are indebted to A. Bosma, R. Braun, V. Cayatte,
R. Kennicutt, J. Knapen, C. Mundell, A. Thean, J. van Gorkom, and
L. van Zee for providing us with data for this study.  This research
is based on the Ph.D. thesis of T.W. and has been supported by
National Science Foundation grants AST 96-13998 and 99-81308 to the
U.C. Berkeley Radio Astronomy Laboratory, and a Bolton Fellowship from
the Australia Telescope National Facility.  Finally, this project
would have been much more difficult without the convenience provided
by NASA's Astrophysics Data System (ADS) Abstract Service and the
NASA/IPAC Extragalactic Database (NED), which is gratefully
acknowledged.

\bibliographystyle{apj}
\bibliography{thesis}

\end{document}